\documentclass[letter,referee]{aa}

\pdfoutput=1

\usepackage{ifdraft} \usepackage{ifpdf}

\usepackage[latin1]{inputenc} \usepackage[english]{babel}
\usepackage{natbib}

\usepackage{times} \usepackage{color} \usepackage{subfigure}
\usepackage{multirow} \usepackage{amsmath} \usepackage{dsfont}
\usepackage{xspace} \usepackage{algorithm} \usepackage{algpseudocode}

\usepackage[normalem]{ulem} 

\ifpdf%
\usepackage[pdftex,final]{graphicx}
\DeclareGraphicsExtensions{.pdf,.jpeg,.png}%
\else%
\usepackage[dvips,final]{graphicx} \usepackage{psfrag}
\DeclareGraphicsExtensions{.eps}%
\fi

\definecolor{rouge}{rgb}{1,0,0} \definecolor{vert}{rgb}{0,1,0}
\definecolor{bleu}{rgb}{0,0,1} \definecolor{rose}{rgb}{1,0,1}

 \def\eR{\mathds{R}} 
\def\ie{i.e.}  \def\eg{e.g.}  \def\cf{cf}

\def\XS{\xspace}
\DeclareMathAlphabet{\mathb}{OML}{cmm}{b}{it}
\def\sbm#1{\ensuremath{\mathb{#1}}}                
\def\sdm#1{\ensuremath{\mathrm{#1}}}               
\def\scu#1{\ensuremath{\mathcal{#1\XS}}}           

\def\Db{{\sbm{D}}\XS}  
  
  \def\fb{{\sbm{f}}\XS}
  
\def\Hb{{\sbm{H}}\XS}

  \def\nb{{\sbm{n}}\XS}
  \def\ob{{\sbm{o}}\XS}
\def\Pb{{\sbm{P}}\XS}  \def\pb{{\sbm{p}}\XS}

\def\Sb{{\sbm{S}}\XS}  \def\sb{{\sbm{s}}\XS}
\def\Tb{{\sbm{T}}\XS}  
  
  \def\vb{{\sbm{v}}\XS}
  
  \def\xb{{\sbm{x}}\XS}
  \def\yb{{\sbm{y}}\XS}
  \def\zb{{\sbm{z}}\XS}

\def\Ec{{\scu{E}}\XS}

\def\Rc{{\scu{R}}\XS}

\def\Xc{{\scu{X}}\XS}

  \def\bD{{\sdm{b}}\XS}
  \def\cD{{\sdm{c}}\XS}
  \def\dD{{\sdm{d}}\XS}

  \def\oD{{\sdm{o}}\XS}
  \def\pD{{\sdm{p}}\XS}

  \def\sD{{\sdm{s}}\XS}
  \def\tD{{\sdm{t}}\XS}

\def\Sv{{\sbv{S}}\XS}

\def\Dbb{{\sbl{D}}\XS}

\def\eR{\Rbb}

  \def\eg{{\textgoth{e}}\XS}
  
\def\XS{\xspace}

\RequirePackage{ifthen}
\RequirePackage{calc}

\newcommand{\taille}[1][\scad]{%
\ifthenelse{#1 = -5}{}{}%
\ifthenelse{#1 = -4}{\tiny}{}%
\ifthenelse{#1 = -3}{\scriptsize}{}%
\ifthenelse{#1 = -2}{\footnotesize}{}%
\ifthenelse{#1 = -1}{\small}{}%
\ifthenelse{#1 = 0}{\normalsize}{}%
\ifthenelse{#1 = 1}{\large}{}%
\ifthenelse{#1 = 2}{\Large}{}%
\ifthenelse{#1 = 3}{\LARGE}{}%
\ifthenelse{#1 = 4}{\huge}{}%
\ifthenelse{#1 = 5}{\Huge}{}}

\newboolean{@serif}
\setboolean{@serif}{true}
\def\scad{-5} 

\newcounter{taille}
\newcommand{\sca}[2][\scad]{\setcounter{taille}{#1}%
  \ifthenelse{\boolean{@serif}}
  {{\taille[\thetaille]\textsc{#2}}}
  {\setcounter{taille}{\value{taille}-1}{\uppercase{\taille[\thetaille]#2}}}}

\def\rem#1{}                    

\def\apost{\textit{a posteriori}\XS}

\def\post{\textit{posterior}\XS}

\def\aprio{\textit{a priori}\XS}

\def\cf{\textit{cf.}\XS}

\def\eg{\textit{e.g.,}\XS}

\def\ie{\textit{i.e.,}\XS}

\def\Toeplitz{T{\oe}plitz\XS}

\def\vs{\textit{vs}\XS}

\def\wrt{w.r.t.\XS}

\newcommand{\cnrs}[1][\scad]{\sca[#1]{cnrs}\XS}

\def\acc#1{\left\{#1\right\}}              
\def\cro#1{\left[#1\right]}

\def\Exp#1{\exp\cro{#1}}

 \def\T{^\tD} \def\+{^\dagger}

\def\nequiv{\not\kern-.05em\equiv}
\def\egal{\kern-.5em=\kern-.5em}        
\def\propt{\kern-.2em\propto\kern-.2em} 
\def\wh#1{\widehat{#1}}                 

\def\argmin{\mathop{\mathrm{arg\,min}}} 
\def\froc#1#2{{#1/#2}}                  

\def\intdouble{\int\kern-0.3em\int}
\def\inttriple{\int\kern-0.3em\int\kern-0.3em\int}

\def\rond#1{\overset{\kern-0.33em~_\circ}{#1}}
\def\rondit[#1]#2{\overset{\kern#1~_\circ}{#2}}

\def\erf{\mathrm{erf}}

\def\erfcx{\mathrm{erfcx}}

\def\conv{\star}
\def\eR{\mathds{R}}

\def\Hcal{\mathcal{H}}

\def\alm{\alpha_{lm}}
\def\blm{\beta_{lm}}
\def\ylm{y_{lm}}
\def\ylmn{y_{lm}^n}

\def\fa{f_{\alpha}}
\def\fb{f_{\beta}}

\def\fe{f_\sD}

\def\pa{p_{\alpha}}
\def\pb{p_{\beta}}
\def\va{v_{\alpha}}
\def\vb{v_{\beta}}

\def\ca{c_{\alpha}}

\def\cbe{c_{\beta}}

\def\Ciel{\phi}

\def\SigCornet{\Ciel_{\cD}}
\def\SigCornet{\Ciel}

\def\RepFiltre{h_{\fD}}
\def\RepFiltre{h_k} 

\def\RepBolo{h_{\bD}}
\def\RepOpt{h_{\oD}}

\def\Ta{T_{\alpha}}
\def\Tb{T_{\beta}}
\def\Ta{\delta_{\alpha}}
\def\Tb{\delta_{\beta}}
\def\Tab{\delta_{\alpha/\beta}}
\def\Te{T_{\sD}}
\def\Fe{F_{\sD}}

\def\cstP{\bar P}

\def\cstT{\bar T}

\def\Vp{V_{\pD}}

\def\Rc{R_{\cD}}

\def\T0{T_{0}}
\def\G0{G_{0}}

\def\sa{\sigma_{\alpha}}
\def\sb{\sigma_{\beta}}
\def\sab{\sigma_{\alpha/\beta}}
\def\Sa{\Sigma_{\alpha}}
\def\Sb{\Sigma_{\beta}}
\def\Sab{\Sigma_{\alpha/\beta}}

\def\oa{o_{\alpha}}
\def\obeta{o_{\beta}}

\def\Smc{\sigma_{\mD\cD}}
\def\Smc{\sigma_{\oD}} 

\def\Sv{\Sigma_v}

\def\hcal{\mathcal{H}}

\def\Hbc{\Hb_{\cD}}

\def\hyperR{\mu} 

\def\xij{x_{ij}}

\def\xijstar{x_{ij}^{*}}
\def\xijprime{x_{i'j'}}

\def\Dba{\Db_{\alpha}}
\def\Dbb{\Db_{\beta}}

  \def\ybB{\yb_{B}}
\def\ybB1{\yb_{B+1}}

\def\Ciel{\Xc} \def\sanepic{SANEPIC\XS} \def\MADmap{MADmap\XS}
\def\ie{i.e.\XS} \def\eg{e.g.\XS} \def\cf{cf.\XS}

\begin{document}

\title{Super-resolution in map-making based on
  a physical instrument model and regularized inversion.\\
  Application to SPIRE/Herschel.}

\titlerunning{Super-resolution: instrument model and regularized
  inversion.}

\author{F. Orieux\inst{1} \and J.-F. Giovannelli\inst{1,2} \and
  T. Rodet\inst{1} \and A. Abergel\inst{3} \and H. Ayasso\inst{3}\and M. Husson\inst{3}}

\offprints{F. Orieux}

\institute{Laboratoire des Signaux et Syst\`emes
  (\cnrs~--~Sup\'elec~--~Univ. Paris-Sud 11), Plateau de Moulon,
  91\,192 Gif-sur-Yvette, France.  \email{orieux,rodet@lss.supelec.fr}
  \and Univ. Bordeaux, IMS, UMR 5218, F-33400 Talence,
  France. \email{Giova@IMS-Bordeaux.fr} \and Institut d'Astrophysique
  Spatiale (\cnrs~--~Univ. Paris-Sud 11), 91\,405 Orsay,
  France. \email{abergel@ias.u-psud.fr}}

\date{Accepted 5 December 2011}

\abstract{~---~ We investigate super-resolution methods for image
  reconstruction from data provided by a family of scanning
  instruments like the Herschel observatory. To do this, we
  constructed a model of the instrument that faithfully reflects the
  physical reality, accurately taking the acquisition process into
  account to explain the data in a reliable manner. The inversion, \ie
  the image reconstruction process, is based on a linear approach
  resulting from a quadratic regularized criterion and numerical
  optimization tools. The application concerns the reconstruction of
  maps for the SPIRE instrument of the Herschel observatory. The
  numerical evaluation uses simulated and real data to compare the
  standard tool (coaddition) and the proposed method. The inversion
  approach is capable to restore spatial frequencies over a bandwidth
  four times that possible with coaddition and thus to correctly show
  details invisible on standard maps. The approach is also applied to
  real data with significant improvement in spatial resolution.

  \keywords{Techniques: image processing, data acquisition modelling,
    inverse problem, deconvolution, super-resolution, regularization,
    image processing. Methods: statistical, numerical. Astronomical
    instrumentation, methods and techniques} }

\maketitle

\section{Introduction}
\label{sec:intro}

Map making is a critical step in the processing of astronomical data
of various imaging instruments (interferometers, telescopes,
spectro-imager, etc.), and two recent special issues have been
published~\citep{IEEE-SPM-ASTRO10,IEEE-SP-ASTRO08} on the
subject. Because the observed sky may contain structures of various
scales, from extended emission to point sources, the challenge is to
design reconstruction methods that deliver maps that are
photometrically valid for the broadest range of spatial frequencies.

For long-wavelength instruments, be they ground based (SCUBA/JCMT,
LABOCA/APEX, etc.), on-board balloons (Archeops, BLAST, etc.) or space
borne (IRAS, ISO, Spitzer, WMAP, Planck, Herschel, etc.), the task is
especially challenging for two reasons. First, the physical resolution
is poor at these wavelengths. Second, the distance between the
detectors of these instruments generally prevents a proper sampling of
the focal plane, given the maximum spatial frequency allowed by the
optical response.  Therefore, specific scanning strategies have to be
defined, which depend on the detector positions and need to be closely
combined with a well designed image reconstruction method.

The Herschel Space Observatory \citep{Pilbratt10} was launched in May
2009 together with the Planck satellite. It continuously covers the
55--672\,$\mu$m spectral range with its very high spectral resolution
spectrometer HIFI \citep{deGraauw10} and its two photometers / medium
resolution spectrometers PACS \citep{Poglitsch10} and SPIRE
\citep{Griffin10}. With a 3.5\,m primary mirror, Herschel is the
largest space telescope launched to date. In order to take full
advantage of the telescope size, the accurate representation and
processing of the highest spatial frequencies presents a particular
challenge. To this end, two step-by-step photometer pipelines have
been developed by the instrument consortia by \citep{Griffin08} for
SPIRE and by \citep{Wieprecht09} for PACS: they produce flux density
timelines corrected for various effects, calibrated and associated
with sky coordinates (level-1 products), then produce maps (level-2
products). An important step is the correction of the $1/f$ noise
components, which can be correlated or uncorrelated between
bolometers. For SPIRE, a significant fraction of the correlated
component is processed using the signals delivered by blind
bolometers. For PACS, it is currently processed using different kinds
of filtering. The glitches caused by the deposit of thermal energy by
ionizing cosmic radiation are flagged or corrected. Finally, the
timeline outputs can be simply coadded on a spatial grid to produce
``naive maps'', with a rounded pointing approximation. Maximum
likelihood approaches with the same coaddition algorithm, namely
\MADmap~\citep{Cantalupo10} and \sanepic~\citep{Patanchon07} have also
been developed to compute maps, using the spatial redundancy to
correct for the $1/f$ noise.

There are several drawbacks to these pipelines.  First, because they
work on a step-by-step basis, the performance of the whole process is
limited by the step with the worst performance. Second, the ultimate
performance of one step is out of reach because only a reduced part of
the available information is handed over from the previous step. This
mean that better perfomances can be achieved by a more global
approach.  More important, the instrument and the telescope properties
(mainly the diffraction) are not taken into account, which is why the
maps are unavoidably smoothed by the Point Spread Function (PSF),
whereas the scanning strategy allows higher spatial frequencies to be
indirectly observed.

To overcome these limitations, we resorted to an inverse problem
approach \citep{idier08} that is based on an instrument model and an
inverson method.
\begin{itemize}
\item It requires an instrument model that faithfully reflects the
  physical reality to distinguish in the observations between what is
  caused by the instrument and what is due to the actual sky. To this
  end, an important contribution of our paper is an analytical
  instrument model based on a physical description of the phenomena as
  functions of continuous variables. Moreover, it includes scanning
  strategy, mirror, wavelength filter, feedhorns, bolometers and
  read-out electronics. The point for the resolution is the
  following. On the one hand, the field of view is covered by
  hexagonally packed feedhorn-coupled bolometers, the sampling period
  is twice the PSF width, which potentially lead to spectral aliasing
  for wide-band objects. On the other hand, the scanning strategy with
  a pointing increment lower than the bolometer spacing introduces an
  higher equivalent sampling frequency. Therefore, it is crucial to
  properly take into account the scanning strategy and the whole
  instrument including irregular sampling to obtain super-resolution
  (see also the analysis in~\citep{orieux09a}). To the best of our
  knowledge, a physical model of the instrument this accurate has
  never been used in a map making method.
\item The inversion of our instrument model constitutes an ill-posed
  problem~\citep{idier08} because of the deficit of available
  information induced by convolution with the instrument PSF.
  Moreover, the ill-posedness becomes all the more marked as the
  resolution requirement increases. The inversion methods must
  therefore exploit other information by regularization to compensate
  for the deficits in the observations. Each reconstruction method is
  therefore specialized for a certain class of maps (point sources,
  diffuse emission, superposition of the two, etc.) according to the
  information that is included. From this standpoint, the present
  paper is essentially devoted to extended emission.

  The method is linear \wrt the data for the sake of simplicity and
  computational burden.  From the methodological point of view, it is
  built within the framework of quadratic
  regularization~\citep{Tikhonov77,Andrews77}.  It relies on a
  criterion involving an adequation measure (observed data vs model
  output) and a spatial smoothness measure. From a numerical
  standpoint, we resort to a gradient-based optimisation
  algorithm~\citep{Nocedal00} to compute the map.

  Moreover, in as much as it relies on two sources of information, the
  method is based on a trade-off tuned by means of an
  hyperparameter. It is empirically set in the present paper and work
  in progress, based on \citep{Robert04} and \citep{orieux10a}, is
  devoted to the question of the hyperparameter and instrument
  parameter auto-calibration (myopic and unsupervised inversion).
\end{itemize}

One of the most striking results of our research is the correct
restoration of small-scale structures (wide-band), which are not
detectable on naive maps. This result is reached thanks to the
developed instrument model together with the used inversion: they
jointly enable the proposed method to reduce instrument effects,
overtake instrument limitations and restore high spatial frequencies.

In the image processing community, these capabilities are referred to
as super-resolution~\citep{Park03} and we were partly inspired by
recent developments in this field. They are usually based on various
(scene or camera) motion or scanning strategy. Some of them account
for possible rotation \citep{Elad99} and/or a magnifying factor
\citep{rochefort06}. Other approaches introduce an edge-preserving
prior \citep{Nguyen01,Woods06}. These works rely on the description of
the unknown object as a function of continuous variables that is
decomposed on pixel indicator basis \citep{Hardie97,Patti97}, on a
truncated discrete Fourier basis \citep{Vandewalle07}, on a family of
regularly shifted Gaussian functions \citep{rodet08}, or spline family
\citep{rochefort06}.  Other approaches have been proposed, based on
shift-and-add step~\citep{farsiu04} followed by a deconvolution
step~\citep{Molina89}. Finally, several contributions are devoted to
the performance of super-resolution approaches
\citep{champagnat09,orieux09a}.

The paper is organized as follows. The instrument model describing the
relationship between the measured data and the unknown sky is
presented in Section\,\ref{sec:le-modele-direct}.
Section\,\ref{sec:inversion} details the method that we propose to
inverse the data and compute high-resolution maps. Finally,
Section\,\ref{sec:resultats} presents experimental results, first on
simulated data (Section\,\ref{sec:donnees-sim}), then on real data
(Section\,~\ref{sec:donnees-reelles}).

\section{Instrument model}
\label{sec:le-modele-direct}

The prime objective of the instrument model is the reproduction of
observed data taking into account the physics of the acquisition. In
addition, the reconstruction algorithms use the instrument model many
times, it is therefore necessary to adopt hypotheses and
approximations to the reduce computational burden. This is one of the
differences to a simulator~\citep{sibthorpe09,sibthorpe06}, which is
designed to be run once per data set.

\subsection{Physical models}
\label{sec:apropr-des-model}

\subsubsection{Mode of observation}
\label{sec:mode-dobservation}

The sky, $\Ciel(\alpha,\beta,\lambda)$, is characterized by two
spatial dimensions $(\alpha,\beta)$ and one spectral dimension
$\lambda$.  To model telescope translations, we used a frame of
reference defined by the instrument. The map at the input is
time-dependent and can be written
\begin{equation}
  \label{eq:1}
  \Ciel(\alpha,\beta,\lambda,t) = \Ciel\big(\alpha - \pa (t),\beta -
  \pb (t),\lambda\big) \,,
\end{equation}
where $\alpha$ and $\beta$ define the central angular position of the
observation and $(p_{\alpha}(t),p_{\beta}(t))$ the translations into
the two directions as a function of time $t$.

Here, we present only the \textit{``Large map''} protocol. Data were
acquired over a complete observation sequence composed of two almost
perpendicular directions and several scans back and forth for each of
the two directions. The pointing acceleration and deceleration phases
were not included in the zone of interest and there was no rotation
during the observation sequence. The pointing functions are therefore
written
\begin{equation}
  \label{eq:2}
  \pa(t) = \va t + \ca \qquad\text{and}\qquad \pb(t) = \vb t + \cbe
\end{equation}
for scanning at a constant velocity $(\va,\vb)$. The pointing accuracy
is of the order of a few seconds of arc. This protocol enables us to
introduce spatial redundancy, which is an essential element for the
reconstruction of a sky at a resolution greater than the detector
spatial sampling period~\citep{orieux09a,champagnat09}.

\subsubsection{Optics}
\label{sec:louverture}

The Herschel Telescope is a classical Cassegrain instrument with a
3.5\,m diameter primary mirror and a 308\,mm diameter secondary
mirror. The SPIRE photometer has three channels for a single field of
view. The light is split by a combination of dichroics and
flat-folding mirrors. The spectral channels are defined by a sequence
of metal mesh filters and the reflection/transmission edges of the
dichroics. They are centred at approximately 250, 350 and 500\,$\mu$m
(noted as PSW, PMW and PLW respectively). We assumed the overall
transmission curves of the wavelength filter $\RepFiltre(\lambda)$,
for $k=1,2,3$, as given by the SPIRE Observers' Manual (no analytical
form is available).

The three detector arrays contain 139 (250\,$\mu$m), 88 (350\,$\mu$m)
and 43 (500\,$\mu$m) bolometers, each coupled to the telescope beam
with hexagonally close-packed circular feedhorns.  The beam solid
angle is apodized by a bell-shaped weight whose width increases with
$\lambda$. Efforts have been made to correctly integrate the feedhorns
in the instrument model but the detailed coupling of feedhorns on
incoming radiation is, to the best of our knowledge~\citep{griffin02},
not fully understood at present.

Our final choice as an effective PSF for the telescope coupled with
feedhorns was a Gaussian shape $\RepOpt(\alpha, \beta, \lambda)$. This
choice has two advantages: (i)~it allows a closed equation for the
instrument model (see Sec.~\ref{sec:calcul-du-modele}), and (ii)~it
agrees with the response measured from observations of Neptune
(Griffin et al. 2010). As a first approach, we assumed isotropic
Gaussians with standard deviations $\Smc(\lambda) = c \lambda$
proportional to the wavelength since the width of the beam varies
almost linearly with the wavelength. The widths obtained are close to
the FWHM measured on the sky with 18.1\arcsec, 25.2\arcsec, and
36.9\arcsec at 250\,$\mu$m, 350\,$\mu$m and 500\,$\mu$m, respectively
(Griffin et al. 2010). The feedhorn diameter is $2F\lambda$, which
introduces a detector spatial sampling period of $2F\lambda$
(50\arcsec for the 350\,$\mu$m array, or equivalently with sampling
frequency $\fe \approx 0.02\,\mathrm{arcsecond}^{-1}$).

The output after each feedhorn is then written as a 2D convolution of
the input $\Ciel(\alpha,\beta, \lambda, t)$ and the effective PSF
$\RepOpt$ in addition to the $\RepFiltre$ wavelength filter
\begin{multline}
  \label{eq:3}
  \SigCornet_k^{lm}(\lambda, t) = \RepFiltre(\lambda) \iint
  \Ciel\left(\alpha,\beta,\lambda,t\right)\\ \RepOpt\left(\alpha - \alm,
    \beta - \blm, \lambda\right) \dD \alpha \dD \beta
\end{multline}
where $(\alm,\blm)$ is the direction pointed at by the feedhorn
$(l,m)$, for $l=1,\dots L$ and $m=1,\dots M$. The $k$ subscript can be
safely removed from $\SigCornet_k^{lm}$ since each spectral band is
processed separately. Finally, the optics was modelled as a linear
invariant system \wrt continuous variable.

\subsubsection{Bolometers}
\label{sec:les-bolomtres-1}

To set up the bolometer model, we took the thermal model of
~\citep{sudiwala02}, which was also used in the simulator developed
by~\citep{sibthorpe09}. Bolometers absorb the entire received
radiation
\begin{equation}
  \label{eq:4}
  P^{lm}(t) = \int_{\lambda} \SigCornet^{lm}(\lambda,t)\, \dD \lambda,
\end{equation}
and this power provides the system excitation. The temperature
$T^{lm}(t)$ determines the system output. The link between the input
$P(t)$ and the response $T(t)$ is described by the differential
equation deduced from a thermal balance,
\begin{equation*}
  C\frac{\dD T}{\dD t} - \frac{ R(T) \Vp^2}{\Rc^2} +
  \frac{\G0}{\T0^{\nu}(\nu + 1)} \left( T^{\nu + 1} - \T0^{\nu + 1}\right) = P,
\end{equation*}
where $C$ is the heat capacity of the bolometer, $R(T)$ is its
resistivity, $\T0$ is the temperature of the thermal bath, $\nu$ is a
physical parameter that depends on the bolometer, $\G0$ is the thermal
conductance (at temperature $\T0$) and $\Vp$ and $\Rc$ are the
polarization voltage and charge. No explicit solution of this equation
is available in the literature. Sudiwala's
approach~\citep{sudiwala02}, which we adopted here, is to linearize
this equation around an operating point $(\cstT,\cstP)$. In the
following, we consider only the variable part of the flux and exclude
the constant part that defines the operating point. All constants are
defined with respect to the operating point.

For SPIRE, most of the observations should be carried out in the
linear regime~\citep{griffin06, griffin07}. We therefore considered
that a development is sufficient to model the bolometer behaviour
correctly. Then, knowing the variations of the resistivity $R(T)$ with
temperature, it is possible to determine the tension at the
terminals. This first-order development models the bolometer as a
first-order, low-pass filter with an impulse response
\begin{equation}\label{eq:5}
  \RepBolo(t) = S \Exp{-t/\tau},
\end{equation}
where the gain $S$ and the time constant $\tau$ depend on the physical
parameters in the differential equation~\citep{sudiwala02}. The values
of these parameters are defined with respect to the operating point
and correspond to the official SPIRE
characteristics~\citep{griffin06,griffin07}. The output voltage around
the operating point can then be written as a function of the incident
flux,
\begin{equation}\label{eq:6}
  \ylm(t) = \int_{t'} \int_{\lambda} \SigCornet^{lm}(\lambda, t')
  \RepBolo(t'- t)\dD t' \, \dD \lambda \,.
\end{equation}

Finally, downstream, we have the read-out electronics, composed of
several stages (modulation, amplification, low-pass filter,
demodulation, quantification). However, except for the low-pass
filters, they seem to have negligible effects to the other elements
and are not included in our model. The equations are nevertheless
available~\citep{griffin07} and it is possible to integrate them into
the model.

The low-pass filters introduce a delay on the data with respect to the
telescope position along the scan. As a trade-off between model
accuracy and computational burden, we have chosen to model the
combination of the low-pass filter and the bolometer as a unique
first-order filter. The time constant\footnote{For the illustration on
  real data in Section~\ref{sec:donnees-reelles}, the correction of
  the low-pass filter was performed using the Herschel Interactive
  Processing Environment \citep{2010ASPC..434..139O}, and the time
  constant of the first-order low-pass filter was set to the time
  constant for the bolometer alone (5.7\,ms).} value ($0.2$ s) is
taken to be representative of the combination.

Finally, we accounted for regular time sampling that takes the values
at times $t = n\Te$ (with a sampling frequency $\Fe=1/\Te\approx
30$\,Hz) and then $\ylmn = \ylm(n\Te)$, for $n=1,\dots N$. Given the
scanning speed of $30\arcsec\text{s}^{-1}$ this induces a spatial
sampling period of $2\arcsec$ between two succesive time samples for
one bolometer, while the detector sampling period is $50\arcsec$ for
the 350\,$\mu$m array.

\subsubsection{Complete model equation}
\label{sec:equation-complete-du}

Adding these elements yields the equation of the acquisition chain.
For a spectral channel $k$, the time signal at the bolometer $(l,m)$
at time $n$ is
\begin{multline}
  \label{eq:8}
  \ylmn = \iint \RepFiltre(\lambda) \iint \Ciel\big(\alpha - \pa(t),
  \beta - \pb(t), \lambda \big) \\ \RepOpt(\alpha - \alm, \beta -
  \blm, \lambda) \,\dD\alpha \,\dD\beta~\RepBolo(n\Te -
  t)\,\dD\lambda\,\dD t.
\end{multline}
This equation introduces four integrals: two from the optics (spatial
convolution), one from the spectral integration, and one from the time
convolution. This is the fundamental equation of the instrument model
since it describes the data $\ylmn$ bolometer by bolometer at each
instant as a function of the sky $\Ciel(\alpha,\beta,\lambda)$. It
should be noted that this model includes the discretization process
(and possible aliasing) in the sense that the data $\ylmn$ is a
discret set and $\Ciel$ is a function of continuous variables.

\subsubsection{Super-resolution sky model}
\label{sec:decomp-du-ciel}

The model of the sky is an important element for the reconstruction
method. As stated in the introduction and presented in
Section~\ref{sec:mode-dobservation}, the sub-pixel scanning strategy
should allow for reverse aliasing and enable to estimate a
super-resolved sky~\citep{orieux09a}. The model of $\Ciel$ must
therefore be suitable for super-resolved reconstruction and, in
particular, allows a fine description of the physical reality and
integration with the instrument model.

Firstly, unlike conventional models of SPIRE
~\citep{sibthorpe09,Cantalupo10}, we considered the sky spectrum
within each channel. The emission observed by SPIRE is mainly caused
by thermal equilibrium (between emission and absorption of UV and
visible photons from incident radiation), and the intensities can be
written
\begin{equation}
  \label{eq_spec}
  I_{\lambda} = \tau_{\lambda_{0}} \times
  \left(\frac{\lambda}{\lambda_0}\right)^{-\beta}\times B_{\lambda}(T),
\end{equation}
where $\tau_{\lambda_0}$ is the optical depth at wavelength
$\lambda_0$, $\beta$ is the spectral index, $B_{\lambda}$ is the
Planck function, and $T$ the dust temperature. The SPIRE data alone do
not allow the proper measurement of the dust temperature (the
combination of SPIRE and PACS is mandatory, \citep{Abergel10}),
consequently we decided to exclude the dust temperature in our sky
model and work in the Rayleigh-Jeans approximation, so that $
B_{\lambda}(T)\propto\lambda^{-4}$. Moreover, we assumed $\beta=2$,
which is the ``standard'' value of the diffuse ISM (e.g.,
\citep{Boulanger1996}).  Finally, we have
\begin{equation}
  \label{Eq:CielLambda}
  \Ciel(\alpha,\beta,\lambda) = \lambda^{-\varrho} \Ciel(\alpha,\beta)
\end{equation}
with $\varrho = 6$. However, as we will see in Section
\ref{sec:calcul-du-modele}, the wavelength integration of the
acquisition model will be performed numerically. In other words, the
spectrum profile can be set adequately with the available knowledge of
the observed sky.

Secondly, $\Ciel(\alpha,\beta)$ was generated onto a family of
functions regularly shifted in space: $\psi_{ij}(\alpha,\beta) =
\psi(\alpha - i\Ta,\beta - j\Tb)$ where $\psi$ is an elementary
function and $(\Ta,\Tb)$ are the shifts between the $\psi_{ij}$ in
$(\alpha,\beta)$. We then obtain
\begin{equation}
  \label{Eq:DecompoCiel}
  \Ciel (\alpha,\beta) = \sum_{ij} \xij \, \psi(\alpha - i\Ta,\beta - j\Tb),
\end{equation}
where $\psi$ is the generating function and $x_{ij}$ are the
coefficients. In addition, the axis $\alpha$ is determined by the
first scan of the observation.

One of the purposes of this model is to describe maps with arbitrary
fine details, that is to say, arbitrary wide band. Within this model,
the usual function $\psi$ is the cardinal sine with shift and width
adapted to the target band. However, cardinal sines require analytical
calculations that cannot be made explicit.  To lighten the
computational burden, we chose Gaussian $\psi$ functions. These
functions are parametrized by their spatial shifts $(\Ta,\Tb)$ and
their standard deviations $(\sa,\sb)$. The parameters $(\Ta,\Tb)$ are
chosen to be equal to the inverse of the target band width as for the
cardinal sines.  In the numerical processing of
Section~\ref{sec:resultats}, the values of $(\Ta,\Tb)$ are equal to
the sampling period of 2\arcsec{} induced by the scanning scheme of
the \textit{``Large map''} protocol~\citep{Orieux09b}. For the
Gaussian function width parameters $(\sa,\sb)$, we determined the
value that minimizes the difference between the width at half-maximum
of the cardinal sine and the Gaussian: $\sab \approx 0.6\,\Tab$ in a
similar manner in $\alpha$ and $\beta$.

\subsection{Explicit calculation of the acquisition model}
\label{sec:calcul-du-modele}

Given the linearity of the instrument model~\eqref{eq:8} and the sky
model~\eqref{Eq:CielLambda}-\eqref{Eq:DecompoCiel}, the instrument
output for a given sky is
\begin{multline}
  \label{eq:11}
  \ylmn = \sum_{ij} \xij \, \int \lambda^{-\varrho}
  \RepFiltre(\lambda) \iint \RepOpt(\alpha - \alm, \beta - \blm,
  \lambda) \\ \psi\big(\alpha-i\Ta-\pa(t),\beta-j\Tb-\pb(t)\big)
  \,\dD\alpha \dD\beta ~ \RepBolo(n\Te - t)\,\dD t\,\dD\lambda \,.
\end{multline}
Thus, to obtain the contribution of a sky coefficient $\xij$ to a data
item $\ylmn$, it is necessary to calculate four integrals, whose
discretization by brute force would result in time-consuming numerical
computations.

Concerning the optics, the convolution of the function $\psi$ with the
optical response $\RepOpt$ appears in Eq.~\eqref{eq:11} and, because
these are Gaussians, the convolution can be made explicit
\begin{multline}
  \label{eq:13}
  \iint \psi\big(\alpha - i\Ta - \pa,\beta - j\Tb - \pb\big)
  \RepOpt(\alpha - \alm, \beta - \blm) \,\dD\alpha \dD\beta\\
  \propto \Exp{-\frac{(\pa + i\Ta - \alm)^2}{2\Sa^2} - \frac{(\pb +
      j\Tb -\blm)^2}{2\Sb^2}}
\end{multline}
with, in a similar manner in $\alpha$ and $\beta$: $\Sab^2 = \sab^2 +
\Smc^2$.

For the integral over time, only the constant velocity phases can be
explicitly described for the \textit{``Large map''} protocol.  To
integrate over time in~\eqref{eq:11}, we used the expressions
of~\eqref{eq:2} for $\pa(t)$ and $\pb(t)$, which gives
\begin{multline}
  \label{eq:15}
  \sum_{ij} \xij \, \int \lambda^{-\varrho}\RepFiltre(\lambda) \int
  \Exp{- \frac{(\va t + \ca + i\Ta - \alm)^2}{2\Sa^2}}\\ \Exp{-
    \frac{(\vb t + \cbe + j\Tb -\blm)^2}{2\Sb^2}} \RepBolo(n\Te - t)
  \,\dD t \,\dD\lambda \,.
\end{multline}
It can be shown that explicit integration can be performed by
including the Gaussians and the bolometer response (see details of the
calculations in appendix~\ref{Ann:ModelCalculExplicite}, and the model
becomes
\begin{multline}
  \label{eq:17}
  \ylmn = \frac{S}{2 \sqrt{2\pi}\Sv} \sum_{ij} \xij \, \int
  \lambda^{-\varrho} \RepFiltre(\lambda) \\ \erfcx \left(
    \frac{\Sa\Sb}{\sqrt{2}\tau \Sv} -\frac{\Sb\va(\oa + n\Te
      \va)}{\sqrt{2}\Sa\Sv} - \frac{\Sa\vb(\obeta +
      n\Te\vb)}{\sqrt{2}\Sb\Sv} \right) \\ \Exp{-\frac{\left( \oa +
        n\Te\va \right)^2}{2 \Sa^2} - \frac{\left( \obeta +
        n\Te\vb\right)^2}{2 \Sb^2}} \,\dD\lambda.
\end{multline}
In this equation, the angles $\oa$ and $\obeta$ are defined by $\oa =
\ca + i\Ta - \alm$ and $\obeta = \cbe + j\Tb - \blm$. Moreover, $\Sv^2
= \Sb^2 \va^2 + \Sa^2\vb^2$.

The data point $\ylmn$ does not depend directly on the ``scanning
time'' $t$ because it is integrated.  It depends on time through the
sampling instant $n$ occuring \emph{only} in $n\Te
v_{\{\alpha,\beta\}}$, \ie a distance.  In practice, the sampling
period along the scans $\Te \sqrt{\va^2 + \vb^2}$ is much shorter than
the sampling period of the detecteor array. Thus, this properly
modelled sampling scheme is a key element for reconstruction with an
higher resolution.

In addition, the time constant of the bolometer and the electronics
$\tau$ appears only in the argument of the function $\erfcx$. It is
consequently through this function that the bolometers and the
electronics influence the \textit{spatial} response.

The dependence on the wavelength through $\Smc(\lambda)$ precludes
explicit integration with respect to $\lambda$. However, the integral
depends neither on the data nor on the unknown object but only on the
protocol. Accordingly, for a given protocol, these integrals can be
calculated once and for all. Finally, the work described above allow
three explicit intergration of the four introduced by the initial model.

Eq.~\eqref{eq:17} models the acquisition of the data item $\ylmn$ at
time $n$ by bolometer $(l,m)$ from the coefficients $\xij$. These
equations can be written
\begin{equation}
  \label{eq:18}
  \ylmn = \sum_{ij} \xij \, \hcal_{lmn}(\psi_{ij}),
\end{equation}
where $\hcal$ is calculated from Eq.~\eqref{eq:17}. The model is
linear and we can therefore write
\begin{equation}
  \label{Eq:ModelInstruDiscretDiscret}
  \yb = \Hb \xb,
\end{equation}
where $\yb$ and $\xb$ are vectors of size $LMN$ and $IJ$, and $\Hb$ is
a $LMN \times IJ$ matrix, each row of which can be deduced
from~\eqref{eq:17} by varying $l,m,n$ for fixed $i,j$.

\subsection{Invariant structure}
\label{Sec:ModelFacto}

Initially, the physical model~\eqref{eq:8} is based on convolutive (so
invariant) transforms \wrt continuous variables. However, the
discretization operation is inhomogeneous, consequently the invariance
property does not hold anymore, which lead to long computational
time. Nevertheless, the trace of this initial invariance can still be
perceived because $\Hb$ is a \emph{sum} of terms at different spatial
positions of the Gaussians (\cf Eq.~(\ref{eq:17})). Because the
problem is now discretized, we seek to bring out an invariance by
\emph{quantified} shifts in
\begin{equation*}
  \ca + i \Ta + n\Te\va - \alm
\end{equation*}
for the $\alpha$ direction, and similarly for $\beta$. With the
approximation that the terms are multiples of a common factor
$\Delta_\alpha$, the continuous shift is
\begin{gather*}
  \oa + n\Te\va = (n_0 + i n_1 + nn_2 - ln_3 - mn_4) \Delta_\alpha \,.
\end{gather*}
The pointed directions are therefore rounded to match the grid of
sky. The \MADmap and \sanepic methods use this idea but there is a
notable difference: they perform the operation on a low-resolution
grid, which limits the map resolution. In contrast, the developments
proposed here exploit the idea of a high-resolution grid, enabling
super-resolution reconstruction. By acting in the same way in the
$\beta$ direction, we have
\begin{multline}
  \ylmn = \sum_{ij} \xij \, \Hcal\Big((n_0 + i n_1 + nn_2 - ln_3 -
  mn_4)\Delta_\alpha , \\ (n_0' + i n_1' + nn_2' - ln_3 - mn_4) \Delta_\beta \Big )
\end{multline}
and by computing the discrete convolution, we obtain
\begin{equation}
  \label{eq:25}
  \tilde y(i',j') = \sum_{ij} \xij \, \Hcal\Big((i - i')
  \Delta_\alpha, (j - j') \Delta_\beta \Big ) \,.
\end{equation}
Therefore, $\ylmn = \tilde y(i',j')$ if, and only if,
\begin{gather}
  i - i' = in_1 + ln_3 + mn_4 - nn_2 - n_0 \label{eq:26}\\
  j - j' = jn_1' + ln_3 + mn_4 - nn_2' - n_0' \,.\label{eq:27}
\end{gather}
In these conditions, the data $\yb$, for a given scanning direction
are computed by discrete convolution~\eqref{eq:25} followed by
(inhomogeneous) down-sampling defined by~\eqref{eq:26}-\eqref{eq:27},
which is much more efficient than using a generic linear
model~\eqref{Eq:ModelInstruDiscretDiscret}. First of all, the
decomposition by convolution then decimation is faster than the direct
calculation and, what is more, the convolution can be computed by
FFT. Finally, given that only the impulse response is necessary, there
is no need to compute and store all elements of the matrix.

\begin{figure}[htbp]
  \centering {\ifdraft{
      \includegraphics[width=0.3\textwidth]{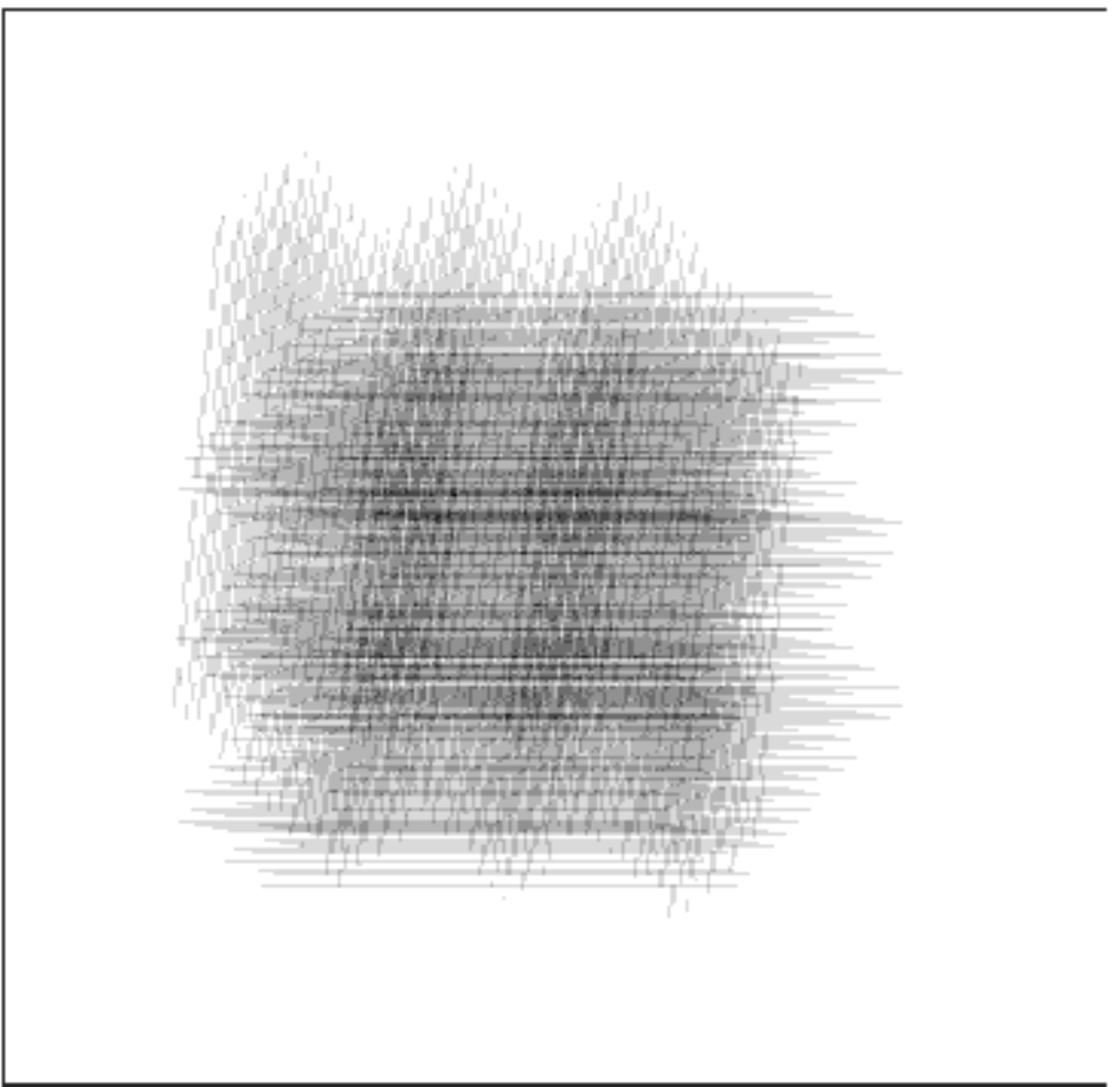}}{
      \includegraphics[width=0.3\textwidth]{./figs1/redondance250}}}
  ~{\ifdraft{
      \includegraphics[width=0.1\textwidth]{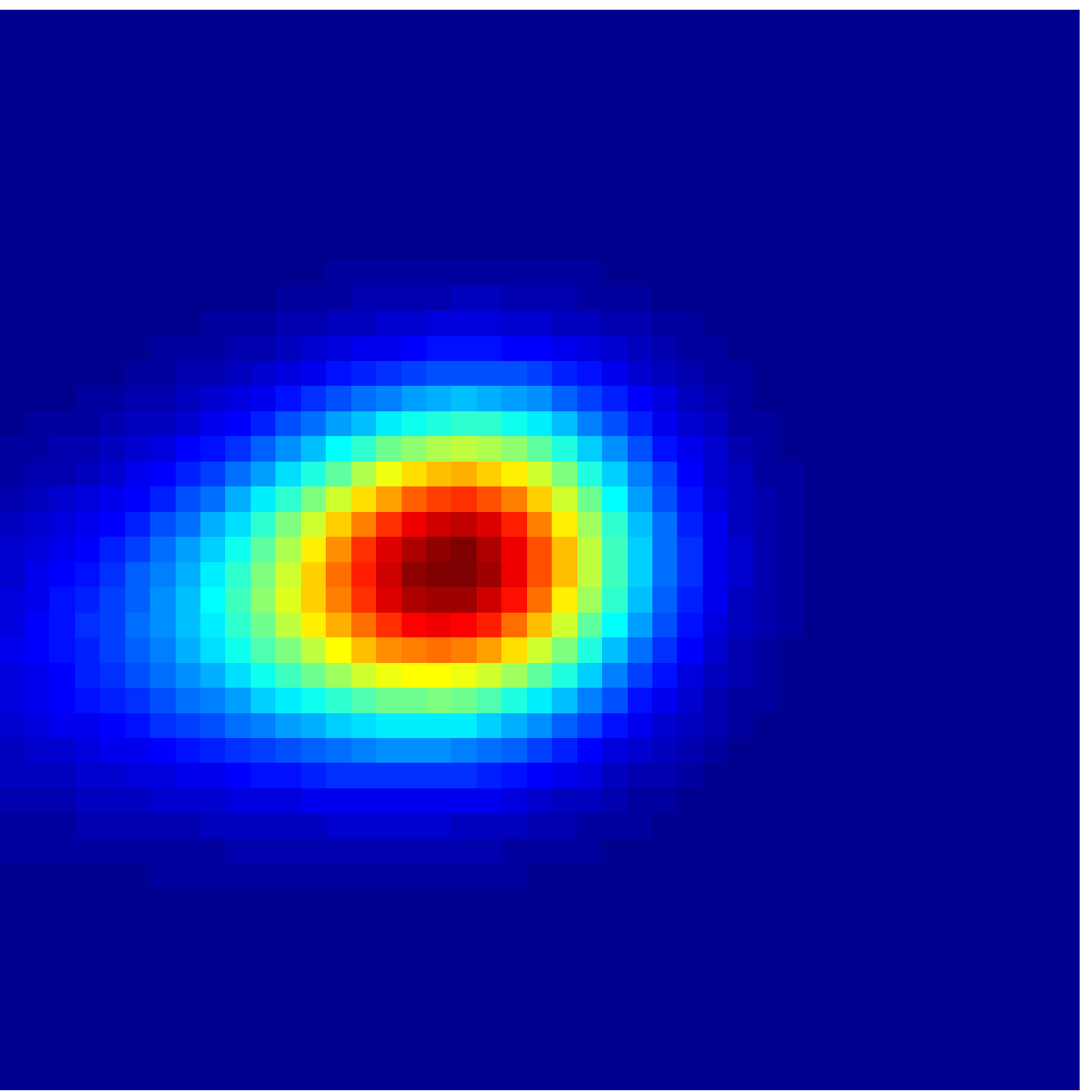}}{
      \includegraphics[width=0.1\textwidth]{./figs1/RI}}}

  \caption{Factorised physical model (PSW detector, velocity of
    30\arcsec/s towards the left): map of spatial redundancies~$\Pb$
    (left) and spatial impulse response~$\Hb_c$ (right). The spatial
    scales are different for better visualisation of the impulse
    response.\label{fig:DirectFacto}}
\end{figure}

In this form, some computations may be made even though they are
useless, because the convolution is performed for all indices, whereas
only some of them are used. In practice, the excess computation is
reduced because we chose shifts $(\Ta,\Tb)$ close to the sampling
period induced by the scanning scheme. Almost all convolution results
are observed, from 1 to 7 times for PSW as illustrated in
Fig~\ref{fig:DirectFacto}.

There is, however, the disadvantage that the bolometer positions are
approximated. Yet these positions are important because they allow to
best exploit the data and to properly manage the information needed to
estimate high frequencies. We chose a step $\Delta$ that is close to
the sampling period along the scan, \ie $\Delta \approx 2\arcsec$. The
error introduced is therefore small. This can be seen to be all the
more valid when we consider the expected level of noise and telescope
pointing errors, which are of the same order of magnitude, $2\arcsec$.

Finally, the initial model~\eqref{Eq:ModelInstruDiscretDiscret} is
decomposed in the discrete convolution defined by~\eqref{eq:25}
following the (inhomogeneous) down-sampling defined
by~\eqref{eq:26}-\eqref{eq:27}, that is to say, $\Hb$ is factorised
and
\begin{equation}
  \label{Eq:ModelInstruFactorise}
  \yb = \Hb \xb = \Pb \Hbc \xb,
\end{equation}
where $\Hbc$ is a convolution matrix and $\Pb$ a pointing matrix that
takes the values observed after convolution. It has one, and only one,
``1'' per row because each data item can only come from one
position. Some columns may be entirely zero because certain
coefficients may not be observed. Conversely, some columns may contain
several ``1'' because certain coefficients may be observed several
times.

To summarize, using an approximation of the pointed direction, we have
separated the model $\Hb$ into two sub-models $\Hb = \Pb \Hb_c$, where
$\Hb_c$ is invariant and $\Pb$ contains the non-invariant
structure. This decomposition is broadly similar to the one generally
found in super-resolution in the field of image processing (see
references in the introduction).

Fig.~\ref{fig:DirectFacto} presents this decomposition for the PSW
detector with a velocity of 30\arcsec/s towards the left: spatial
redundancy contained in $\Pb$ (the blacker the pixel, the more often
it was observed) and spatial impulse response (the time response of
the bolometer and the electronics is clearly visible as the spatial
extent of the Gaussian lobe).

\subsection{Conclusion}
\label{Sec:ConcluDirect}

We have constructed a linear instrument model from the physical
description of the phenomena involved during acquisition: scanning,
optics, filters, bolometers, and electronics were taken into account,
together with a description of the sky in continuous variables in the
three dimensions. We next explicitly described certain calculations
and approximated the model in a factorised form to lighten the
numerical computational burden.

The proposed model differs from those currently used in
\sanepic~\citep{Patanchon07} or \MADmap~\citep{Cantalupo10} in that it
includes the physics of acquisition. Moreover, unlike monochromatic
models~\citep{sibthorpe09}, the sky model extends spectrally across
the whole channel. Again, unlike~\citep{sibthorpe09}, our bolometer
model is linearized, which simplifies the developments and allows the
bolometer time response to be made explicit.

Finally, the consistent, global definition of the acquisition allows
the over-sampling to be directly exploited and a processing method to
be designed that uses these properties to estimate the sky at higher
resolution than the detector sampling period.

\section{Data inversion for high-resolution maps}
\label{sec:inversion}

The previous section was dedicated to the instrument model and we
deduced the relationship between the measured data $\zb$ and the
unknown sky $\Ciel$ or its coefficients $\xb$ through
\begin{equation}
  \label{eq:16}
  \zb = \Hcal \Ciel +\ob+ \nb = \Hb \xb +\ob+ \nb \,.
\end{equation}
The matrix $\Hb$ is relatively complex and high-dimensional, but the
forward model~\eqref{Eq:ModelInstruDiscretDiscret} remains linear. The
terms $\ob$ and $\nb$ account for measurement and modelling errors and
quantify the data uncertainties. The term $\ob$ is the noise mean
(offset) and $\nb$ is a zero-mean white and stationnary Gaussian noise
with variance $\sigma_n^2$. We assumed that each bolometer denoted $b$
is affected by an unknown offset
$o_b$. Eq.~\eqref{Eq:ModelInstruDiscretDiscret} can be rewritten for
the bolometer $b$
\begin{equation}
  \label{eq:7}
  \zb_b = \Hb_b \xb + o_b + \nb_b,
\end{equation}
where $\zb_b$ contains data from bolometer $b$, $\Hb_b$ is the
corresponding part of the instrument model and $(\nb_b,o_b)$ accounts for
errors of the bolometer $b$. This section presents the method to
estimate the unknown $\xb$ and the offsets $\ob$ from the data $\zb$.

We tackled the map-making question in an inverse problem framework.
Abundant literature is available on the subject
\citep{idier08,demoment89,Tikhonov77,Twomey62}. As presented in the
previous section, the instrument model embeds convolutions and
low-pass systems. The inverse problem is ill-posed \citep{Park03} and
this is particularly true when super-resolution is intended. In this
context, a naive inversion, such as a least-squares solution, would
lead to an unacceptably noisy and unstable solution.

A usual class of solutions relies on regularization, \ie the
introduction of prior information on the unknown object $\xb$ to
compensate for the lack of information in the data. A consequence of
regularization is that reconstruction methods are specific to a class
of (sky) maps, according to the introduced information. From this
standpoint, the present paper considers extended sources and
relatively spatially regular maps.

Since it is defined as a function of continuous variables, the
regularity can be measured by the squared energy\footnote{As an
  alternative, a non-quadratic norm of the derivative, \eg convex
  penalty, could also be used. Its interest is less penalization of
  high gradients in the map. Unfortunately, the measure on
  coefficients is no more explicit.}  of derivatives of $\Ciel$. For
first derivatives in both directions, it can be shown (see
appendix~\ref{Sec:AnnexeRegularite}) that
\begin{equation}
  \label{eq:700}
  \left\| \frac{\partial \Ciel(\alpha,\beta)}{\partial \alpha} \right\|^2 +
  \left\| \frac{\partial \Ciel(\alpha,\beta)}{\partial \beta} \right\|^2
  = \xb^t \left( \Dba + \Dbb \right)\xb,
\end{equation}
where $\Db = \Dba + \Dbb$ is obtained from the sum of the
autocorrelation of the derivative of $\psi$ with respect to $\alpha$
and $\beta$ and is similar to a discrete gradient operator. This
relation illustrates the equivalence between the measure on the
continuous function $\Ciel$ and the measure on coefficient $\xb$,
thanks to the use of a Gaussian generating function.

With the regularity measure~\eqref{eq:700} and the white Gaussian
model hypothesis for $\nb$, the regularized least-squares criterion is
\begin{multline}
  \label{Eq:CritMCRContinu}
  J_{\Ciel}(\Ciel, \ob)= \|\zb - \Hcal\Ciel - \ob\|^2 + \\ \mu \left(
    \left\| \frac{\partial \Ciel(\alpha,\beta)}{\partial \alpha}
    \right\|^2 + \left\| \frac{\partial \Ciel(\alpha,\beta)}{\partial
        \beta} \right\|^2 \right).
\end{multline}
Another consequence of ill-posedness and regularization is the need to
tune the compromise between different sources of information. The
hyperparameter $\mu$ tunes this trade-off.  With Eq.~\eqref{eq:16}
and~\eqref{eq:700}, we obtain a regularized least-squares criterion
that depends only on the coefficients
\begin{equation}
  \label{Eq:CritMCRDiscret}
  J_{\xb}(\xb,\ob) = \|\zb - \Hb\xb - \ob\|^2 + \mu \xb^t\Db\xb,
\end{equation}

The desired map is defined as the minimizer
\begin{equation*}
  \wh \xb, \wh \ob = \argmin_{\xb, \ob} J_{\xb}(\xb,\ob).
\end{equation*}
As a consequence $\wh \Ciel(\alpha, \beta) = \sum_{ij} \wh \xij \,
\psi(\alpha - i\Ta,\beta - j\Tb)$, is the optimum of the criterion
Eq. \eqref{Eq:CritMCRContinu}.

\begin{remark}\label{RQ:Bayes} A Bayesian interpretation of
  criterion~\eqref{Eq:CritMCRDiscret} is a Gaussian \post law with
  Gaussian iid likelihood, Gaussian correlated prior and flat prior
  law for $\ob$.
  An advantage of the Bayesian interpretation is the ability to derive
  an uncertainty around the maximum through the variance (see
  Sec. \ref{sec:resultats}) of the \post law. Another important
  advantage of the Bayesian interpretation deals with the estimation
  of hyperparameter and instrument parameters~\citep{orieux10a}.
\end{remark}

The proposed algorithm for the computation of $\wh \xb$ and $\wh \ob$
is an alternating minimization algorithm: after an initialization, the
following two steps are iterated
\begin{enumerate}
\item Find $\wh \xb$ for fixed $\ob$
  \begin{equation}
    \label{eq:12}
    \wh \xb^{k+1} = \argmin_{\xb} \|\zb - \Hb\xb - \wh \ob^k\|^2 + \mu
    \xb^t\Db\xb
  \end{equation}

\item Find $\wh \ob$ for fixed $\xb$
  \begin{equation}
    \label{eq:14}
    \wh \ob^{k+1} = \argmin_{\ob} \|\zb - \Hb\wh\xb^{k+1} - \ob\|^2
  \end{equation}

\end{enumerate}
until a criterion is met. For fixed $\xb$, the solution is
straightforward and $\wh o_b$ is the empirical mean of the residual
$\zb_b - \Hb_b\xb$ for each bolometer separately. For fixed $\ob$, the
solution Eq.~\eqref{eq:12} is unique and explicit
\begin{equation}
  \label{eq:23}
  \wh \xb = \left(\Hb^t \Hb + \mu \Db\right)^{-1}\Hb^t(\zb - \ob) \,.
\end{equation}
The estimator is linear \wrt data $\zb$. Unfortunately, since $\Hb$ is
not circulant, $\wh \xb$ cannot be computed with a ``brute force''
algorithm: the practical inversion of the Hessian matrix $\Hb^t \Hb +
\mu \Db$ is impossible (the size of this matrix is the square of the
number of coefficients $\xb$). The proposed solution relies on an
iterative conjugate gradient descent
algorithm~\citep{Nocedal00,Shewchuck94}. The most expensive part is
the computation of the product between the matrix $\Hb^t\Hb$ and the
current point $\xb_k$ and it can be efficiently computed based on FFT,
decimation, and zero-padding (see appendix
\ref{Ann:CalculModelFacto}).

\section{Experimental results}
\label{sec:resultats}

This part illustrates the improvement that our approach can bring \wrt
to the standard approach based on
coaddition 
first using simulated data and then with actual data transmitted by
Herschel.

\subsection{Simulated data}
\label{sec:donnees-sim}

\subsubsection{Experimental protocol}

We chose three $20\arcmin \times 20\arcmin$ maps used by the SPIRE
consortium to assess reconstruction methods \citep{clements06}: a map
of galactic cirrus (Fig.~\ref{fig:cirrus}) complying with the \aprio
regularity model, a map of galactic cirrus superimposed on point
sources (Fig.~\ref{fig:cirrusDot}), and a galaxy map
(Fig.~\ref{fig:Galaxie}).

We studied the PMW channel and the \textit{``Large Map''} protocol
with three scans in each direction and a velocity of 30\arcsec/s. The
data were generated using a simulated map of coefficients $\xb$ and
\citep{clements06} the instrument
model~\eqref{Eq:ModelInstruDiscretDiscret}, considering for this
simulation part that the bolometers are not affected by any offset.
We used a flat spectrum ($\varrho = 0$ in Eq.\,\eqref{Eq:CielLambda})
for the simulations and the inversions.  The noise is zero-mean white
Gaussian with three levels characterized by their standard deviation
$\sigma_n$ (``standard noise'' hereafter), $10\,\sigma_n$ (``high
noise'') and $0.1\,\sigma_n$ (``low noise''). The standard deviation
is the same for all bolometers and, unless stated otherwise, all data
sets were generated with the same noise realization.

The proposed reconstruction for the $20\arcmin \times 20\arcmin$ maps
performed using $\Ta =\Tb=2\arcsec$, \ie maps of $600 \times 600$
coefficients. We compare our results with the map obtained by
coaddition, with $6\arcsec$ as pixel size.

In agreement with Section~\ref{sec:inversion}, the map was
reconstructed as the minimizer of
criterion~\eqref{Eq:CritMCRContinu}-\eqref{Eq:CritMCRDiscret} and the
minimization was performed by a conjugate gradient algorithm with
optimal step size.  The value of the criterion decreases at each
iteration and a few tens of iterations appear to be sufficient to
reach the minimum.

In the simulated cases, the original map (the ``sky truth'') is known,
accordingly, we can quantitatively assess the reconstruction through
an error measure defined by
\begin{gather}
  \label{eq:28}
  \Ec = \sum_{i,j} |\xijstar - \wh \xij | ~/~ \sum_{i,j} |\xijstar|,
\end{gather}
where $\xb^{*}$ and $\wh \xb$ are the coefficients of the true map and
the reconstructed map.

\begin{figure}[htbp]
  \centering
  \includegraphics[width=0.25\textwidth]{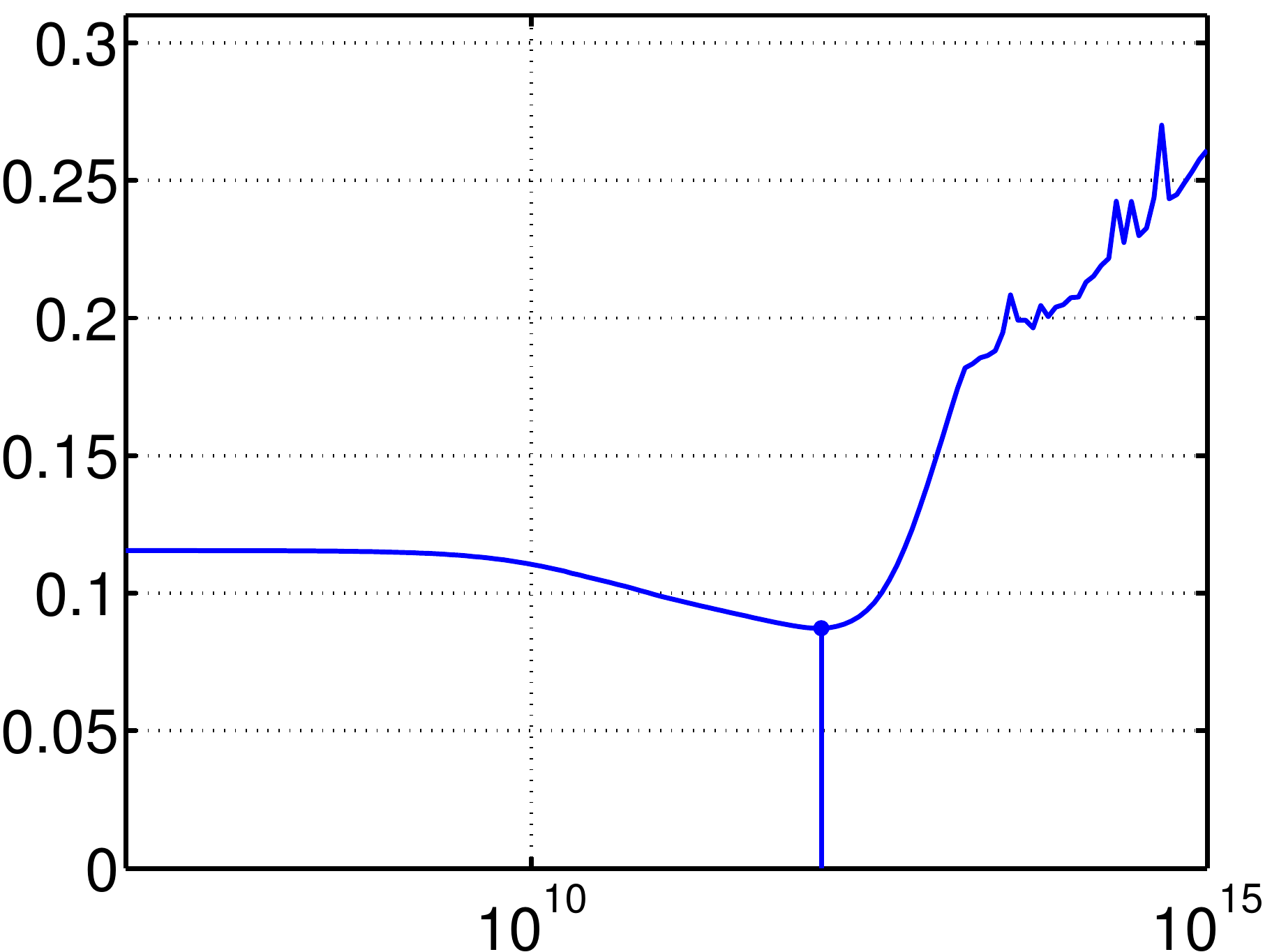}
  \caption{Reconstruction error $\Ec$ \vs regularization parameter
    $\hyperR$ for of cirrus with ``standard noise''. The minimum error
    is $\Ec_\textrm{min} = 0.08$ for the proposed method, while $\Ec =
    0.12$ for $\hyperR=0 $.\label{fig:hyperR}}
\end{figure}

The estimate $\wh \xb$ depends on the regularization parameter, as
illustrated in Fig.~\ref{fig:hyperR}. A non-zero optimum value
$\hyperR_\text{opt}$ appears (here $\sim 10^{12}$) for which $\Ec$ is
a minimum, which confirm the interest of the regularization. A value
lower than $10^{11}$ leads to an under-regularized map and a value
greater than $10^{13}$ to an over-regularized one. In the following,
it is, of course, the optimal value that is used to reconstruct the
maps.  Also, it appears empirically that $\mu$ needs to vary by a
factor 2 around $\hyperR_{\text{opt}}$ to obtain a noticeable
modification of the map.  This result is confirmed in
Fig.~\ref{fig:hyperR}, where the minimum is not very marked compared
to the horizontal scale.

Fig.~\ref{fig:hyperR} also illustrates the improvement provided by the
re\-gu\-la\-ri\-za\-tion: the errors for the non-regularized and
optimum-regularized maps are $0.12$ and $0.08$ respectively.

\subsubsection{Restoration of galactic cirrus}
\label{sec:cirrus}

\begin{figure*}[htbp]
  \centering

  \subfigure[True map]{\includegraphics[width=0.3\textwidth]
    {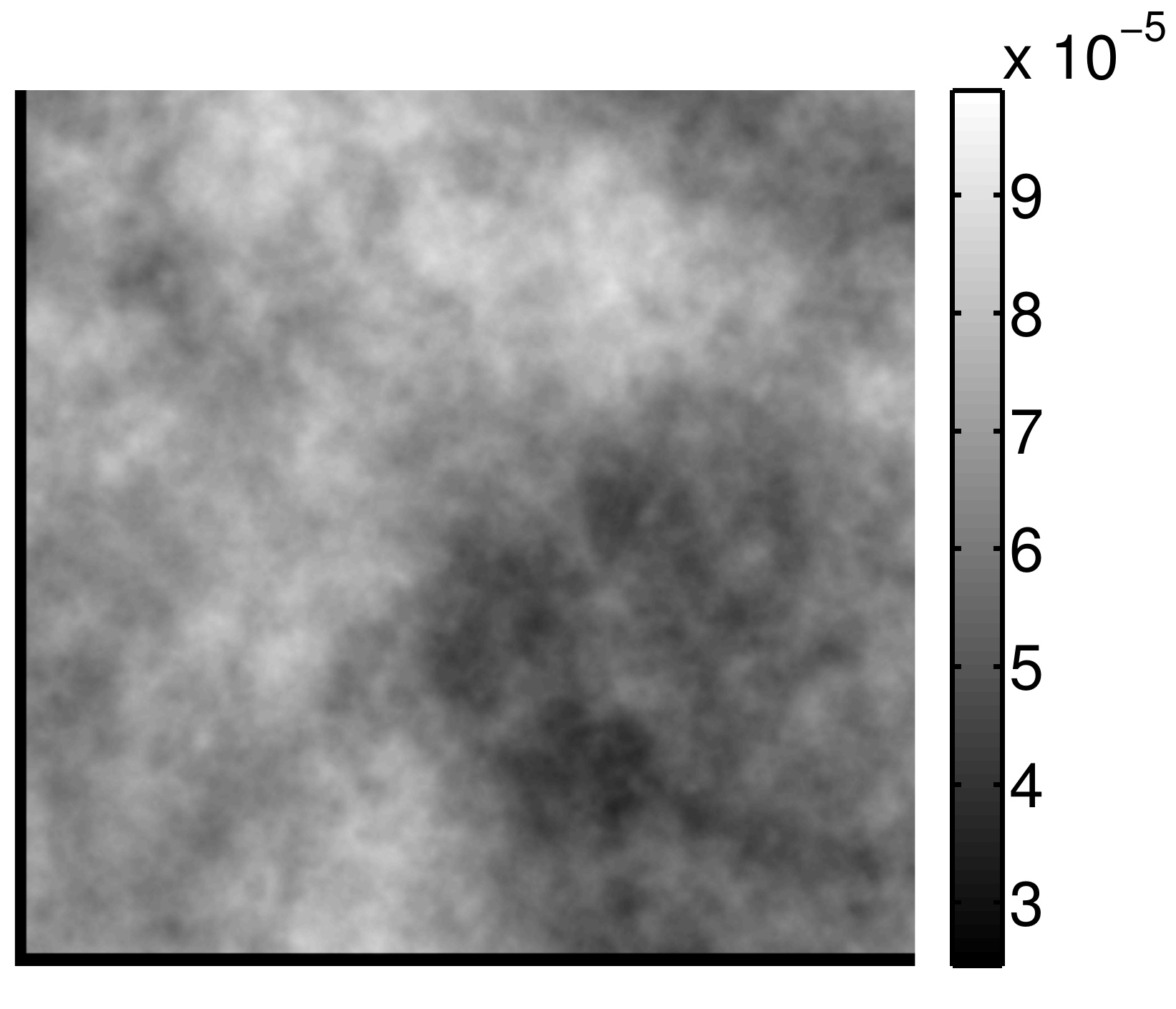}\label{fig:cirrusTrue}}
  \subfigure[Proposed map]{\includegraphics[width=0.3\textwidth]
    {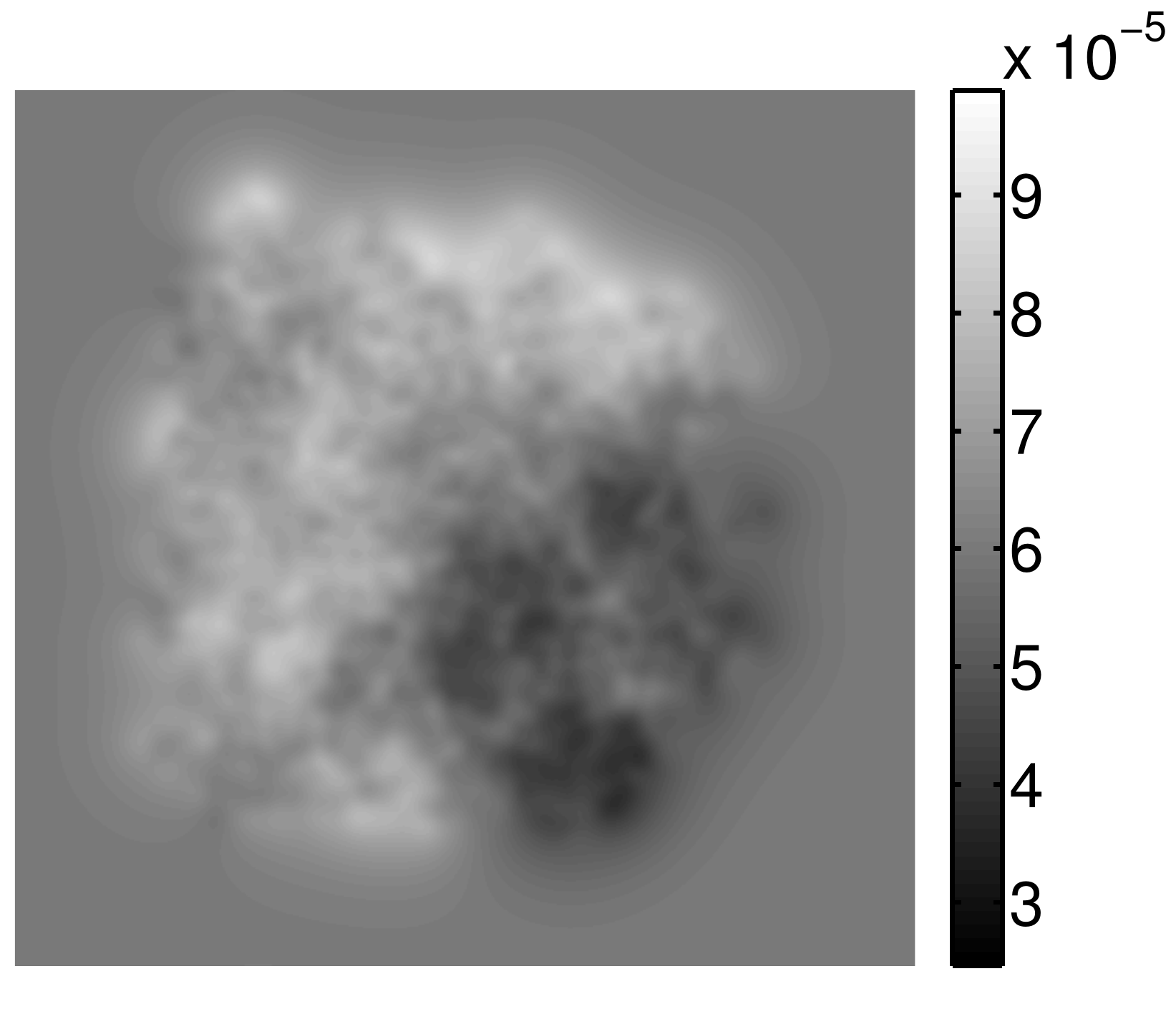}\label{fig:cirrusDiffRegNormal}}%
  \subfigure[Coaddition]{\includegraphics[width=0.3\textwidth]
    {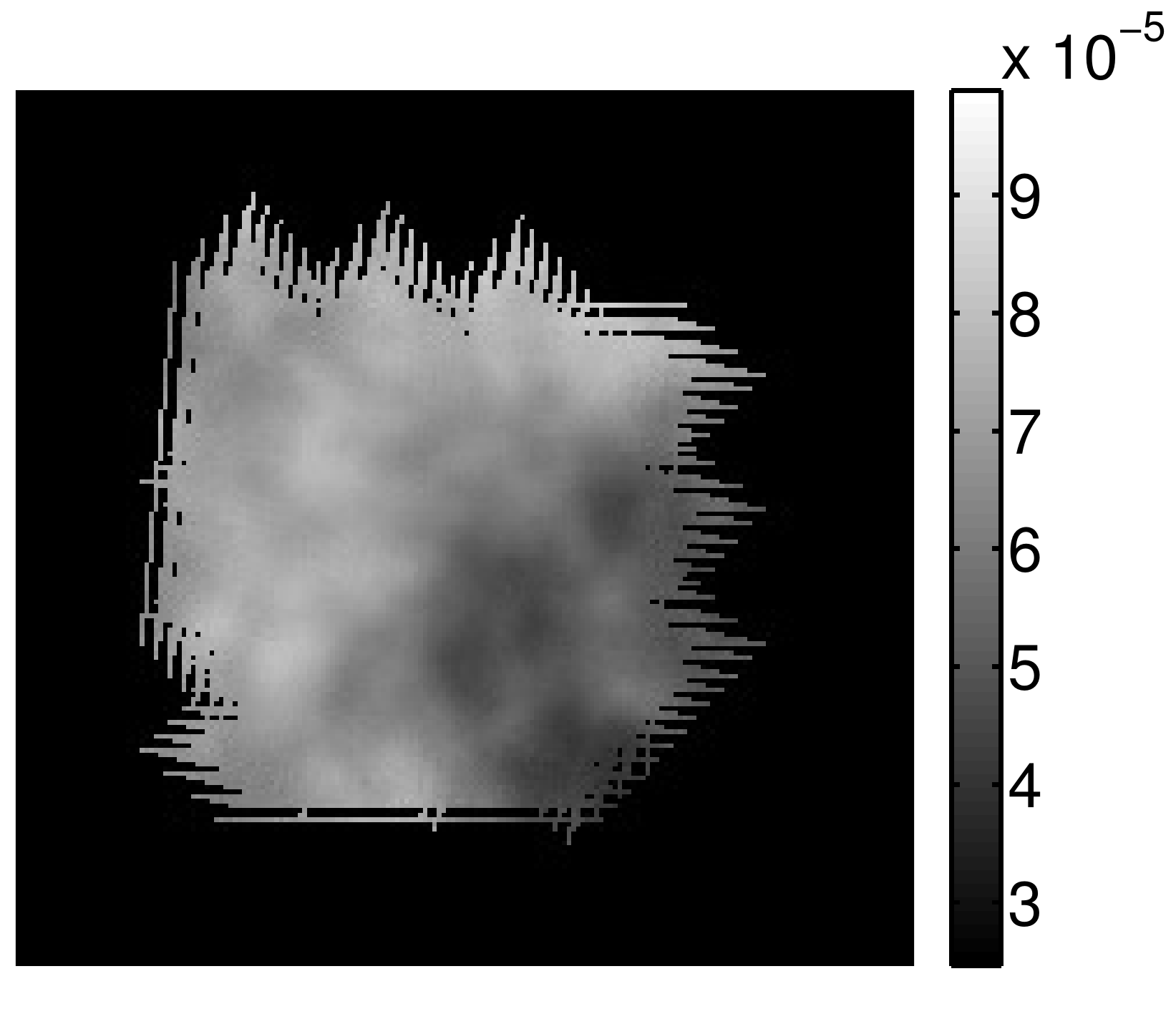}\label{fig:cirrusCoAdd360}}%

  \subfigure[one horizontal
  profile]{\includegraphics[width=0.25\textwidth]
    {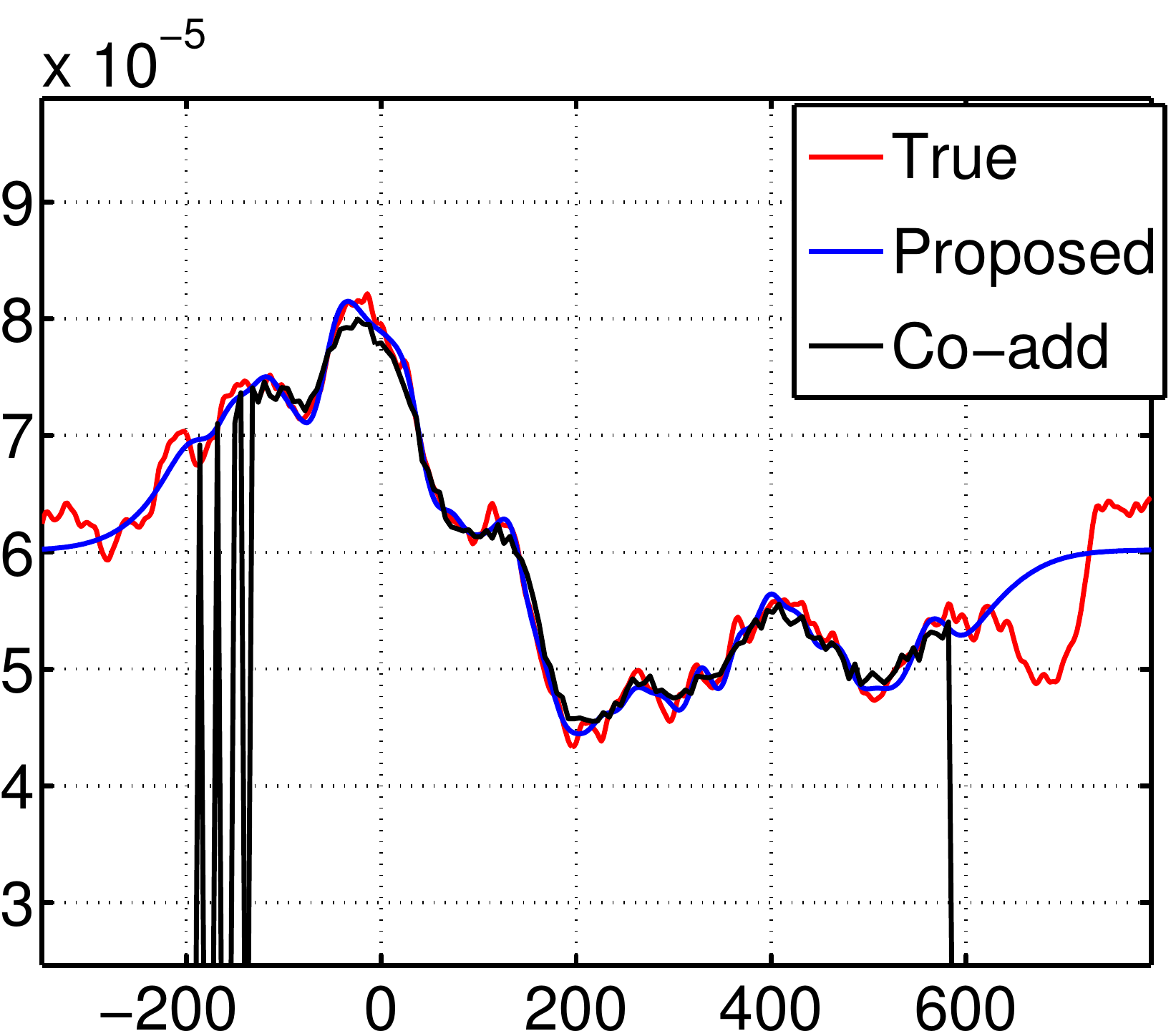}\label{fig:sliceBestCirrus}}
  \subfigure[one horizontal profile
  (zoom)]{\includegraphics[width=0.25\textwidth]
    {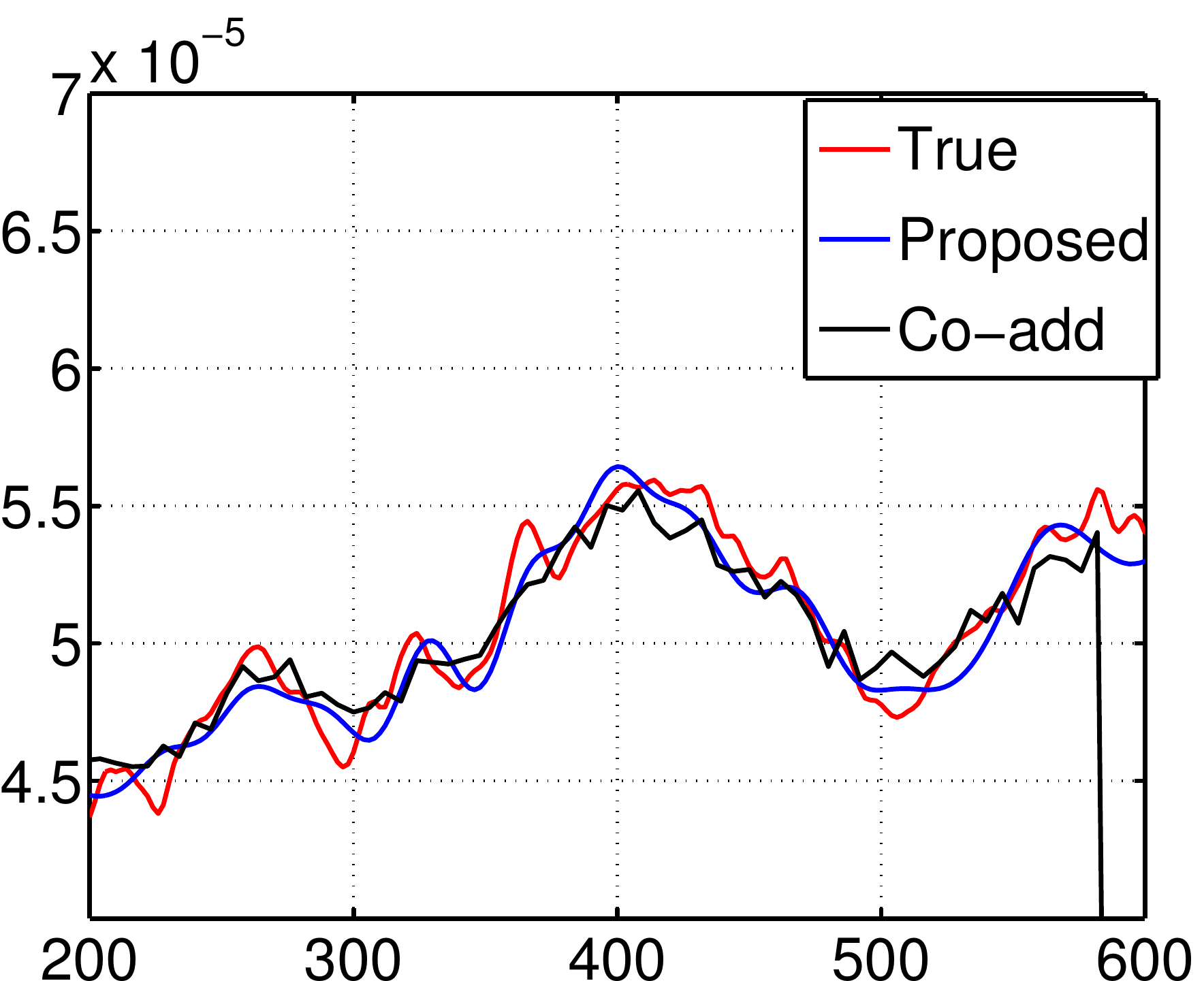}\label{fig:sliceBestCirrusZoom}}

  \caption{Comparison of results. Fig.~\ref{fig:cirrusTrue} shows the
    true map, Fig.~\ref{fig:cirrusDiffRegNormal} presents the proposed
    map and Fig.~\ref{fig:cirrusCoAdd360} the coaddition. A horizontal
    profile is shown in Fig.~\ref{fig:sliceBestCirrus} and
    Fig.~\ref{fig:sliceBestCirrusZoom} gives a
    zoom.\label{fig:cirrus}}

\end{figure*}

Fig.~\ref{fig:cirrus} summarises the results concerning the cirrus in
the ``standard noise'' case. The proposed map is very close to the
true one. Our method restores details of small spatial scales (with
spectral extension from low to high frequency) that are invisible on
the coaddition but present on the true map (see the profiles in
Figs.~\ref{fig:sliceBestCirrus} and~\ref{fig:sliceBestCirrusZoom},
especially the fluctuations around pixels 250 and 350). In addition,
our method also correctly restores large-scale structures, which
correspond to low-frequencies down to the null frequency (mean level
of the map). We conclude that our method properly estimates the
photometry.

\begin{remark} Moreover, the reconstruction method is linear with
  respect to the data (see Section~\ref{sec:le-modele-direct}), which
  means that the use of arbitrary units is valid. \end{remark}

\begin{figure*}[htbp]
  \centering

  \subfigure[Standard noise]{\includegraphics[width=0.3\textwidth]
    {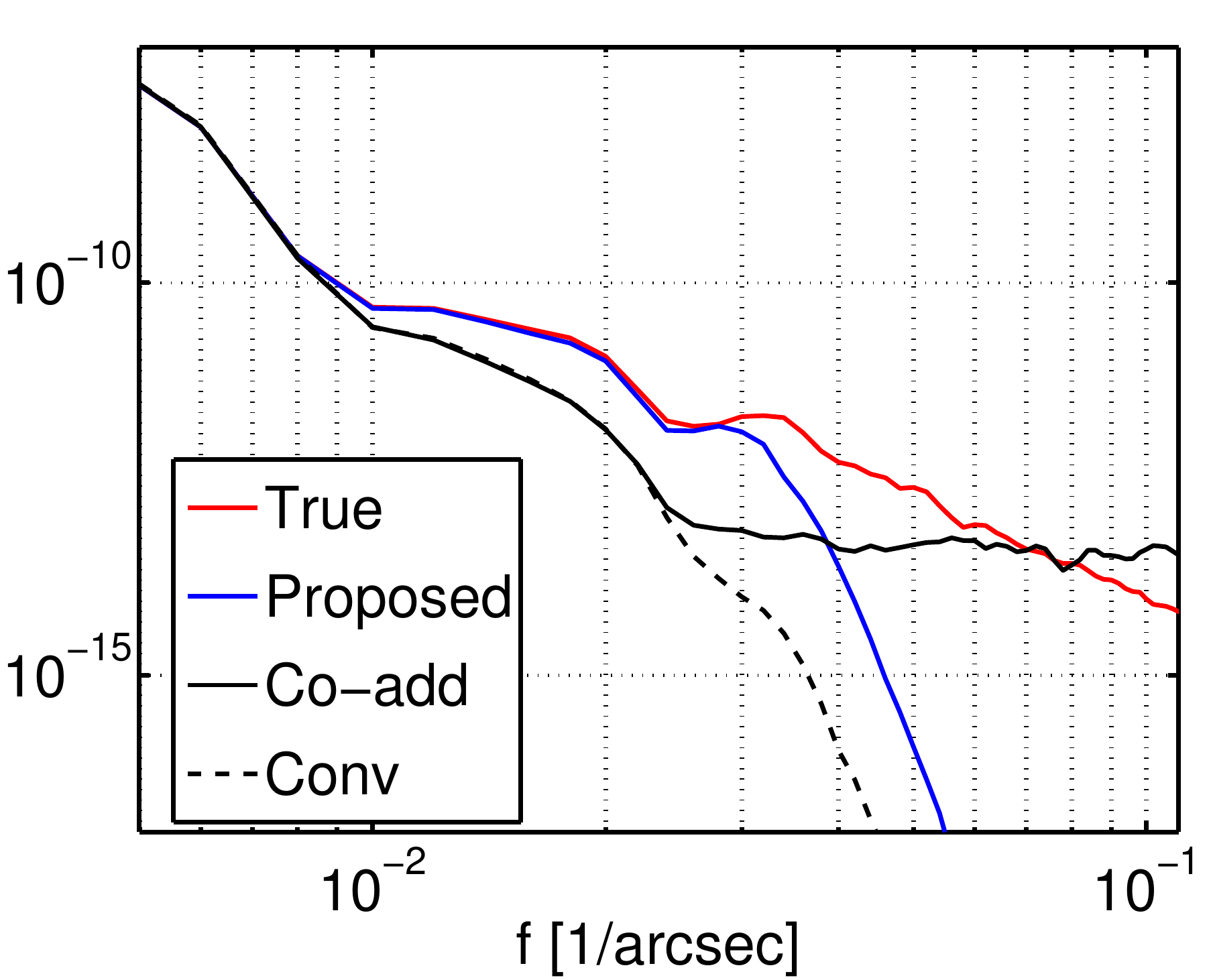}\label{fig:psdcirrusDiffRegNormal}} \hfill
  \subfigure[High noise]{\includegraphics[width=0.3\textwidth]
    {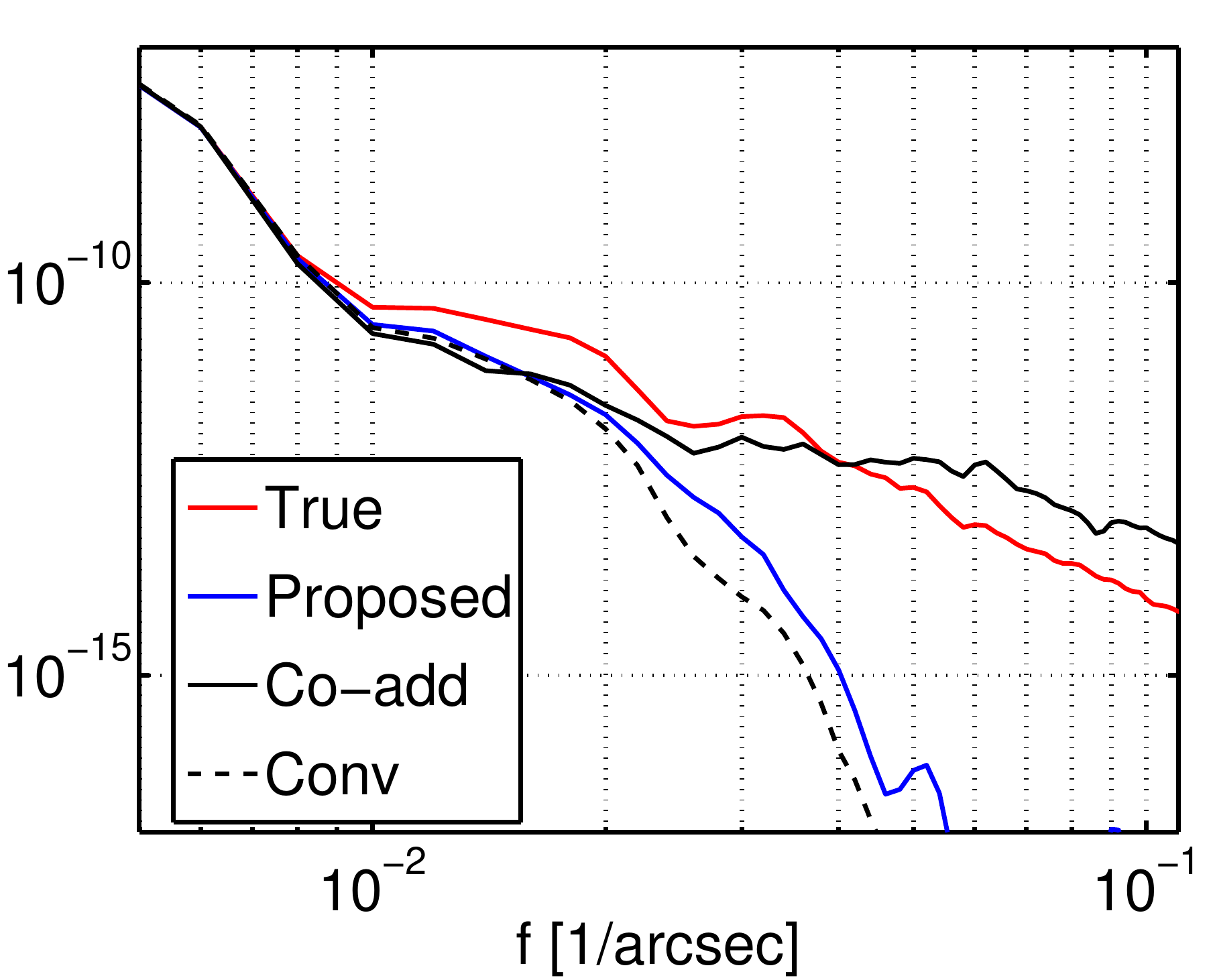}\label{fig:psdcirrusHighNoise}} \hfill
  \subfigure[Low noise]{\includegraphics[width=0.3\textwidth]
    {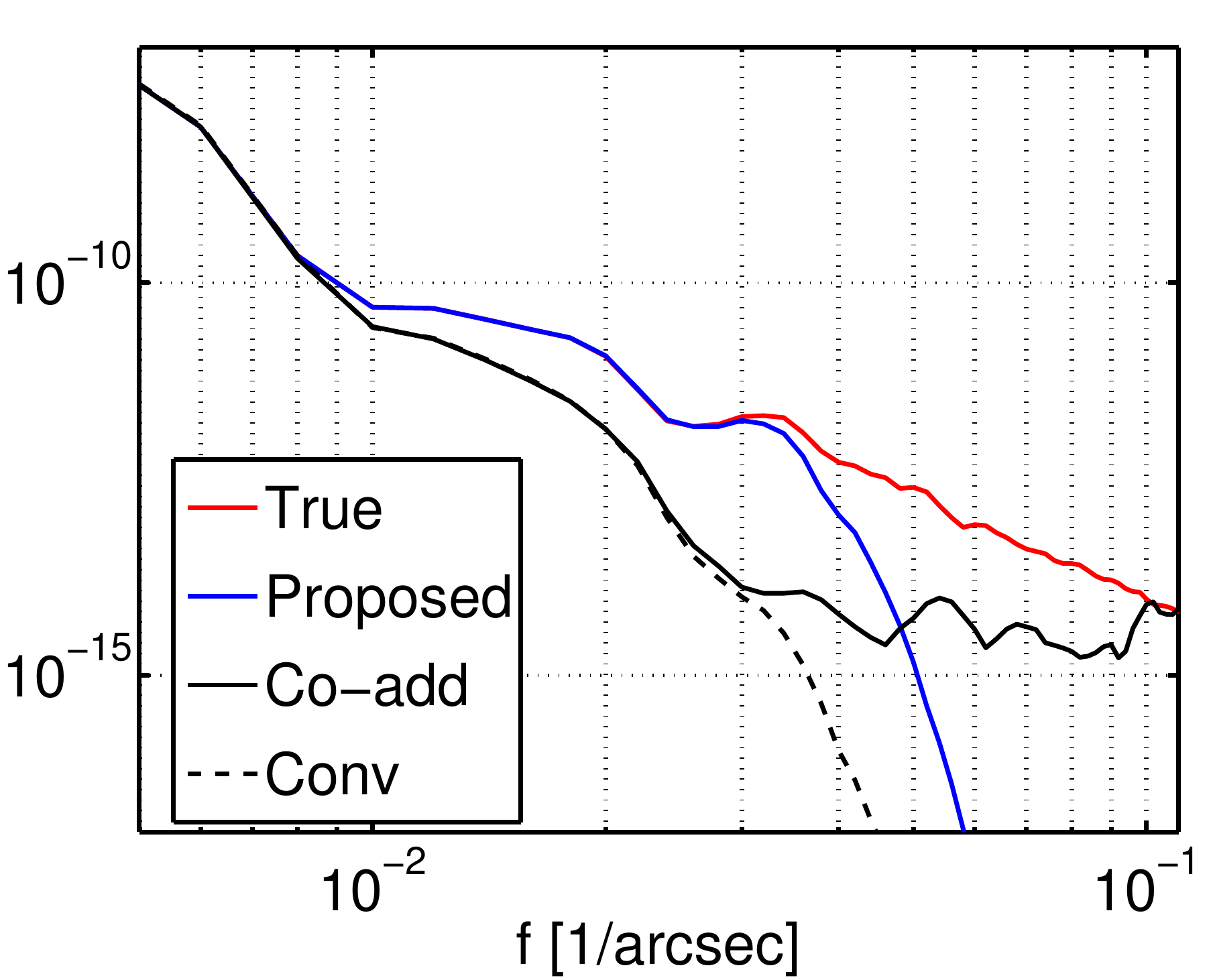}\label{fig:psdcirrusLowNoise}}

  \caption{Circular means of power spectra for the three levels of
    noise (standard deviations: $\sigma_n$, $10\,\sigma_n$ and
    $0.1\,\sigma_n$). The parameter $\hyperR$ is chosen to be optimal
    each time from the point of view of the error $\Ec$
    (Eq.~\eqref{eq:28}\label{fig:cirrusDiffRegSlicePSD}).}

\end{figure*}

To quantify the gain in correctly restored bandwidth, we considered
the power spectra of the maps (Fig.~\ref{fig:cirrusDiffRegSlicePSD})
for the true sky, the sky convolved with the PSF $\RepOpt$ (see
Sect. \ref{sec:louverture}), the coaddition, and the proposed sky. As
mentioned in Section~\ref{sec:louverture}, the sampling frequency of
the detector is $\fe \approx
0.02\,\mathrm{arcsecond}^{-1}$. Consequently the acquired data during
one integration cannot correctly represent frequencies above $\fe/2
\approx 0.01\,\mathrm{arcsecond}^{-1}$. We have also seen in
Section~\ref{sec:louverture} that the FWHM of the PSF is 25.2\arcsec
at 350\,$\mu$m, i.e. a cutoff frequency of the optical transfer
function of $\approx 0.04\,\mathrm{arcsecond}^{-1}$. The attenuation
effect of the convolution by the PSF on the true map is visible the
power spectra of the convolved and coaddition maps for all frequencies
above $\approx 0.008\,\mathrm{arcsecond}^{-1}$
(Fig.~\ref{fig:cirrusDiffRegSlicePSD}).

The power spectrum of the proposed map perfectly follows the power
spectrum of the true map, from the null frequency up to a limit
frequency that depends on the noise level. In the ``standard noise''
case (Fig.~\ref{fig:psdcirrusDiffRegNormal}) this limit is
$0.025\,\mathrm{arcsecond}^{-1}$, that is to say, almost three times
the limit frequency of each integration ($\fe/2 \approx
0.01\,\mathrm{arcsecond}^{-1}$). It illustrates that our method also
takes full advantage of the high-frequency temporal sampling. In any
case and compared to the coaddition, we have multiplied the spectral
bandwidth by a factor $\approx 4$ (starting from the null frequency)
where frequencies attenuated by the optical transfer function are
accurately inverted.

\begin{figure*}[htbp]
  \centering

  \subfigure[$\sigma$
  map]{\includegraphics[width=0.3\textwidth]{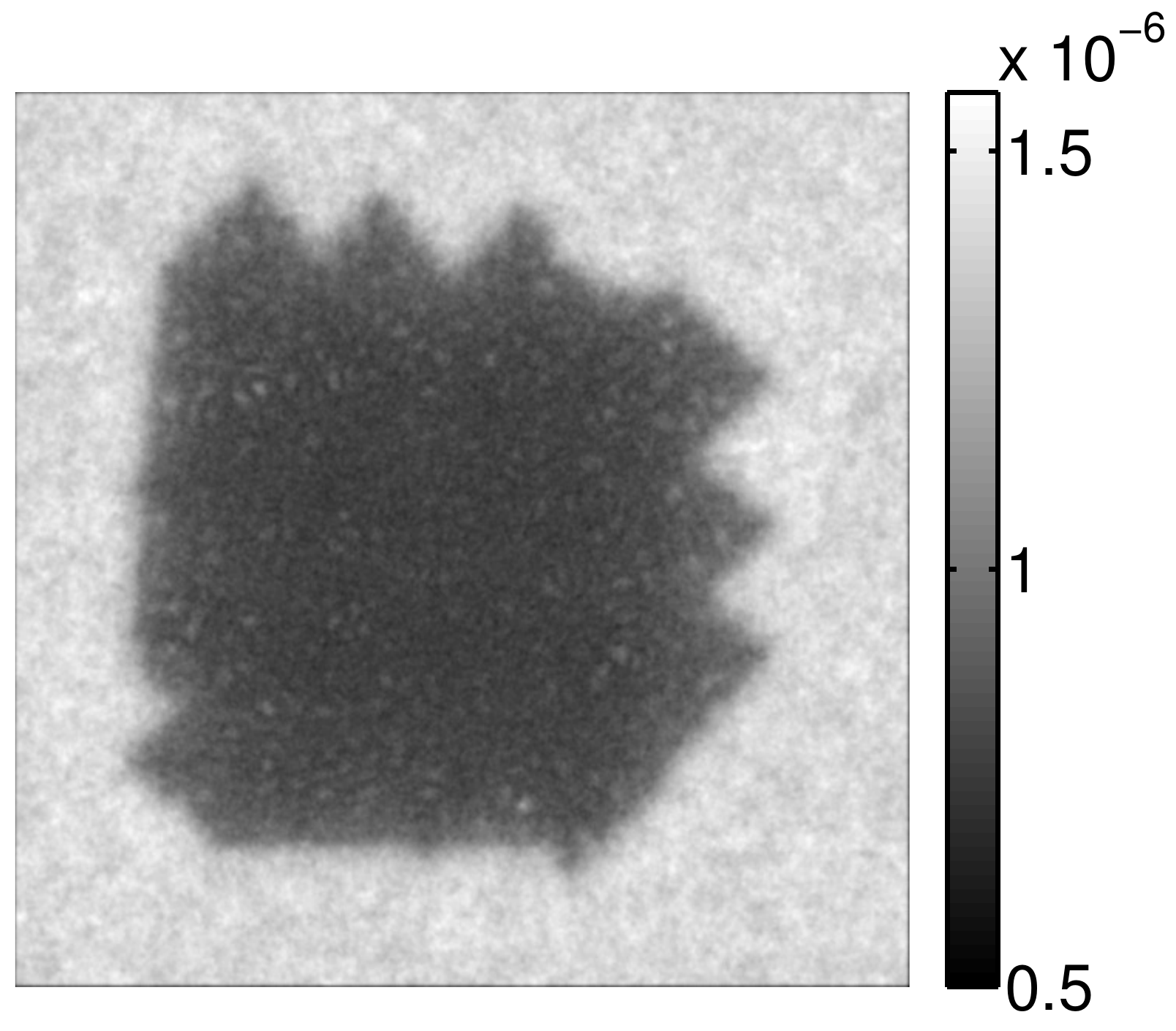}
    \label{fig:CarteVariance}}%
  ~~~~~~~~~~
  \subfigure[$\sigma$ map
  profile]{\includegraphics[width=0.3\textwidth]{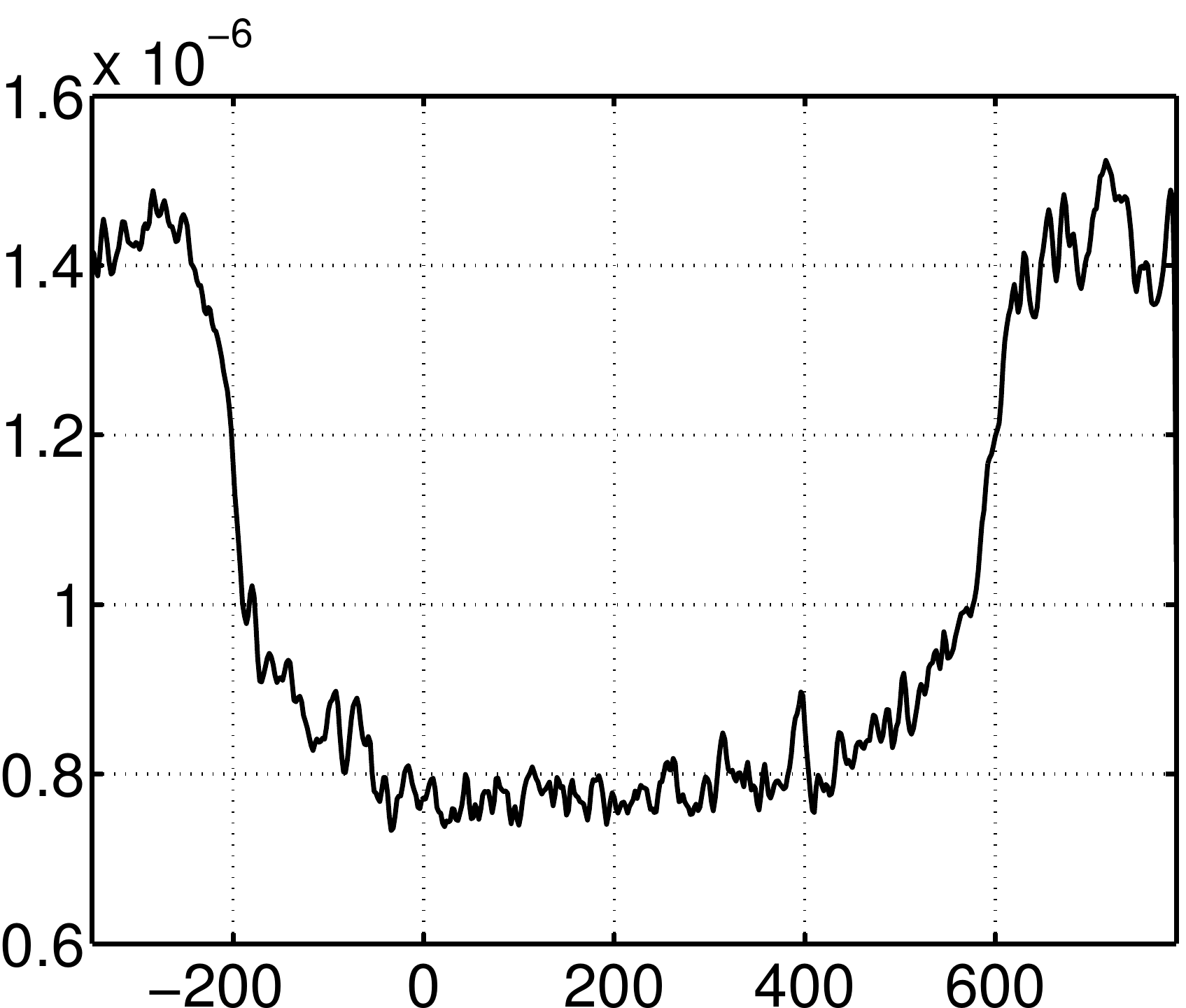}
    \label{fig:CarteVarianceCoupe}}

  \subfigure[Map profile $\pm 3 \wh
  \sigma$]{\includegraphics[width=0.3\textwidth]{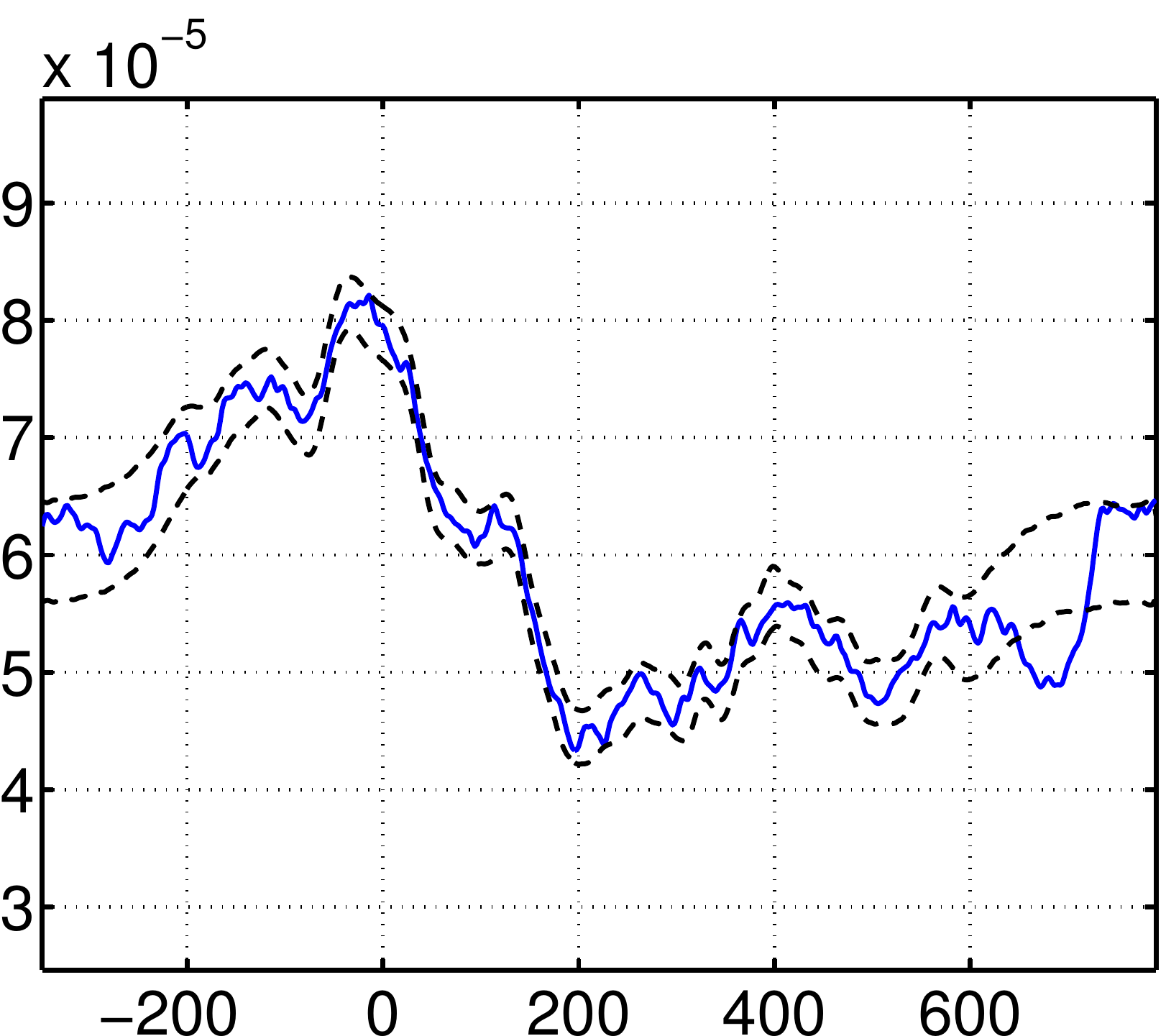}
    \label{fig:CartePlusMoins}}%
  ~~~~~~~~~~
  \subfigure[Power spectra $\pm 3 \wh
  \sigma$]{\includegraphics[width=0.3\textwidth]{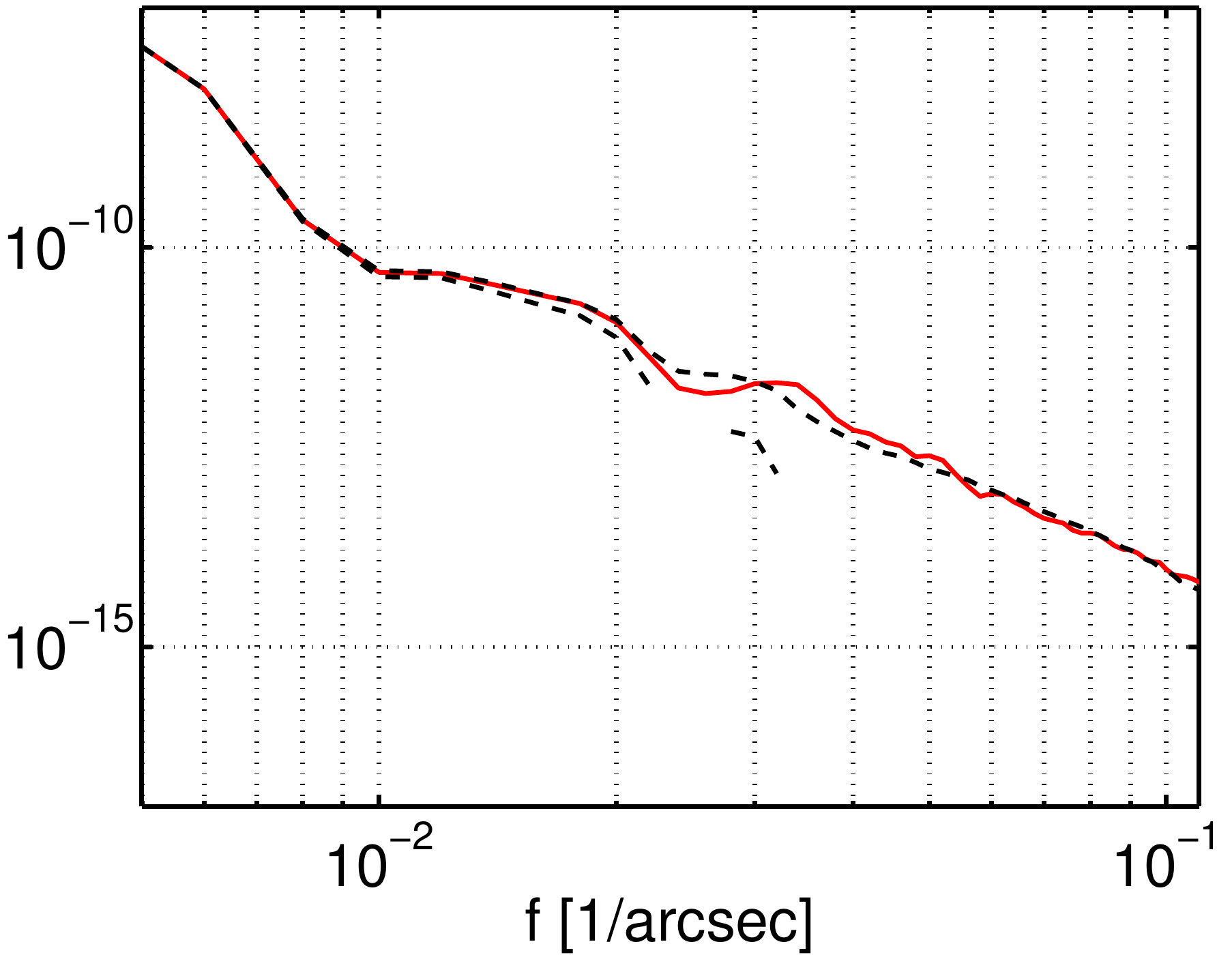}
    \label{fig:CartePlusMoinsRond}}

  \caption{Uncertainty provided by the \apost standard deviation $\wh
    \sigma$. Fig.~\ref{fig:CarteVariance} shows the map of the
    standard deviation for each pixel and
    Fig.~\ref{fig:CarteVarianceCoupe} gives a
    profile. Fig.~\ref{fig:CartePlusMoins} shows a profile of the
    \textsl{true map} as a solid line and the two dashed lines give a
    $\pm 3 \wh \sigma$ interval around the \textsl{estimated
      map}. Fig.~\ref{fig:CartePlusMoinsRond} shows the power spectrum
    of the \textsl{true map} as a solid red line and the two dashed
    lines give a $\pm 3 \wh \sigma$ interval around the
    \textsl{estimated power spectrum} in the ``standard noise''
    case.\label{fig:var}}

\end{figure*}

Our method also yields the uncertainties through the Bayesian
interpretation (see remark~\ref{RQ:Bayes}), from the standard
deviation $\wh \sigma$ of the \apost law
(Figs.~\ref{fig:CarteVariance} and \ref{fig:CarteVarianceCoupe}). The
uncertainties increase as we move away from the centre of the map
because the data contain less information. Moreover, we see in
Fig.~\ref{fig:CartePlusMoins} that the true map is inside a $\pm 3 \wh
\sigma$ interval around the estimated map. In the Fourier space
(Figs.\,\ref{fig:CartePlusMoinsRond}), up to the
$0.03\,\mathrm{arcsecond}^{-1}$, the true power spectrum is inside a
$\pm 3 \wh \sigma$ interval around the estimated power spectrum.

The possibilities of restoring frequencies obviously depend on the
noise levels, as illustrated in the spectra shown on
Fig.~\ref{fig:cirrusDiffRegSlicePSD}. When the noise level is lower,
it is possible to restore slightly higher frequencies: up to
$0.03\,\mathrm{arcsecond}^{-1}$ for ``low noise'', compared to
$0.025\,\mathrm{arcsecond}^{-1}$ for ``standard noise''. Conversely,
in the case of ``high noise'', our method no longer restores the
frequencies attenuated by the optical transfer function
Fig.~\ref{fig:psdcirrusHighNoise}. The deconvolution effect is reduced
and the essential effect is one of denoising. Nevertheless, the
proposed method gives better (or equivalent) results than coaddition
in all cases.

\subsubsection{Other types of sky}

\begin{figure*}[htbp]
  \centering

  \subfigure[True map]{\includegraphics[width=0.3\textwidth]
    {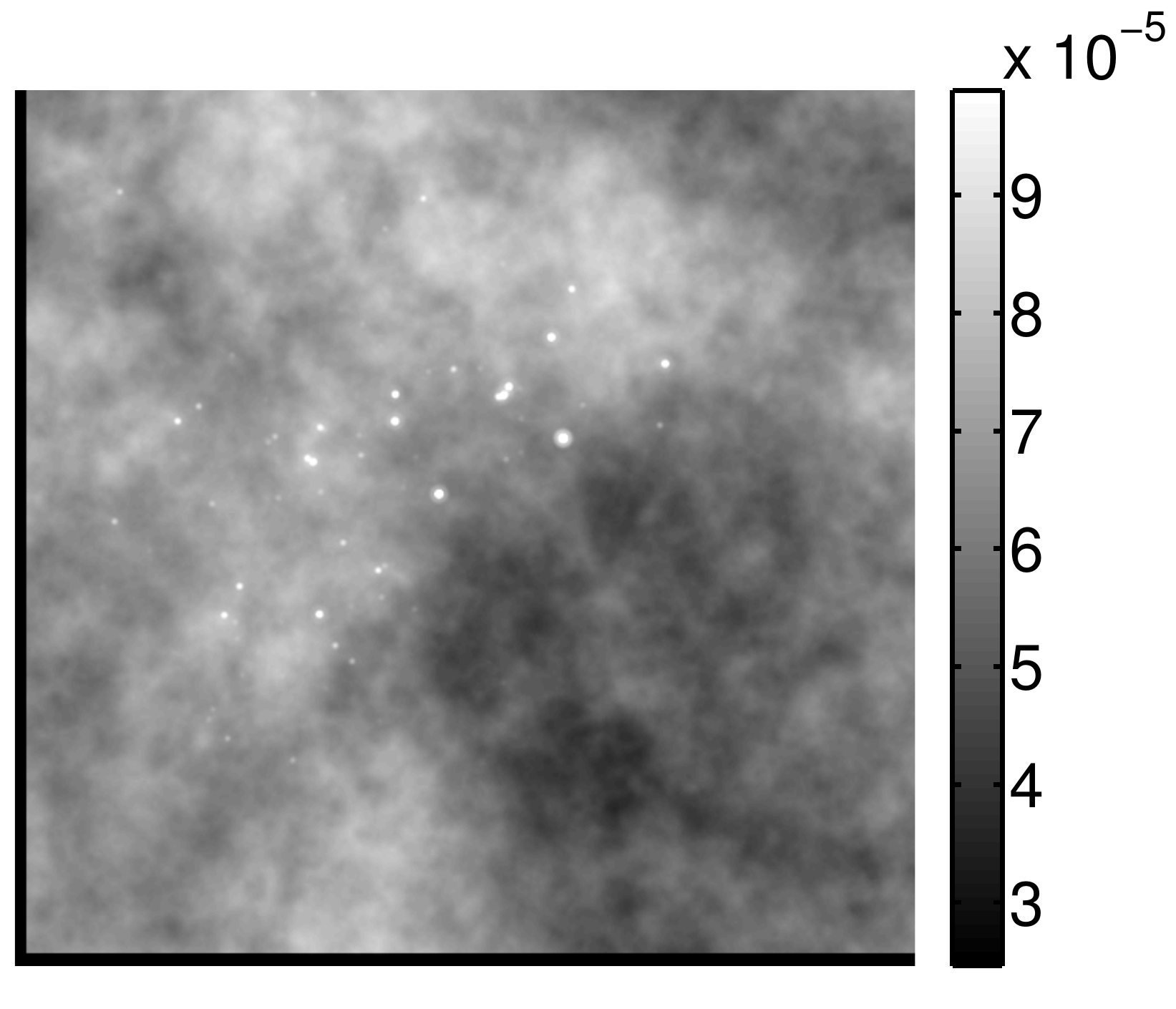}\label{fig:cirrusDotTrue}}
  \subfigure[Proposed]{\includegraphics[width=0.3\textwidth]
    {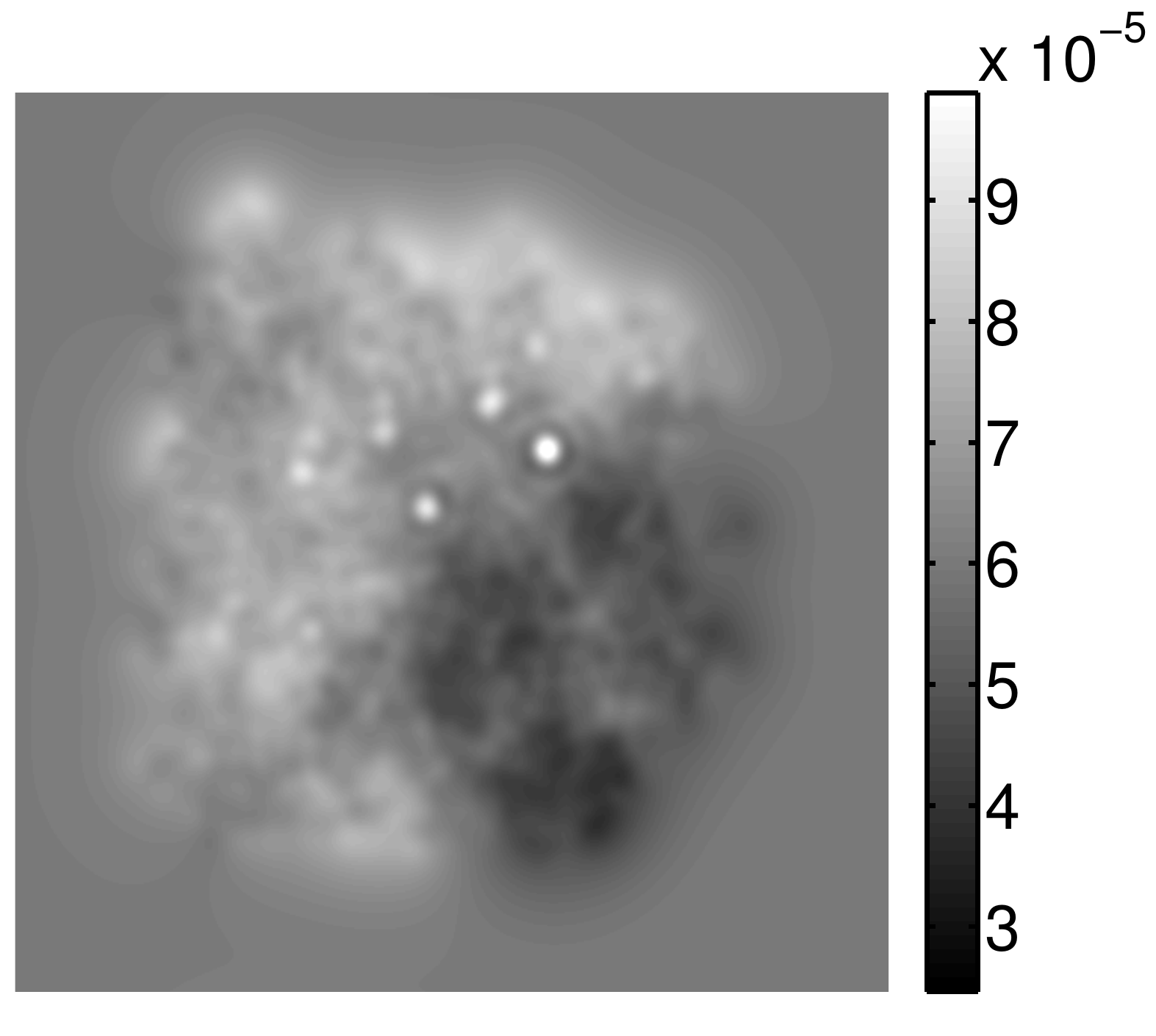}\label{fig:bestCirrusDot}}
  \subfigure[Coaddition]{\includegraphics[width=0.3\textwidth]
    {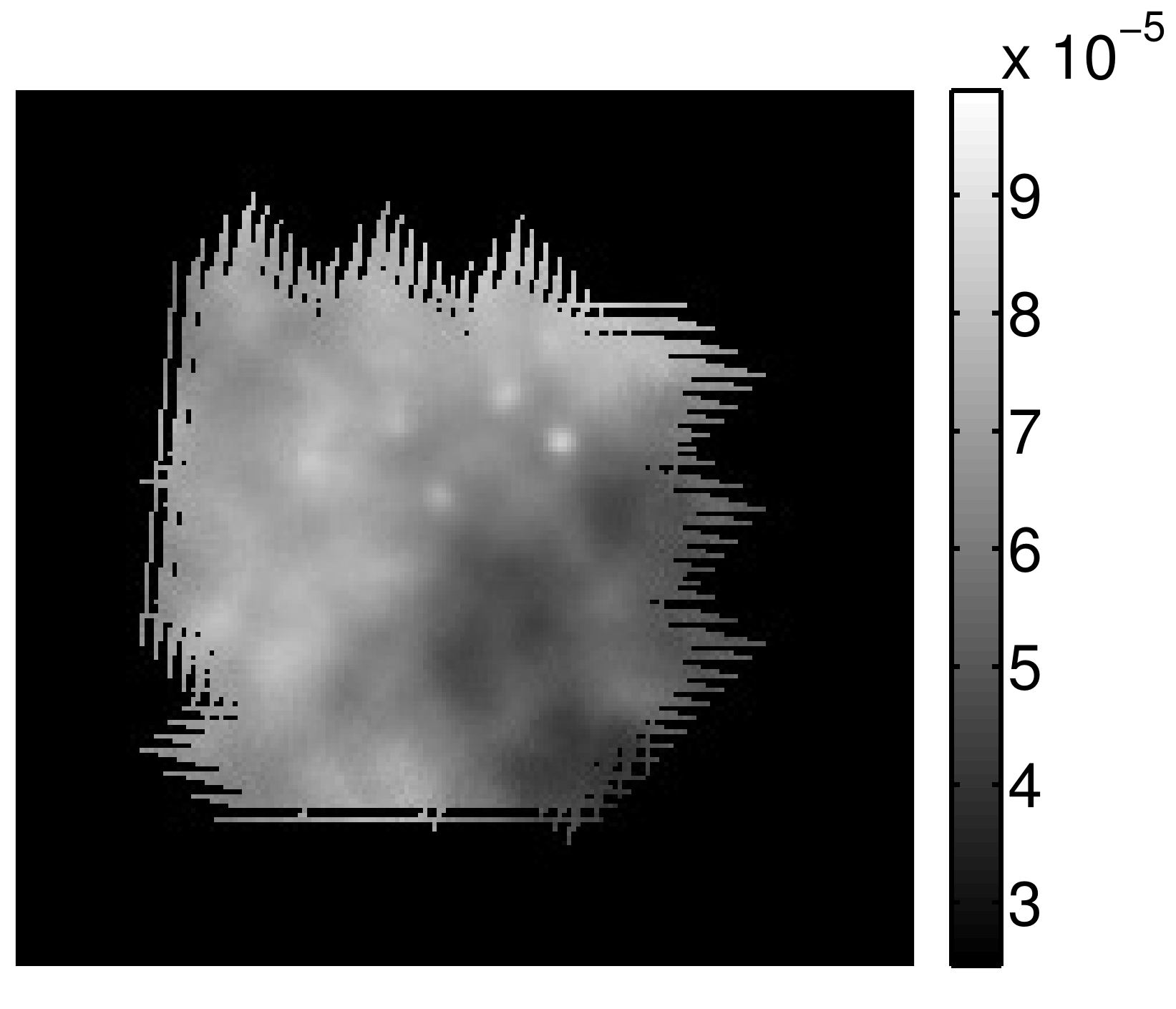}\label{fig:coaddCirrusDot}}

  \subfigure[one horizontal profile]{\includegraphics[width=0.3\textwidth]
    {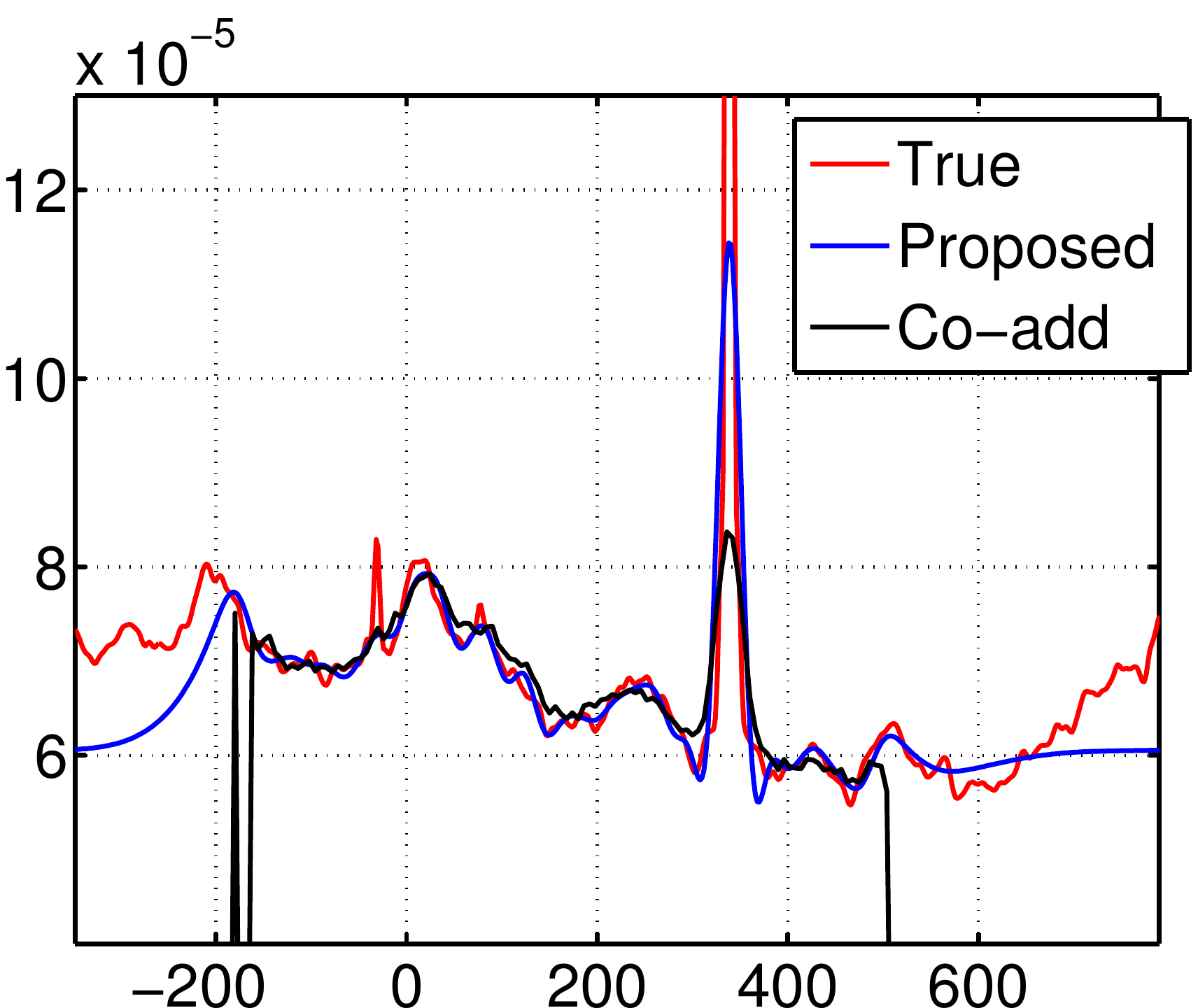}\label{fig:sliceBestCirrusDot}}
  \subfigure[one horizontal profile (zoom)]{\includegraphics[width=0.3\textwidth]
    {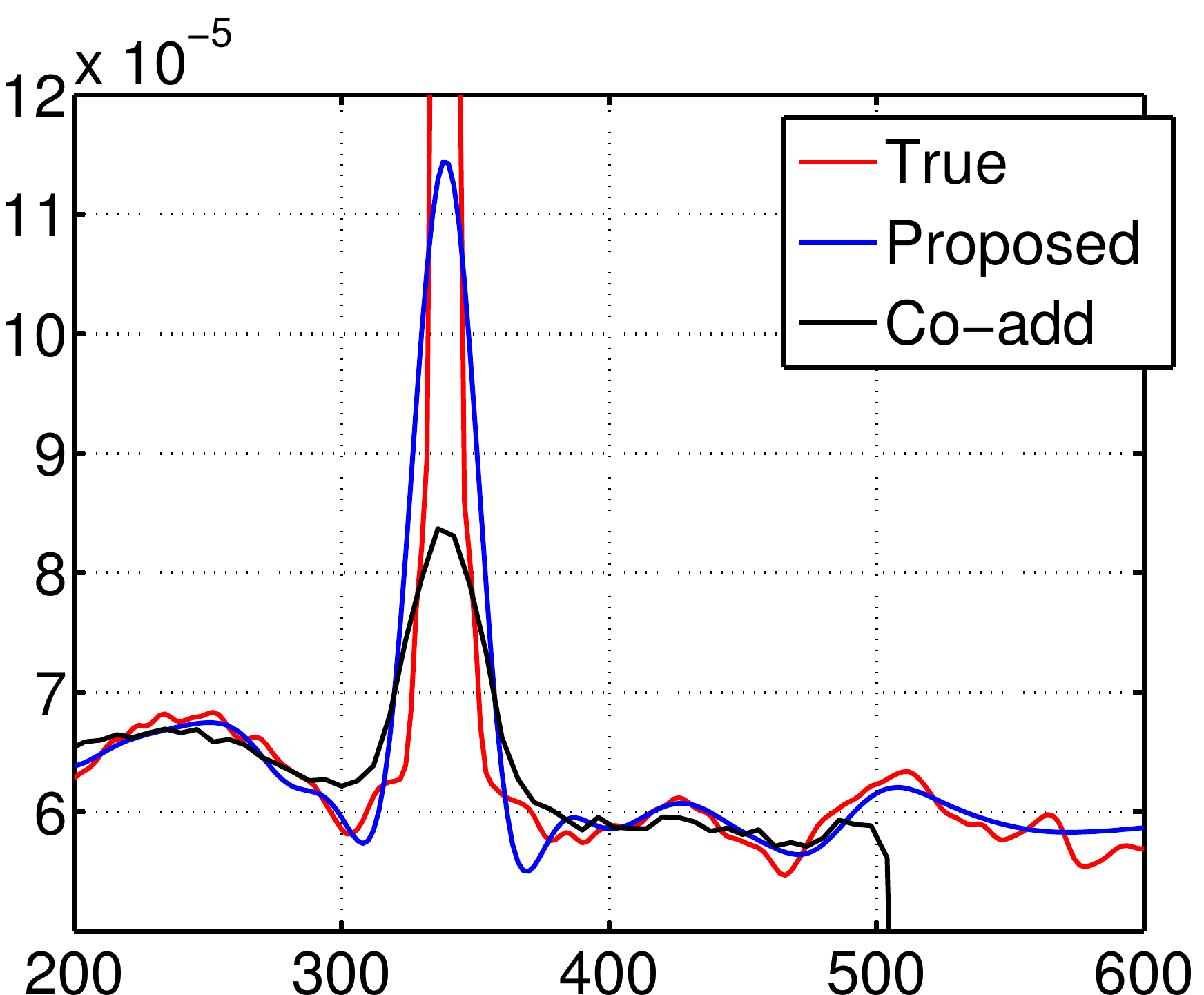}\label{fig:sliceBestCirrusDotZoom}}

  \caption{Restoration of cirrus superimposed on point
    sources.\label{fig:cirrusDot}}

\end{figure*}

\begin{figure*}[htbp]
  \centering

  \subfigure[True map]{\includegraphics[width=0.3\textwidth]
    {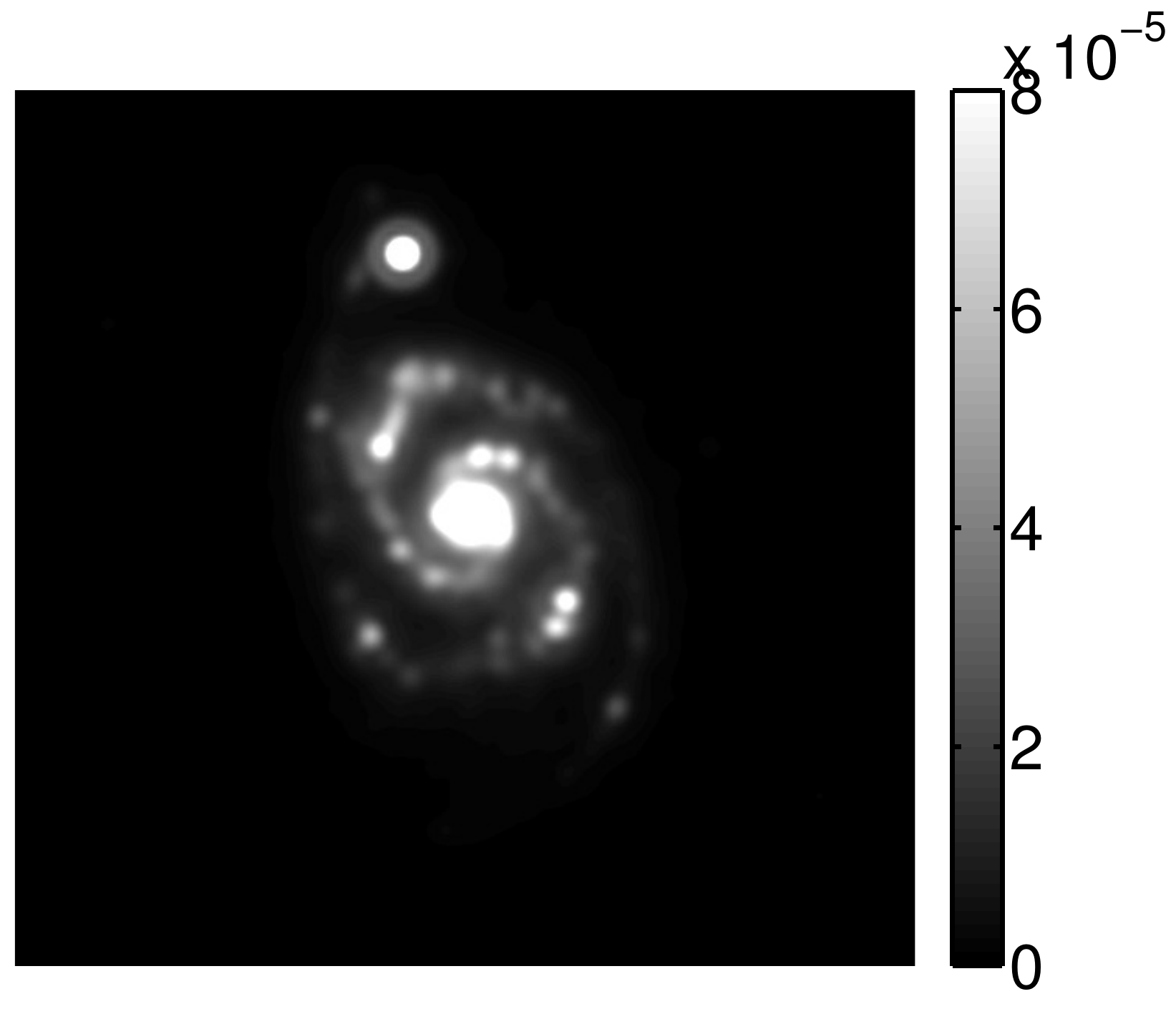}\label{fig:galSatTrue}}
  \subfigure[Proposed]{\includegraphics[width=0.3\textwidth]
    {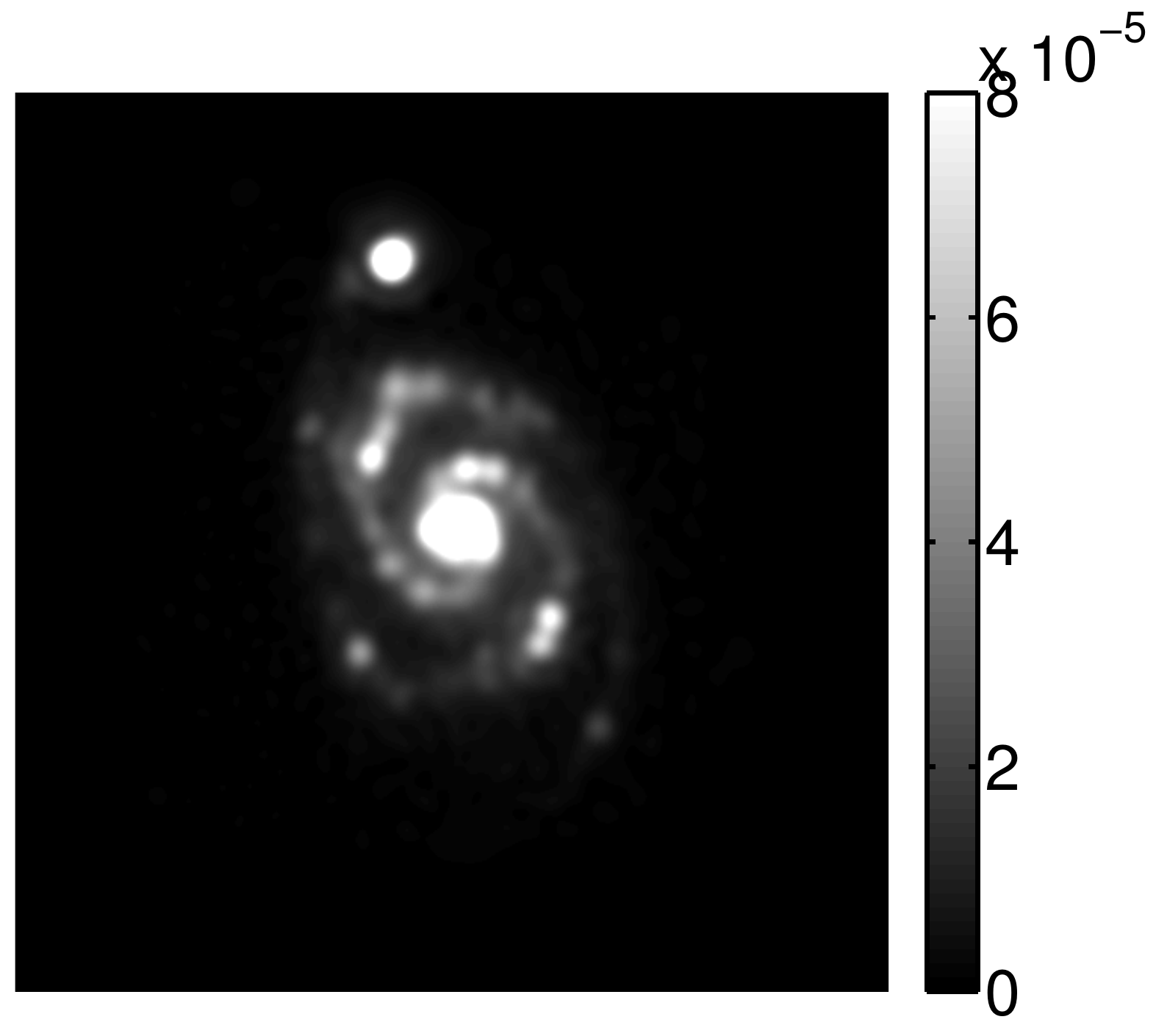}\label{fig:bestGalaxie}}
  \subfigure[Coaddition]{\includegraphics[width=0.3\textwidth]
    {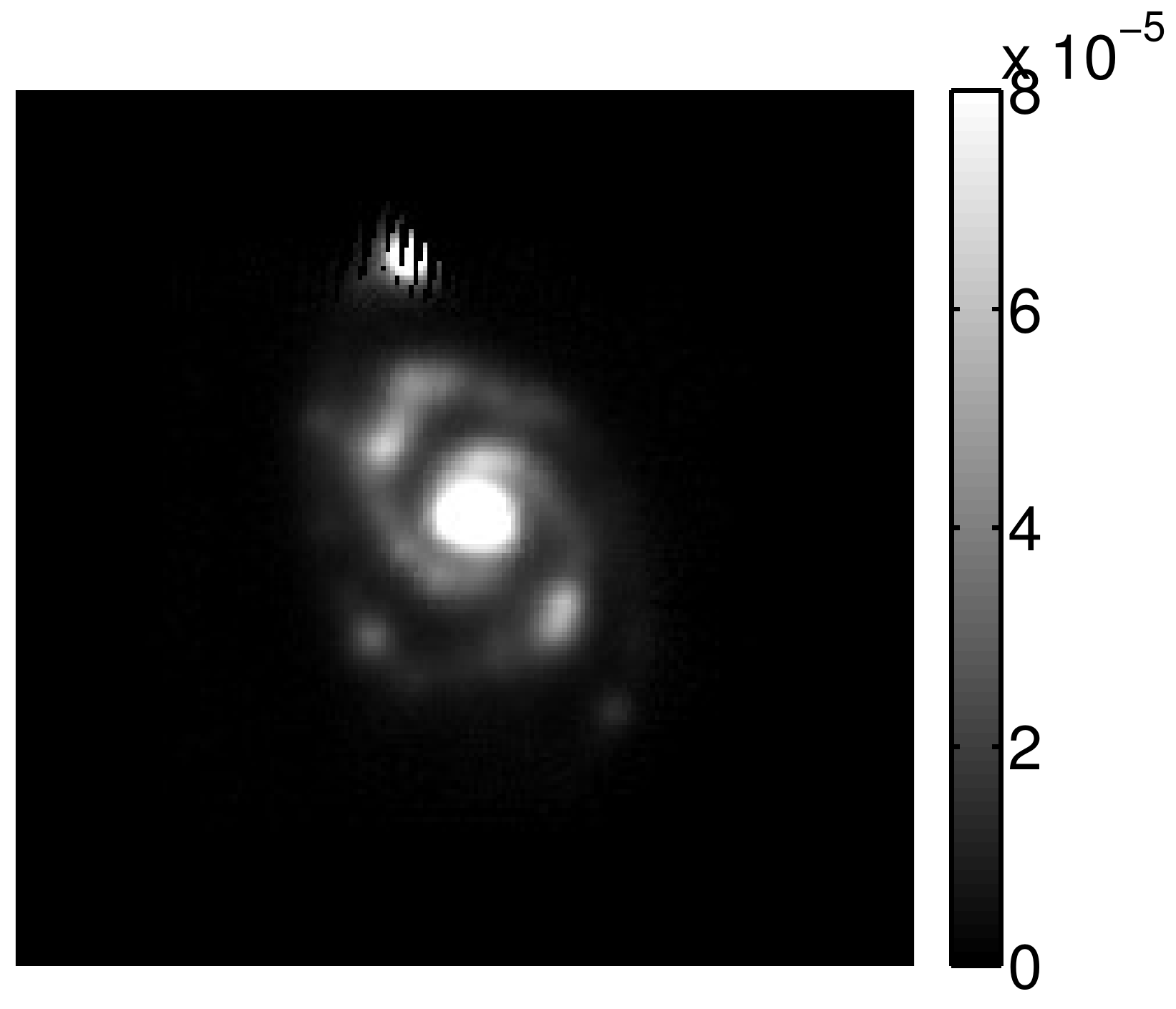}\label{fig:coaddGalaxie}}

  \subfigure[one horizontal profile]{\includegraphics[width=0.3\textwidth]
    {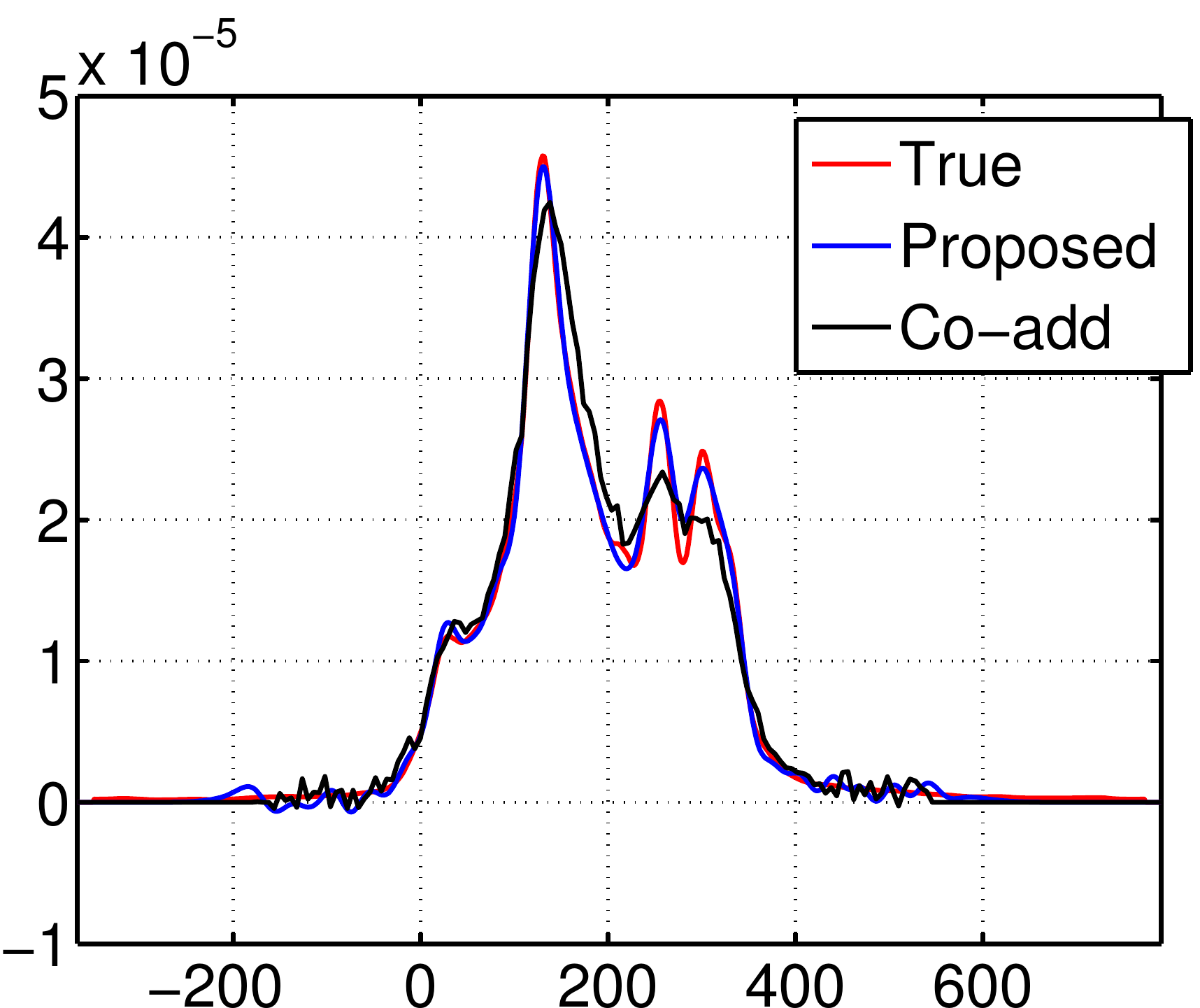}\label{fig:sliceBestGalaxie}}
  \subfigure[one horizontal profile (zoom)]{\includegraphics[width=0.3\textwidth]
    {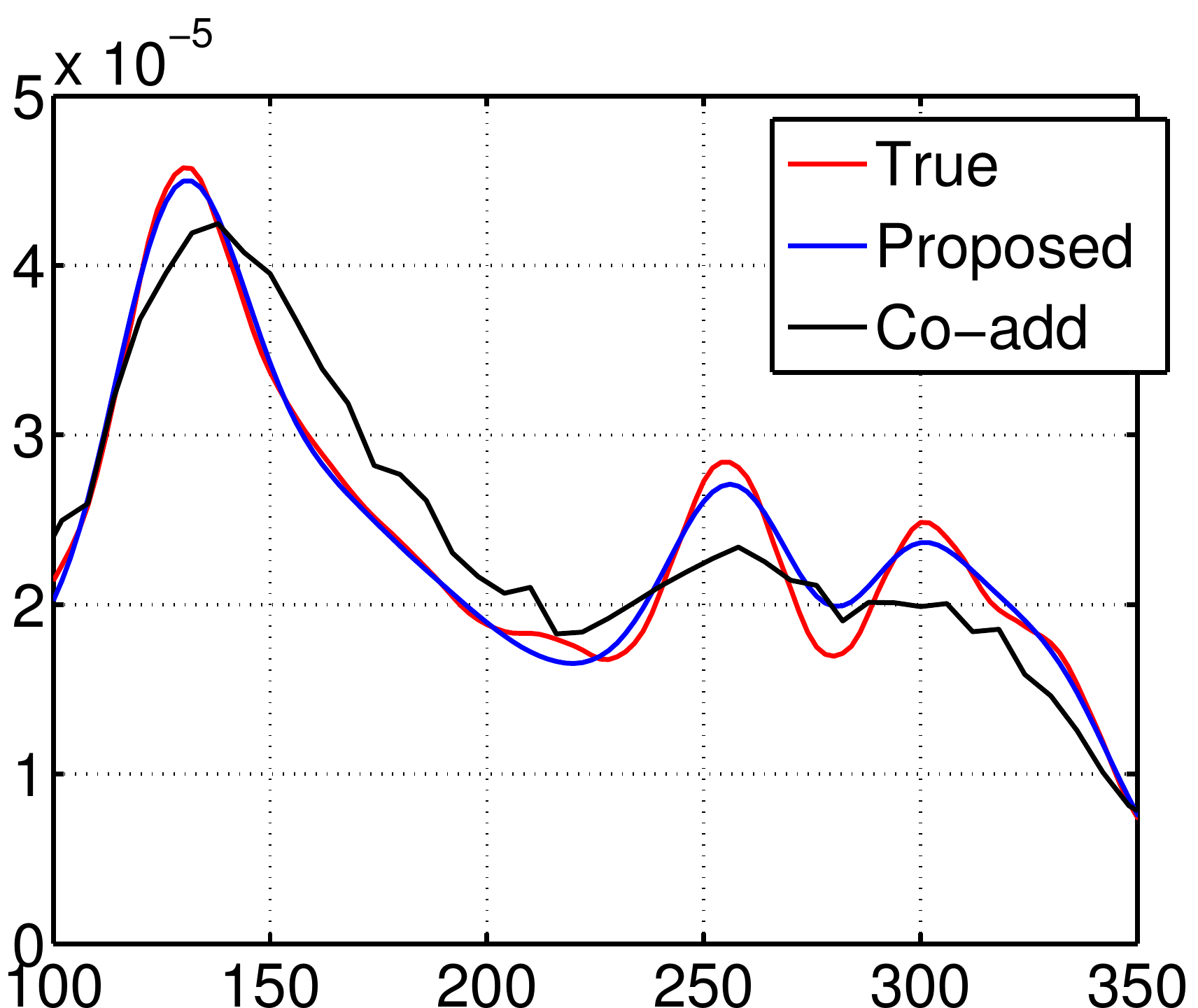}\label{fig:sliceBestGalaxieZoom}}

  \caption{Restoration of galaxy. \label{fig:Galaxie}}

\end{figure*}

Our method is based on spatial regularity information but to assess
its robustness as it is, we tested it with two other types of sky in
which the spatial regularity is less pronounced: galactic cirrus
superimposed on point sources, and a galaxy image
(Figs.~\ref{fig:cirrusDot} and~\ref{fig:Galaxie}).

The coaddition map (Fig.~\ref{fig:coaddCirrusDot}) is smoother than
the proposed one (Fig.~\ref{fig:bestCirrusDot}), and several point
sources are visible on the proposed map but not on the coaddition one.
The amplitude of point sources is underestimated but markedly less so
by the proposed method than by coaddition
(Figs.~\ref{fig:sliceBestCirrusDot} and
\ref{fig:sliceBestCirrusDotZoom}). Rebounds also appear around the
point sources, a feature characteristic of linear deconvolution
(resulting from a quadratic criterion).

The galaxy does not contain any point source but has spatial
structures that are more complex than the galactic cirrus. These
structures are considerably better restored by our method than by
coaddition (Fig.~\ref{fig:Galaxie}) and it is particularly clear
around pixels 250 and 300.

In conclusion, the proposed method is flexible and shows a good
restoration capacity for various types of maps. In particular, it
possesses a certain robustness compared to an input sky presenting
characteristics that are poorly taken into account by the \aprio model
based on regularity information. It provides a sky that is closer to
the real one than that obtained by coaddition, even in the least
favourable cases.

\subsection{Processing real data}
\label{sec:donnees-reelles}

\begin{figure*}[htbp]
  \centering
  \includegraphics[width=0.6\textwidth,angle=90]{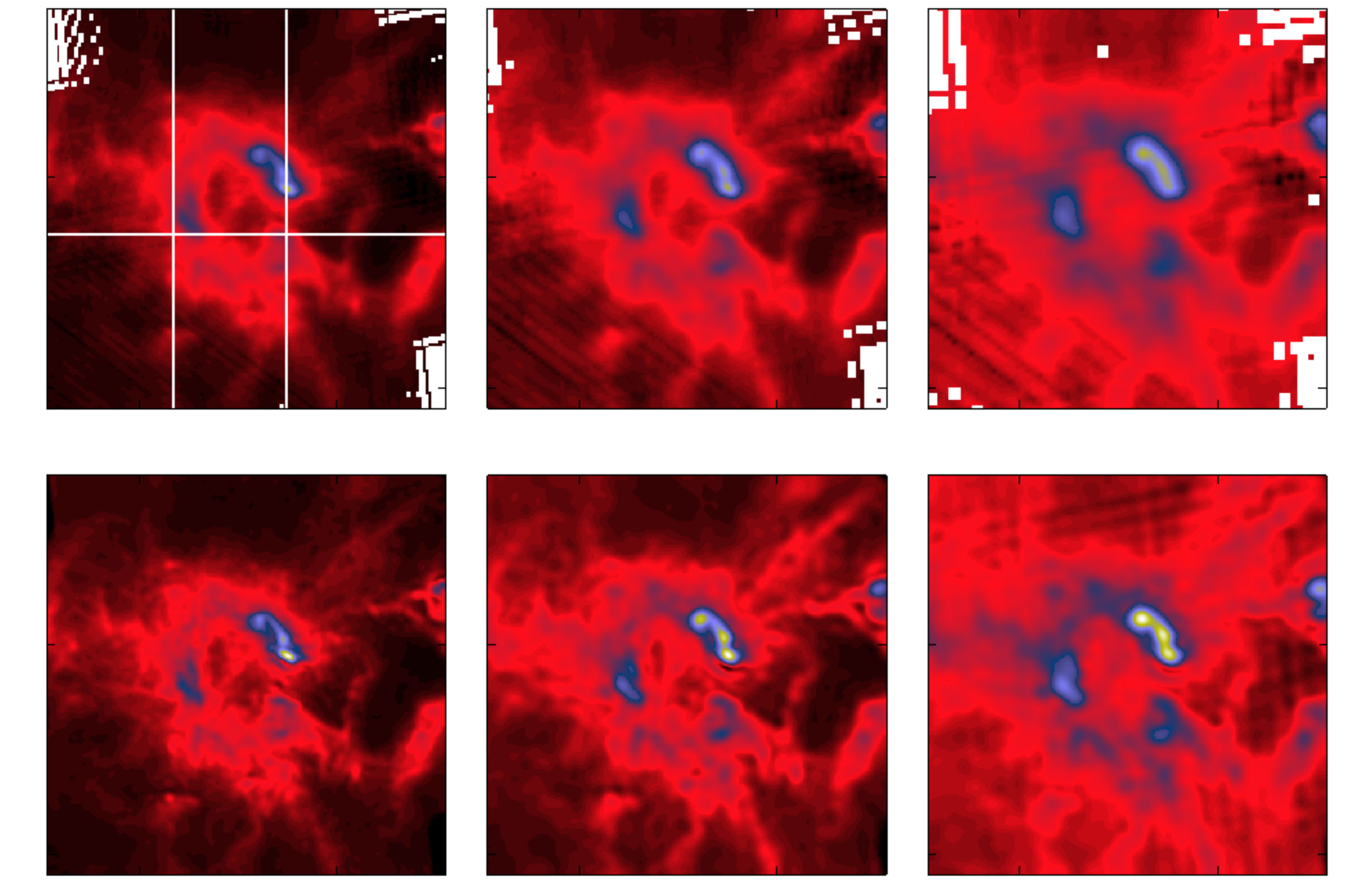}
  \caption{Central part (23\arcmin$\times$23\arcmin) of NGC\,7023 in
    the three channels PSW, PMW and PLW (left, middle and right,
    respectively). Top panels: coadded maps; bottom panels: proposed
    maps.}
  \label{fig:ngc7023_3bands_res}
\end{figure*}

\begin{figure*}[htbp]
  \centering

 \subfigure{\includegraphics[width=0.3\textwidth,angle=180]{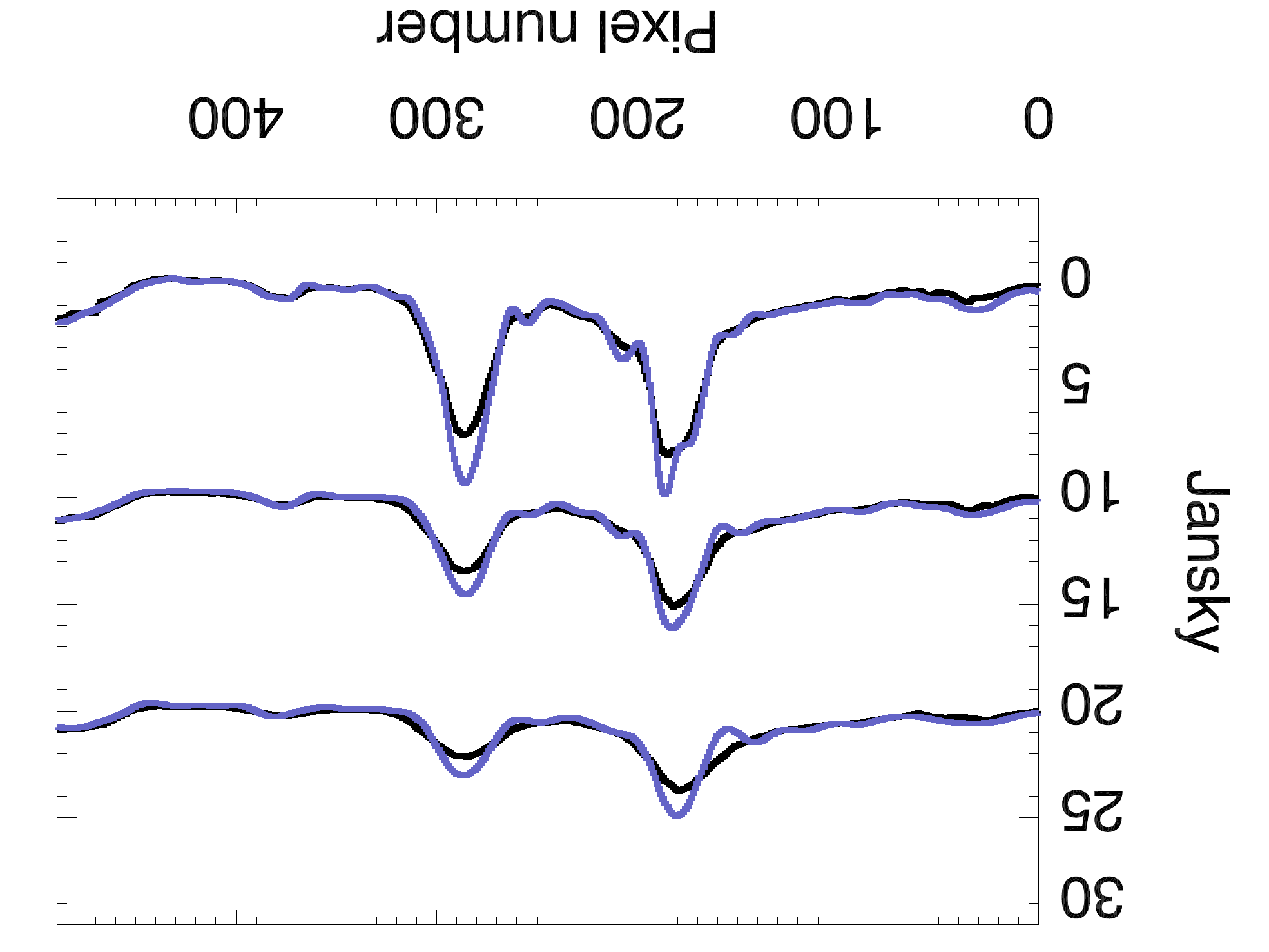}
    \label{fig:ngc7023_profiles_resB1}}%
  \subfigure{\includegraphics[width=0.3\textwidth,angle=180]{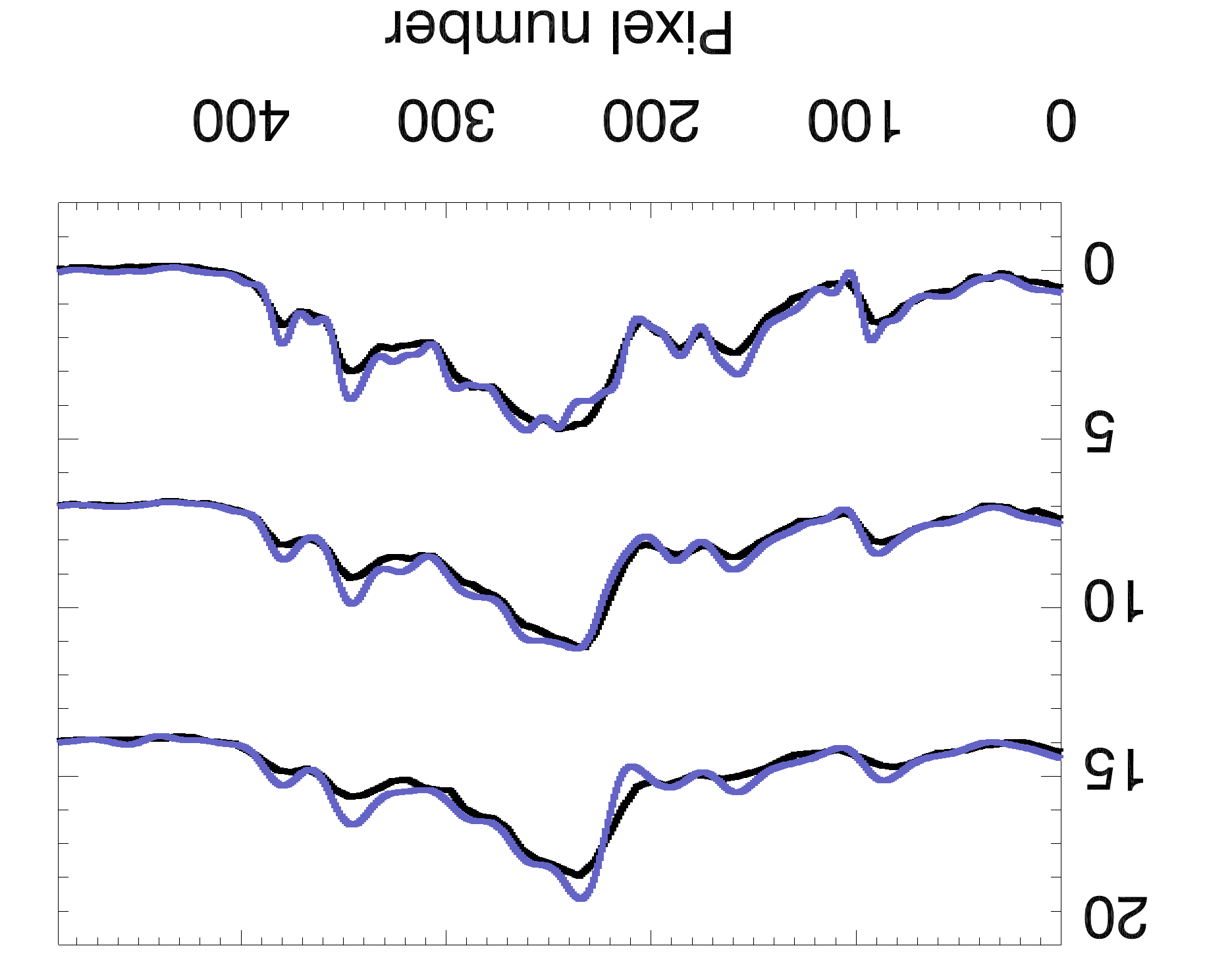}
    \label{fig:ngc7023_profiles_resB2}}%
  \subfigure{\includegraphics[width=0.3\textwidth,angle=180]{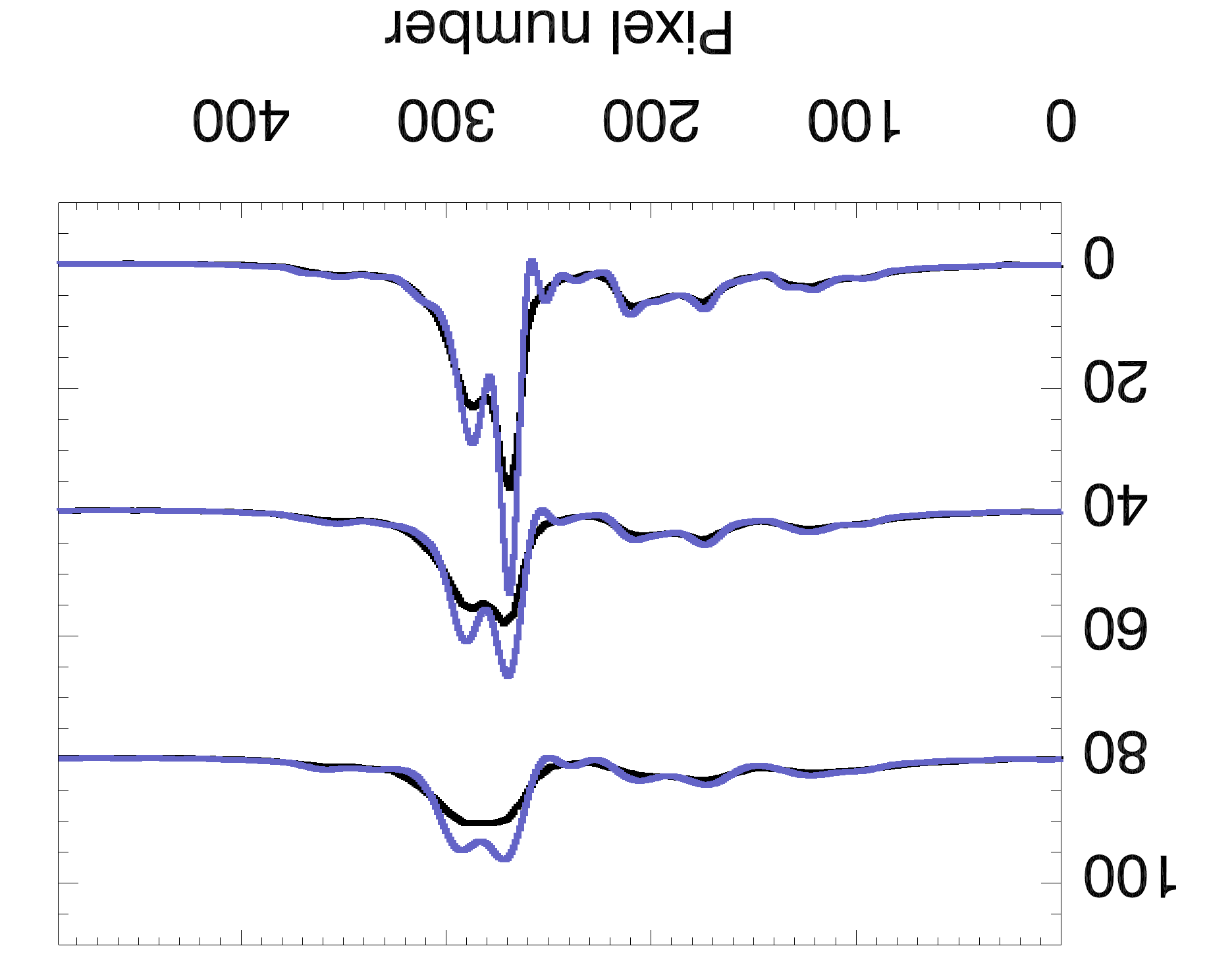}
    \label{fig:ngc7023_profiles_resB3}}

  \caption{Profiles along the three sections shown in the top left
    panel of Fig.~\ref{fig:ngc7023_3bands_res}. Each panel shows the
    profiles within the PSW, PMW and PLW channels, offset for clarity
    from bottom to top, respectively. Left panel: horizontal profile;
    central and right panels: vertical profiles. Black: coadded maps,
    blue: proposed maps.}
  \label{fig:ngc7023_profiles_res}
\end{figure*}

\begin{figure*}[htbp]
  \centering
  \includegraphics[width=0.9\textwidth,angle=0]{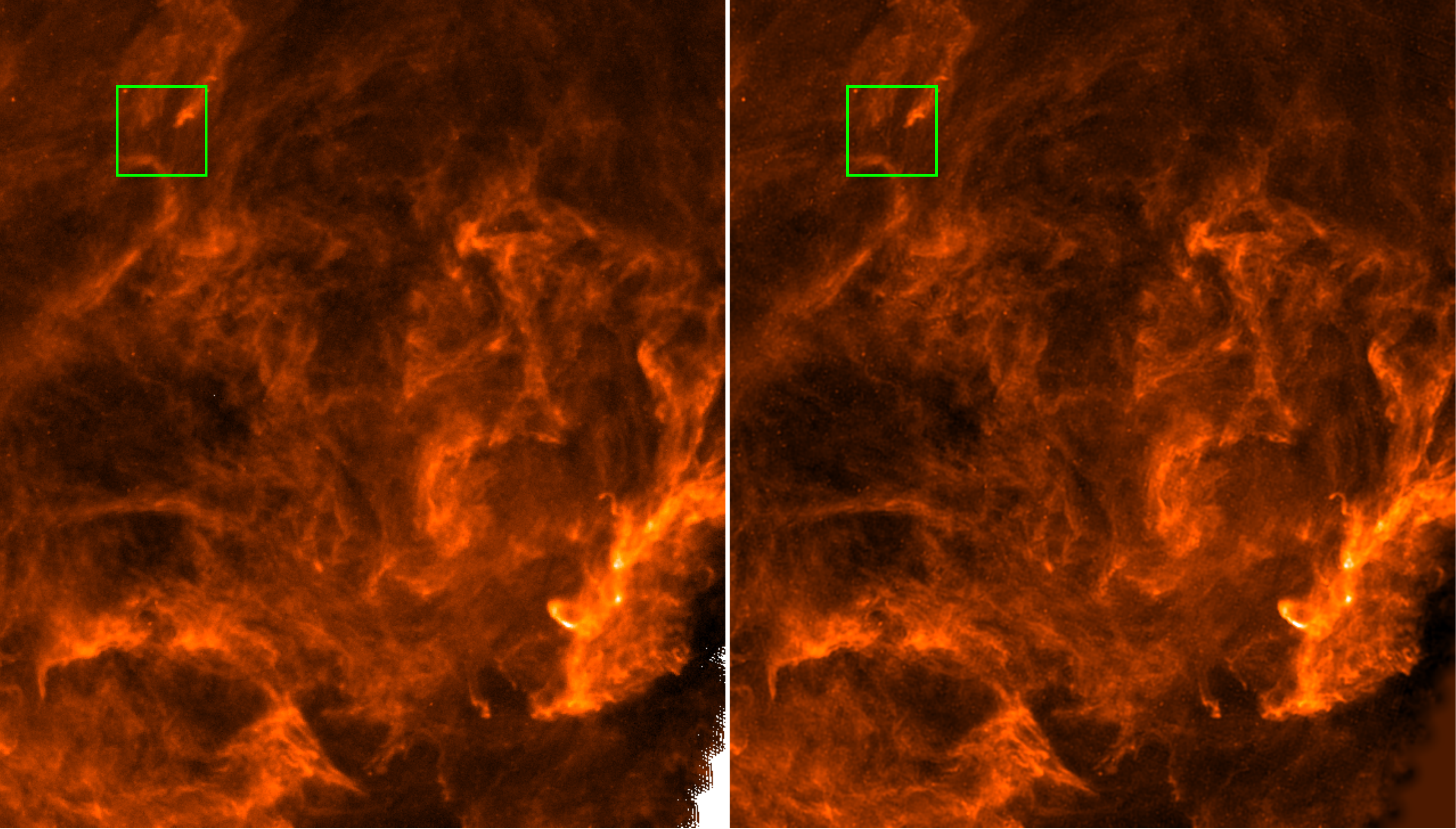}
  \caption{85\arcmin$\times$110\arcmin\ rectangle in the Polaris flare
    for the PSW channel (the total field observed during the science
    demonstration phase of Herschel is presented in
    \citep{MAMD10}). Left panel: coadded map. Right panel: proposed
    result.}
    \label{fig:polaris_images}
\end{figure*}

\begin{figure*}[h]
  \centering
  \includegraphics[width=0.7\textwidth,angle=0]{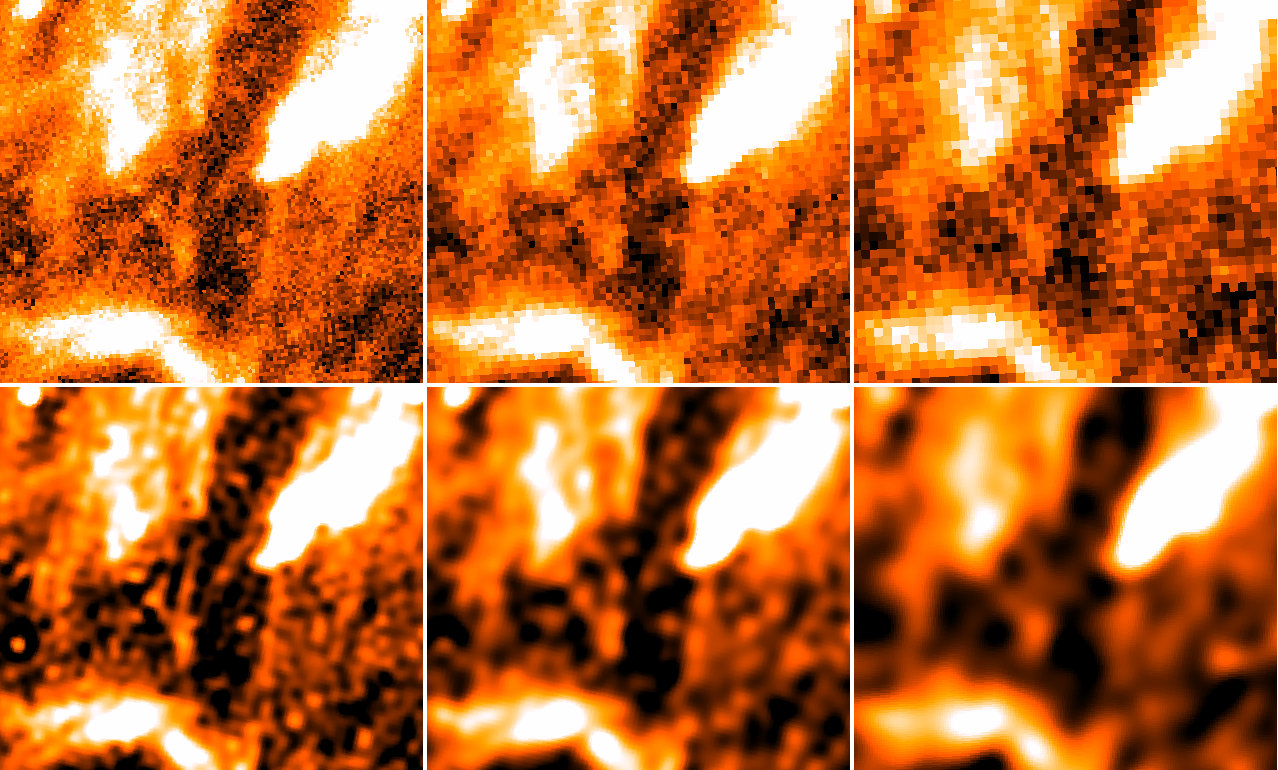}
  \caption{Zoom on the 10\arcmin$\times$10\arcmin\ green square seen
    in Fig\,\ref{fig:polaris_images}. Top panels and from left to
    right: coadded maps in the PSW, PMW and PLW channels,
    respectively; bottom panels: proposed maps in the three channels.}
    \label{fig:polaris_images_zoom}
\end{figure*}

\begin{figure*}[htbp]
  \centering

  \subfigure{\includegraphics[width=0.31\textwidth]
    {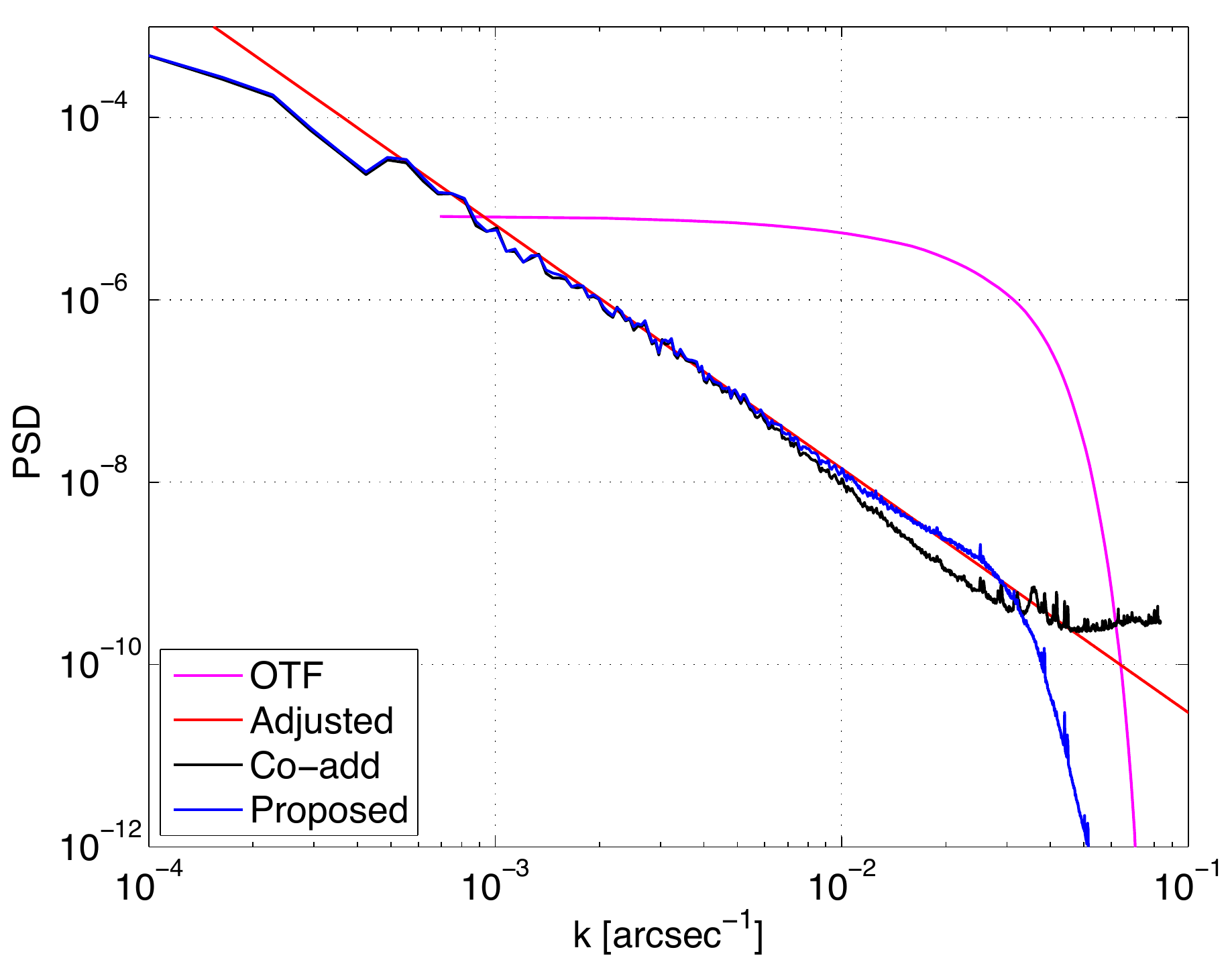}\label{fig:Polaris_PSW_all}}%
  \hfill%
  \subfigure{\includegraphics[width=0.31\textwidth]
    {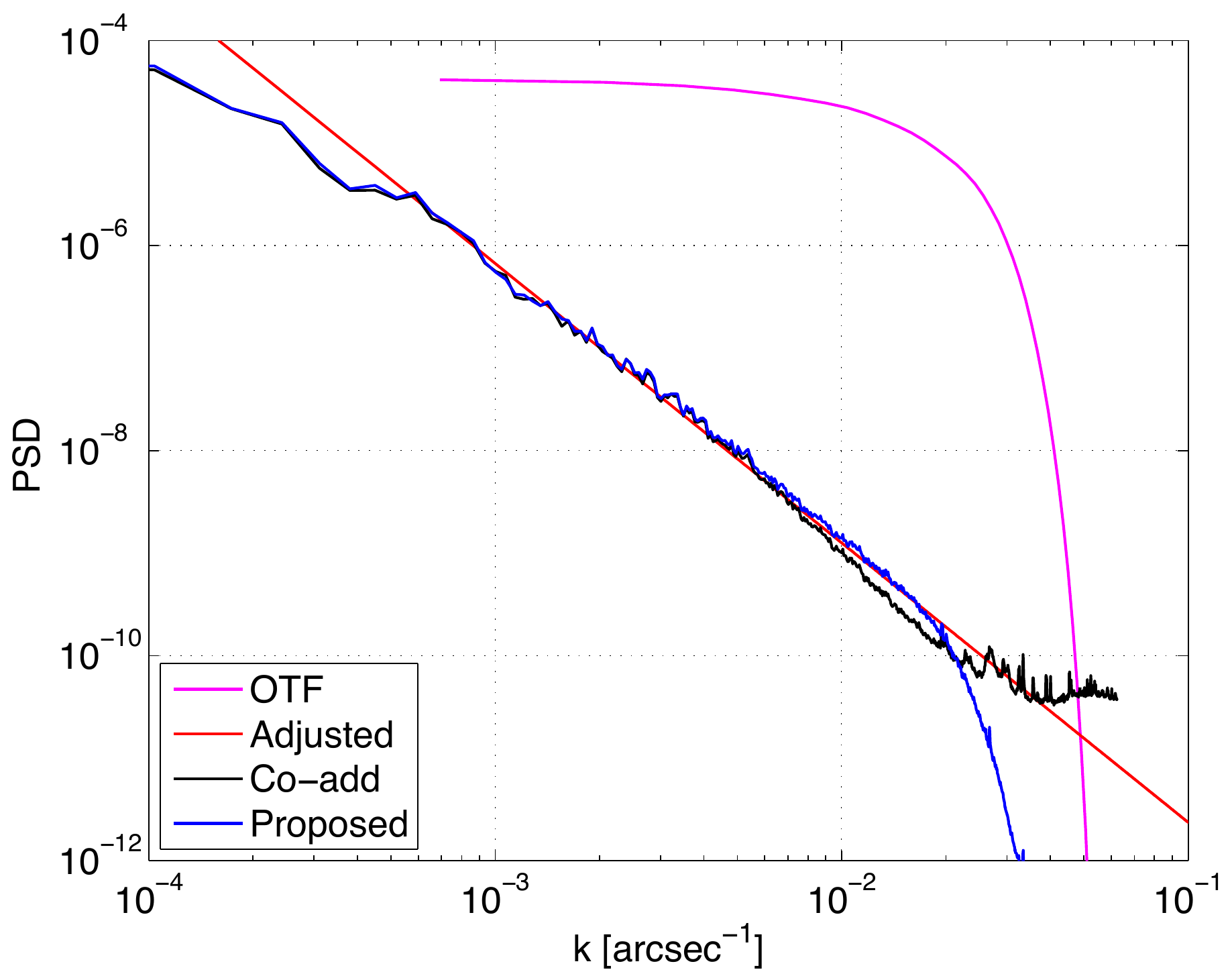}\label{fig:Polaris_PMW_all}}%
  \hfill%
  \subfigure{\includegraphics[width=0.31\textwidth]
    {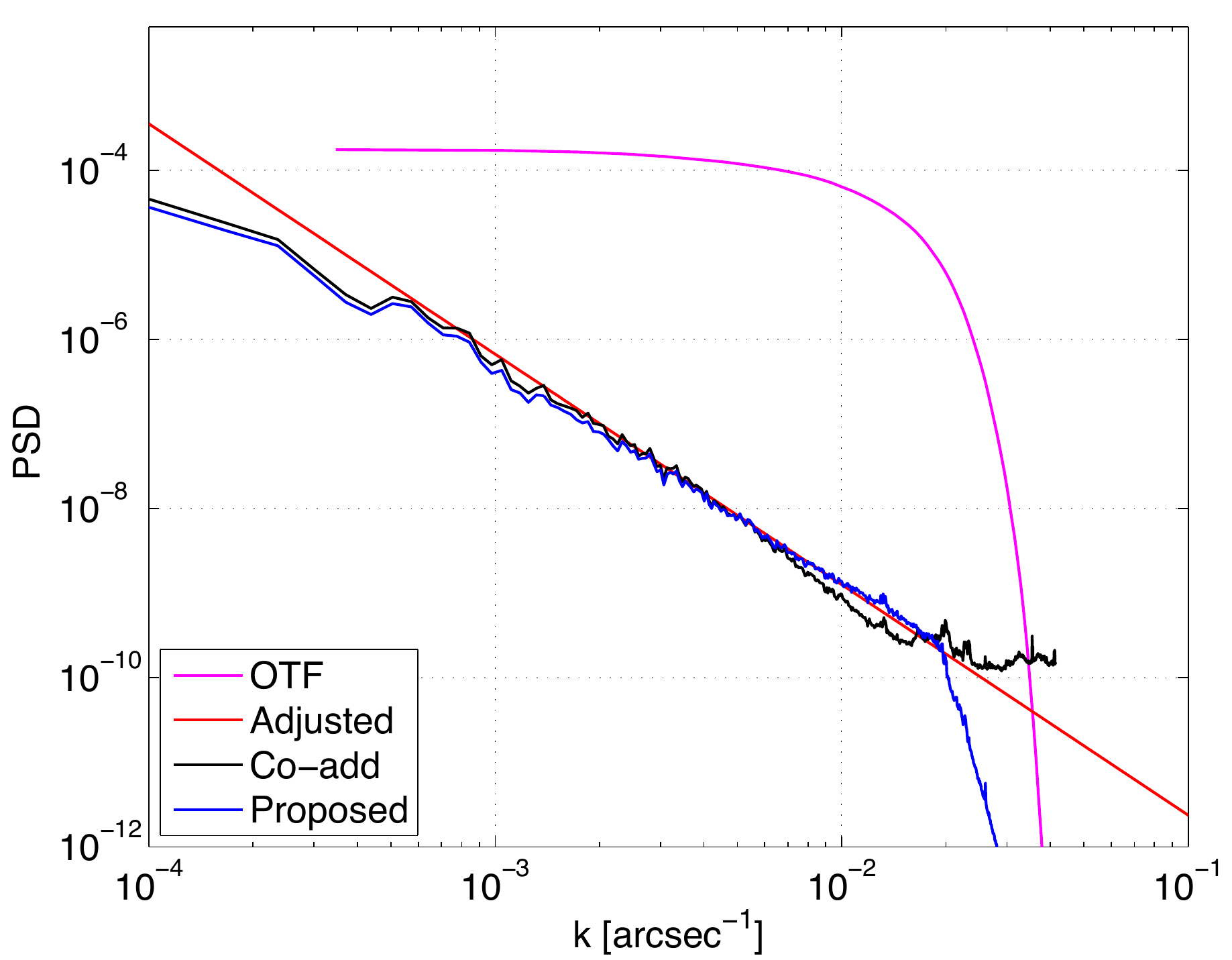}\label{fig:Polaris_PLW_all}}

  \subfigure{\includegraphics[width=0.31\textwidth]
    {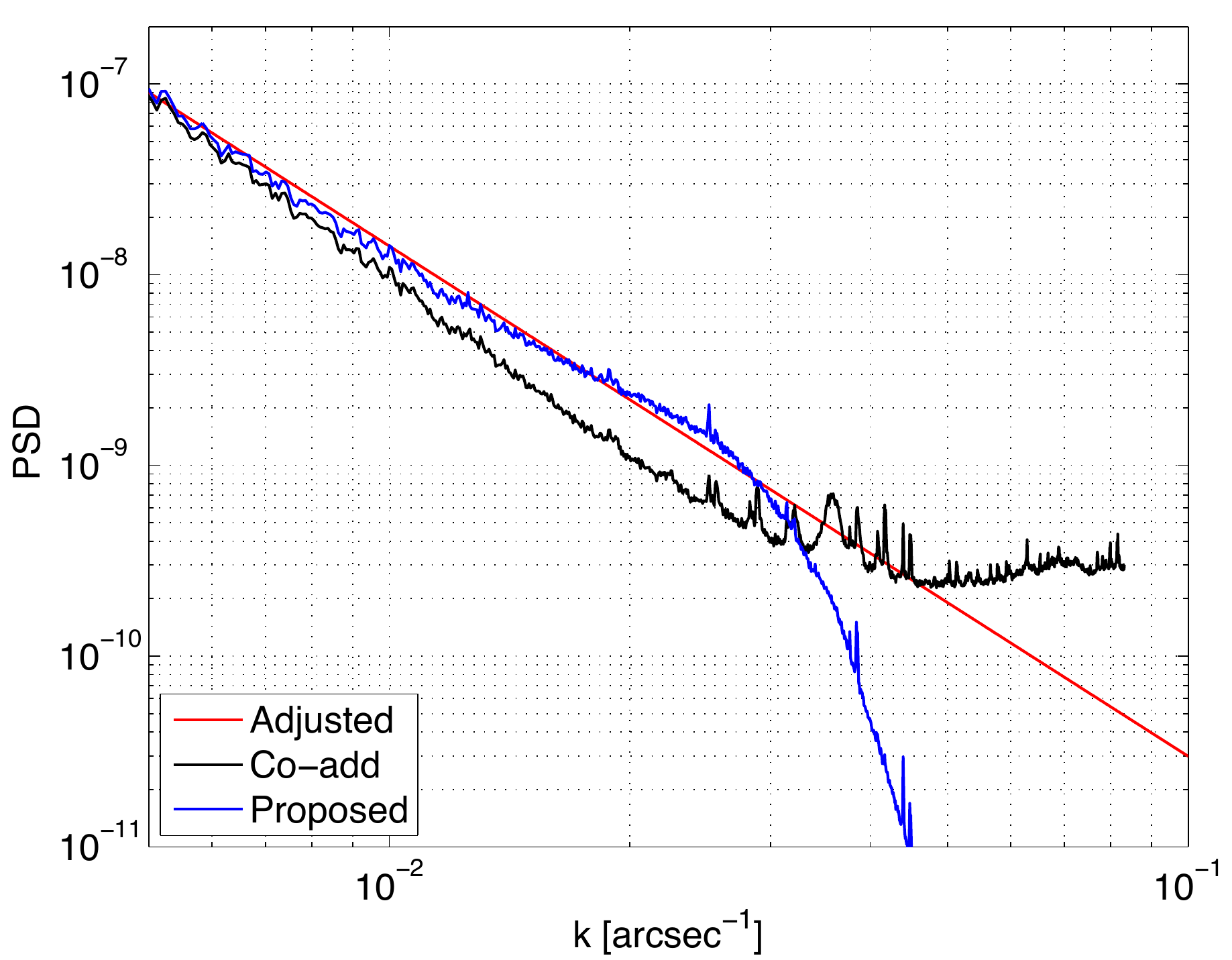}\label{fig:Polaris_PSW_small}}%
  \hfill%
  \subfigure{\includegraphics[width=0.31\textwidth]
    {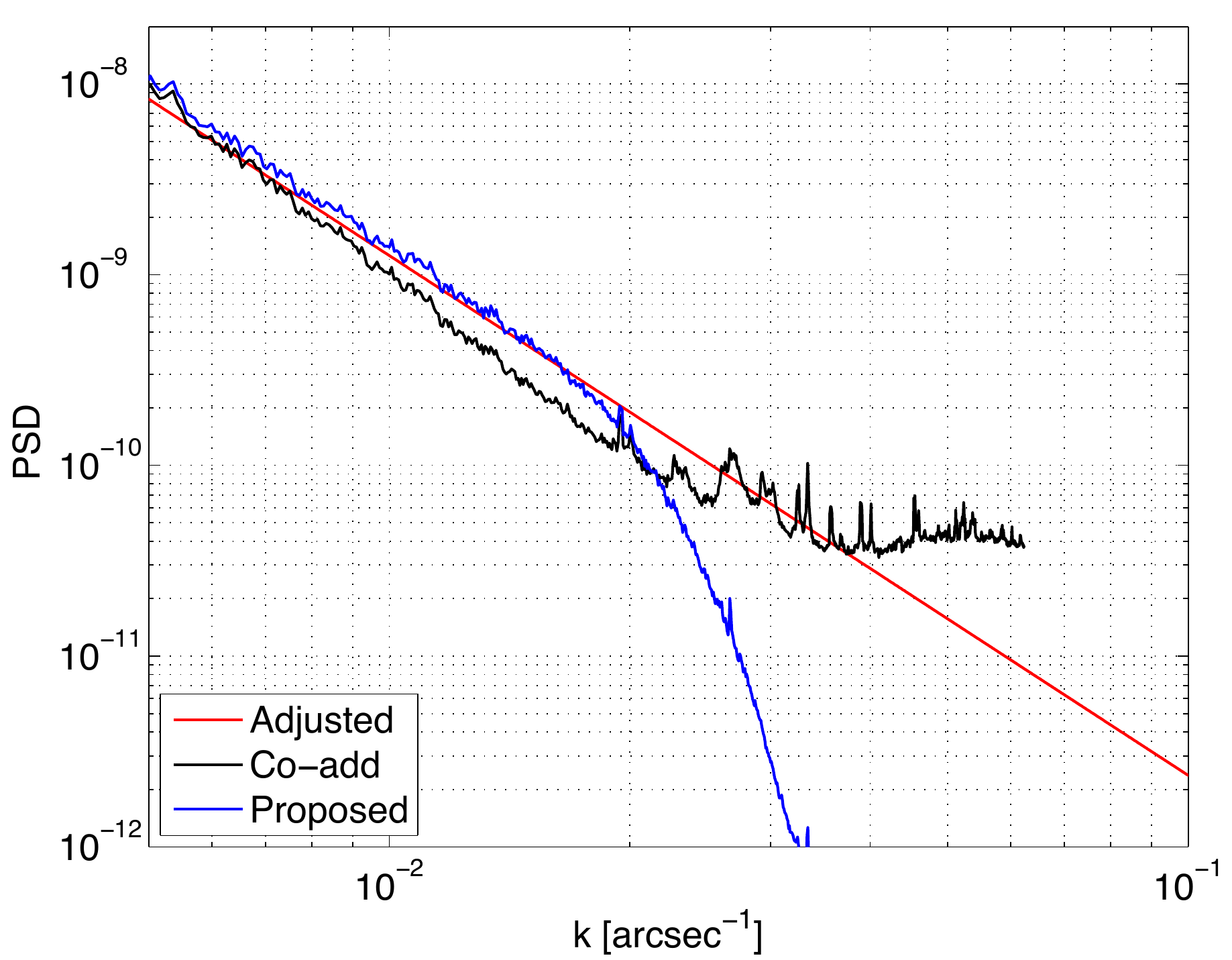}\label{fig:Polaris_PMW_small}}%
  \hfill%
  \subfigure{\includegraphics[width=0.31\textwidth]
    {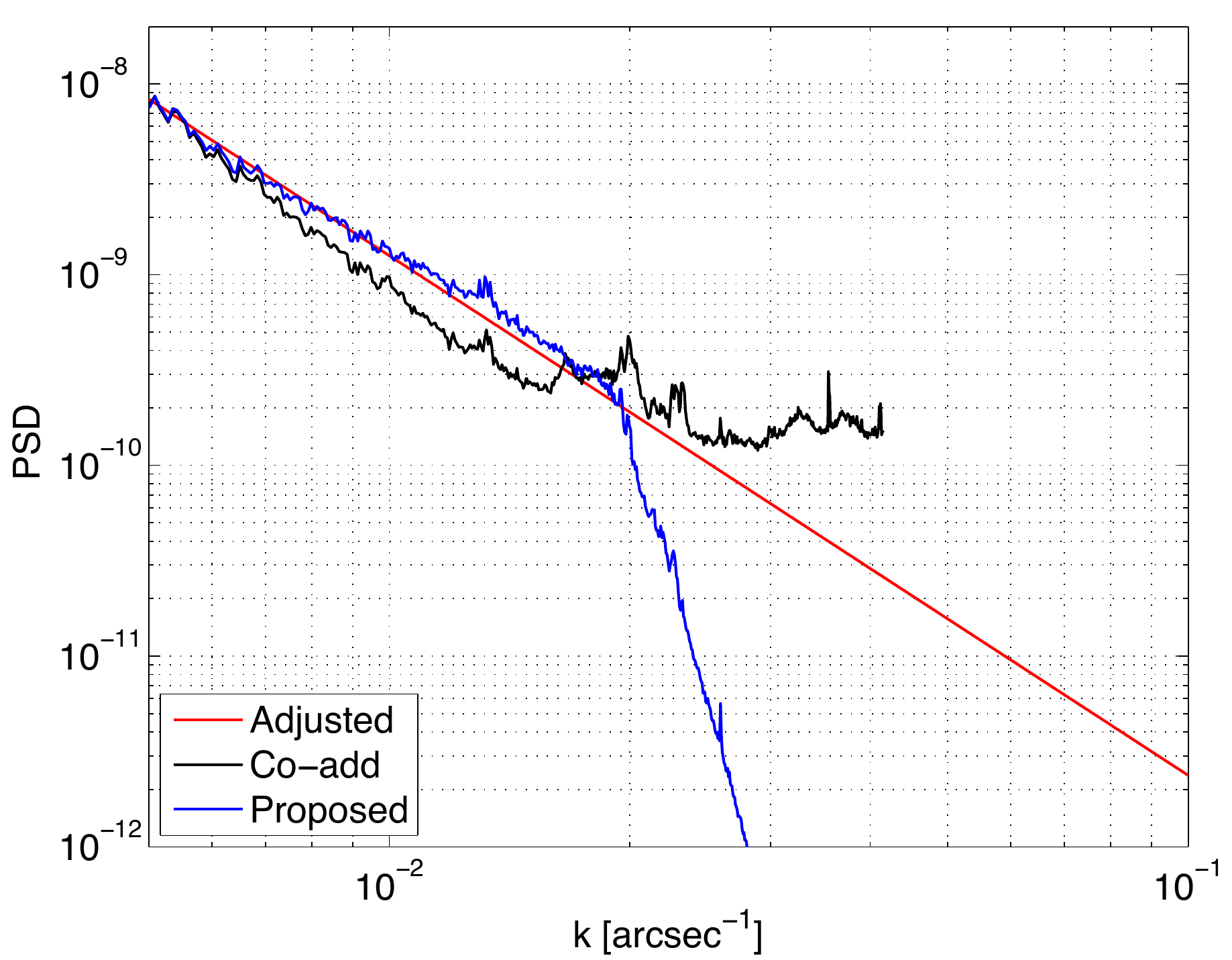}\label{fig:Polaris_PLW_small}}

  \caption{Circular means of the power spectrum of the Polaris flare
    in the PSW (left panels), PMW (middle panels) and PSW (right
    panels) channels. The bottom panels present plots on the frequency
    range from $5\times 10^{-3}$ to $10^{-1}\,\mathrm{arcsecond}^{-1}$. The red
    lines show the power law $P(k)\propto\,k^\gamma$ adjusted in a
    frequency range from $10^{-3}\,\mathrm{arcsecond}^{-1}$ to
    $3\times 10^{-2}\,\mathrm{arcsecond}^{-1}$, with $\gamma=
    -2.7$. The pink solid lines show the optical transfer functions
    (OTF) for each band.\label{fig:polaris_PSD_all}}

\end{figure*}

We conducted tests with real data of the reflection nebula NGC\,7023
and of the Polaris flare (which is a high Galactic latitude cirrus
cloud) performed during the science demonstration phase of Herschel
and already presented in \citep{Abergel10} and \citep{MAMD10},
respectively. In order to run our algorithm, we took the level-1 files
processed using HIPE. The true sky is not known, of course, so the
value of the regularization parameter was fixed for each of the
spectral channels by a trade-off between the gain in spectral
resolution and the amplification of the noise.

Figs.~\ref{fig:ngc7023_3bands_res} to \ref{fig:polaris_PSD_all}
illustrate the results for NGC\,7023 and the Polaris flare. The gain
in spatial resolution is spectacular in the three channels. It is
interesting to note that the map of NGC\,7023 obtained by our method
in the longest wavelength channel (500\,$\mu$m, PLW channel) shows
spatial structures that are not visible in the coaddition but are real
since they are visible at shorter wavelengths (250\,$\mu$m, PSW
channel), as illustrated for instance in the right panel of
Fig.~\ref{fig:ngc7023_profiles_res}. The same panel also shows that
negative rebounds appear on the sharpest side of the brightest
filament of NGC\,7023. This filament is the narrowest structure of the
map and its width is comparable to the width of the PSF. Similar
rebounds were also seen in our simulations with point sources
(Fig.\,\ref{fig:cirrusDot}). The Polaris flare does not contain
comparable bright and narrow filament, so the proposed map does not
present this kind of artefact. The zoom on a
10\arcmin$\times$10\arcmin\ square
(Fig.\,\ref{fig:polaris_images_zoom}) illustrates the gain in angular
resolution for faint structures.

Fig.~\ref{fig:polaris_PSD_all} shows that the power spectra of the
proposed maps of the Polaris flare in the three channels follow the
expected power law that is typical of the infrared emission of high
Galactic cirrus $P(k)\;\alpha\;k^\gamma$ with $\gamma=-2.7$ (\eg,
\citep{MAMD10}) on a frequency range from $10^{-3}$ to $3\times
10^{-2}\,\mathrm{arcsecond}^{-1}$. For the simulated data of our
Section~\ref{sec:donnees-sim}, the attenuation effect of the
convolution by the PSF is accurately inverted up to the frequency
where the noise is dominant. Thanks to this correction, the contrast
of the small-scale structures is enhanced
(Figs.\,\ref{fig:polaris_images} and \ref{fig:polaris_PSD_all}) \wrt
the coaddition, since the energy of each structure is put in a shorter
number of pixels than for the coaddition.  At smaller frequencies,
\citep{MAMD10} have shown that the SPIRE spectra are attenuated
compared to IRAS, which is likely owing to the correction of $1/f$
noise attributed to thermal drifts in the preprocessing of the data.

\section{Conclusion}
\label{sec:conclusion}

We have proposed a new method for super-resolved image reconstruction
for scanning instruments and its application to the SPIRE instrument
of the Herschel observatory.

The first key element is an instrument model that describes the
physical processes involved in the acquisition. To explain the data in
a reliable way, our model combines the descriptions of three elements:
(i)~the sky as a function of continuous variables in the three
dimensions (two spatial and one spectral), (ii)~the optics and the
rest of the instrumentation (bolometer, electronics, etc.) and
(iii)~the scanning strategy. We thus arrived at a linear model in
integral form (Eq.~\eqref{eq:8}). We then wrote it in a matrix form
(Eq.~\eqref{Eq:ModelInstruDiscretDiscret}) by making certain
calculations explicit. Next, by coming close to the pointed directions
(on a fine grid), we decomposed it into a convolution followed by
inhomogeneous down-sampling
(Eq.~\eqref{Eq:ModelInstruFactorise}). This model provides a faithful
link between the data, the sky actually observed, and the instrument
effects.

On the sole basis of this instrument model and the data, the inversion
is an ill-posed problem, especially if resolution enhancement is
desired.  The lack of information brought by the data, considering the
limitations of the instrument, leads to instability of the inversion,
which is all the more noticeable when the target resolution is
high. This difficulty is overcome by a standard regularization method
that constitutes the second key element. The method relies on spatial
regularity information introduced by quadratic penalisation and on a
quadratic data attachment term, the trade-off being governed by a
regularization parameter.  Thus, the inversion is based on a
relatively standard linear approach and its implementation uses
standard numerical optimization tools (conjugate gradient with optimal
step).

The presented results for the SPIRE instrument illustrate, for
simulated and real data, the potential of our method. Through the use
of the accurate instrument model and \aprio regularity information, we
restored spatial frequencies over a bandwidth $\sim4$ times that
obtained with coaddition. In all channels, the attenuation by the
optical transfer function is accurately inverted up to the frequency
where the noise is dominant.  The photometry is also properly
restored.

A future work will focus on the question of hyperparameter and
instrument parameter estimation, that is to say, unsupervised and
myopic problems. We have a work in progess about this problem and it
is developed in a Bayesian framework and resorts to an Markov Chain
Monte-Carlo algorithm. Moreover, an estimation of the correlation
matrix parameters (cutoff frequency, attenuation coefficients,
spectral profile, etc.) could be achieved for the object or the noise
(typically for the $1/f$ noise).

From another perspective, quadratic prior is known for possible
excessive sharp edge penalisation in the restored object. The use of
convex $\mbox{L}_2-\mbox{L}_1$
penalisation~\citep{Kunsch94,Charbonnier97,Mugnier04,Thiebaut08} can
overcome this limitation, if needed. Moreover, the proposed method can
be specialized to deal with construction/separation of two
superimposed components: (i) an extended component together with (ii)
a set of point sources~\citep{Giovannelli05}.

Finally, another relevant contribution could rely on the introduction
of the spectral dependence between the different channels in the data
inversion.  The conjunction with a PACS direct model and the joint
inversion of SPIRE and PACS data would greatly improve the map
reconstruction.

\section*{Acknowledgment}
\label{sec:acknowledgment}

The authors would like to thank M. Griffin (\textsc{cspa} -- Cardiff
University), B. Sibthorpe (\textsc{ukatc} -- Royal Observatory
Edinburgh), G. Le Besnerais (\textsc{dtim -- onera}) and F. Champagnat
(\textsc{dtim -- onera}), for fruitful discussions, and \textsc{cnes}
and the \textsc{astronet} consortium for funding. The authors also
thank K. Dassas for her help in the preprocessing of SPIRE data.

\appendix{}

\section{Energy spectral density}
\label{Sec:AnnexeRegularite}

This appendix gives the details of the calculations concerning the
regularity measure used in Section~\ref{sec:inversion} and its
frequency interpretation. Based on Eq.~\eqref{Eq:DecompoCiel}, the
energy of the first derivative can be written
\begin{align*}
  \left\| \frac{\partial \Ciel}{\partial \alpha}\right\|^2 & =
  \iint_{\eR^2} \left( \frac{\partial \Ciel}{\partial \alpha}
  \right)^2 \, \dD \alpha \,\dD \beta \\
  & = \sum_{ij i'j'} \xij\,\xijprime \iint_{\eR^2}
  \left(\frac{\partial}{\partial \alpha}\psi_{i'j'}\right) \left(
    \frac{\partial}{\partial \alpha} \psi_{ij} \right) \, \dD \alpha\,
  \dD \beta.
\end{align*}
By noting the derivative
$\psi_{\alpha}'=\froc{\partial\psi}{\partial\alpha}$, we obtain the
autocorrelation $\Psi_{\alpha} = \psi_{\alpha}' \conv \psi_{\alpha}'$
of the first derivative of the generating function and we have
\begin{align}
  \left\| \frac{\partial \Ciel}{\partial \alpha}\right\|^2 & =
  \sum_{ij i'j'} \xij\,\xijprime \iint_{\eR^2} \psi_{\alpha}'
  \Big(\alpha - i'\Ta, \beta - j'\Tb\Big) \nonumber \\
  & \hspace{2cm} \psi_{\alpha}' \Big(\alpha -
  i\Ta , \beta - j\Tb\Big)\, \dD \alpha\, \dD \beta \nonumber \\
  & = \sum_{ij i'j'} \xij\,\xijprime \left[\psi_{\alpha}' \conv
    \psi_{\alpha}' \right]
  \acc{ (i' - i)\Ta, (j' - j)\Tb } \nonumber\\
  & = \sum_{ij i'j'} \xij\,\xijprime \Psi_{\alpha} \acc{ (i' - i)\Ta,
    (j' - j)\Tb }. \label{Eq:RegulariteFinale}
\end{align}
As there is a finite number of coefficients $\xij$, the measure can be
put in the form of a quadratic norm
\begin{equation*}
  \left\| \frac{\partial \Ciel}{\partial \alpha} \right\|^2 = \xb^t
  \Dba \xb
\end{equation*}
where the matrix $\Dba$ is obtained from $\Psi_{\alpha}$. Considering
the invariant structure of~\eqref{Eq:RegulariteFinale}, the matrix
$\Dba$ has a \Toeplitz structure. The calculation is performed by
discrete convolution and can be computed by FFT.

By introducing the dimension $\beta$,
\begin{equation*}
  \left\| \frac{\partial \Ciel}{\partial \alpha} \right\|^2
  + \left\| \frac{\partial \Ciel}{\partial \beta} \right\|^2
  = \xb^t \Dba \xb + \xb^t \Dbb \xb.
\end{equation*}
The quadratic regularity measure on the function $\Ciel$ with
continuous variables is expressed through a quadratic regularity
measure on the coefficients $\xb$.

The autocorrelation Fourier transform (FT) is the energy spectral
density, \ie the squared modulus of the FT of~$\psi_{\alpha}'$
\begin{align*}
  \rond \Psi_{\alpha}(\fa, \fb) & = \iint_{\eR^2}
  \Psi_{\alpha}(\alpha,\beta) e^{-2j\pi
    (\alpha\fa + \beta\fb) } \,\dD\alpha \,\dD\beta \\
  & = \left| \iint_{\eR^2} \psi_{\alpha}'(\alpha,\beta) e^{-2j\pi
      (\alpha\fa + \beta\fb)} \,\dD\alpha \,\dD\beta \right|^2 \\
  & = 4\pi^2 \fa^2 \left| \rond \psi(\fa,\fb) \right|^2,
\end{align*}
where $\rond \psi$ is the FT of $\psi$. When the dimension $\beta$ is
introduced, the \aprio energy spectral density for the sky naturally
has circular symmetry
\begin{equation}
  \label{eq:29}
  \rond \Psi(\fa, \fb) = 4\pi^2 \left(\fa^2 + \fb^2\right) \left|
    \rond \psi(\fa,\fb) \right|^2.
\end{equation}
This calculation brings out the frequency structure introduced \aprio
for the sky according to the chosen function $\psi$. This is a
high-pass structure since the factor $\fa^2 + \fb^2$ tends to cancel
$\rond \Psi$ around zero, which is consistent with a regularity
measure.

\section{Explicit calculation of the model}
\label{Ann:ModelCalculExplicite}

In order to integrate over time in~\eqref{eq:15}, we use the
expressions of~\eqref{eq:2} for $\pa(t)$ and $\pb(t)$, which give
\begin{multline*}
  \frac{1}{2 \pi} \frac{1}{\Sa \Sb} \int_t
  \Exp{-\frac{1}{2} \frac{(\va t + \ca + \alpha^{ij} - \alm)^2}{\Sa^2}}\\
  \Exp{-\frac{1}{2} \frac{(\vb t + \cbe + \beta^{ij} -\blm)^2}{\Sb^2}}
  \RepBolo(n\Te - t) \,\dD t \,.
\end{multline*}
With the bolometer response
\begin{equation*}
  \RepBolo(n\Te - t) = \mathds{1}_{[0~+\infty[}(n\Te -
  t)S\Exp{-\frac{n\Te - t}{\tau}} \,,
\end{equation*}
we have
\begin{multline}
  \label{eq:1000}
  \frac{1}{2 \pi} \frac{S}{\Sa \Sb}
  \Exp{-\frac{n\Te}{\tau}}\int^{n\Te}_{-\infty} \\ \Exp{-\frac{1}{2}
    \frac{(\va t + \oa)^2}{\Sa^2}} \Exp{-\frac{1}{2} \frac{(\vb t +
      \obeta)^2}{\Sb^2}} \Exp{\frac{t}{\tau}} \,\dD t
\end{multline}
with $\oa = \ca + \alpha^{ij} - \alm$ and $\obeta = \cbe + \beta^{ij}
- \blm$. This is the integration of a truncated Gaussian since the
argument of the exponential is a quadratic form \wrt~$t$.

\subsection{Calculation of the argument}
\label{sec:calcul-de-largument}

Here, we express the quadratic form in question
\begin{equation*}
  n(t) = \tau \Sb^2 (\va t + \oa)^2 + \tau \Sa^2 (\vb t + \obeta)^2
  - 2\Sa^2\Sb^2t.
\end{equation*}
Expanding and factorizing the numerator $n(t)$ gives
\begin{align*}
  n(t) & = \tau \Sb^2 \left(\va^2 t^2 + 2\va\oa t + \oa^2\right) + \\
  & \hspace{2cm} \tau \Sa^2 \left(\vb^2 t^2 + 2\vb\obeta t + \obeta^2 \right) - 2\Sa^2\Sb^2t \\
  & = \left( (t + a)^2 + b - a^2 \right) / \Sigma^2
\end{align*}
with the constants $a = \Sigma^2 \left( \tau \Sb^2\va\oa + \tau
  \Sa^2\vb\obeta - \Sa^2\Sb^2 \right)$, $b = \tau
\Sigma^2\left(\Sb^2\oa^2 + \Sa^2\obeta^2 \right)$ and $\Sigma^{-2} =
\tau \left( \Sb^2\va^2 + \Sa^2\vb^2 \right)$. Putting this
$t$-quadratic form into the integral, we obtain
\begin{multline}
  \label{Eq:AnnexIntermediaire}
  \frac{1}{2 \pi} S \frac{\sqrt{\pi\tau} \Sigma}{\sqrt{2}}
  \Exp{-\frac{n\Te}{\tau}-\frac{1}{2}\frac{b -
      a^2}{\Sigma^2\Sa^2\Sb^2\tau}} \\ \left( 1 + \erf \left(\frac{n\Te +
        a}{\sqrt{2 \tau}\Sigma\Sa\Sb} \right)\right),
\end{multline}
where the function $\erf$ is defined by
\begin{equation}
  \erf(x) = \frac{2}{\sqrt{\pi}}\int_0^x e^{-\theta^2}\,\dD \theta = -
  \erf(-x). \nonumber
\end{equation}
This expression can be simplified by using the function $\erfcx(x) =
\exp(x^2) (1 - \erf(x))$.

\subsection{Argument of the exponential}
\label{sec:argum-de-lexp}

For the sake of notational simplcity, let us note $S =
\Sigma^2\Sa^2\Sb^2$. The argument of the function $\exp$ then is
\begin{multline*}
  \frac{n\Te}{\tau} - \frac{b - a^2}{2\Sigma^2\Sa^2\Sb^2\tau} = -
  \frac{n\Te}{\tau} - \frac{b - a^2}{2S \tau} \\ +
  \left(\frac{n^2\Te^2}{2S\tau} - \frac{n^2\Te^2}{2S\tau} \right) +
  \left(\frac{2n\Te a}{2S\tau} - \frac{2n\Te a}{2S\tau} \right)
\end{multline*}
and then
\begin{multline*}
  \frac{n\Te}{\tau} - \frac{b - a^2}{2\Sigma^2\Sa^2\Sb^2\tau} = -
  \frac{n\Te}{\tau} - \\ \frac{b + 2n\Te a + n^2\Te^2}{2S \tau} + \left(
    \frac{n\Te + a}{\sqrt{2S\tau}}\right)^{2}.
\end{multline*}
So, by injecting this expression in~\eqref{Eq:AnnexIntermediaire}, the
function $\erfcx$ appears
\begin{multline*}
  \Exp{- \frac{n\Te}{\tau} - \frac{b - a^2}{2S\tau}} \left( 1 + \erf
    \left( \frac{n\Te + a}{\sqrt{2S\tau}} \right) \right) = \\
  \Exp{ - \frac{n\Te}{\tau} -\frac{b + 2n\Te a + n^2\Te^2}{2S \tau} }
  \erfcx \left( -\frac{n\Te + a}{\sqrt{2 S \tau}} \right).
\end{multline*}
The values of $S, a$ and $b$ can be replaced. First of all, the
argument of the exponential is
\begin{multline}
  \label{eq:3000}
  - \frac{n\Te}{\tau} - \frac{ \tau \Sigma^2\left(\Sb^2\oa^2 +
      \Sa^2\obeta^2 \right) }{2 \Sigma^2\Sa^2\Sb^2 \tau} - \\ \frac{2n\Te
    \Sigma^2 \left( \tau \Sb^2\va\oa + \tau \Sa^2\vb\obeta -
      \Sa^2\Sb^2 \right) }{2 \Sigma^2\Sa^2\Sb^2 \tau} -
  \frac{n^2\Te^2}{2S\tau} = \\ - \frac{n\Te}{\tau} -
  \frac{\oa^2}{2\Sa^2} - \frac{\obeta^2 }{2\Sb^2 } - \frac{2n\Te
    \va\oa}{2 \Sa^2} - \frac{2n\Te\vb\obeta }{2\Sb^2} + \\
  \frac{n\Te}{\tau} - \frac{n^2\Te^2}{2 \Sigma^2\Sa^2\Sb^2 \tau} \,,
\end{multline}
and the terms $n\Te/\tau$ simplify.  We then use the expression for
$\Sigma^{2}$
\begin{equation*}
  \frac{n^2\Te^2}{2 \Sigma^2\Sa^2\Sb^2 \tau} =
  \frac{n^2\Te^2 \va^2 }{ 2 \Sa^2} + \frac{n^2\Te^2 \vb^2}{ 2 \Sb^2} \,,
\end{equation*}
to obtain two perfect squares. Finally the argument of the
exponential~\eqref{eq:3000} in~\eqref{Eq:AnnexIntermediaire} is
\begin{equation}
  -\frac{\left( \oa + n\Te\va \right)^2}{2 \Sa^2} - \frac{\left( \obeta
      + n\Te\vb\right)^2}{2 \Sb^2}. \label{eq:4000}
\end{equation}
which is exactly the argument of a bivariate Gaussian. We again find
the same standard deviations $\Sa$ and $\Sb$. However, the response of
the optics, initially $(\oa,\obeta)$, is now shifted by $(n\Te
\va,n\Te \vb)$, \ie the pointing difference between two successive
time samples.

\subsection{Argument of the function erfcx and final expression}
\label{sec:pour-la-fonction}

Another term is needed to know the global response. It comes
from the function $\erfcx$, which corresponds to the influence of the
bolometer. The argument of the function $\erfcx$ is
\begin{multline}
  - \frac{n\Te + a}{\sqrt{2 \tau}\Sigma\Sa\Sb} =
  \frac{\Sa\Sb}{\sqrt{2}\tau \Sv} - \\ \frac{\Sb\va(\oa + n\Te
    \va)}{\sqrt{2}\Sa\Sv} - \frac{\Sa\vb(\obeta +
    n\Te\vb)}{\sqrt{2}\Sb\Sv}, \label{eq:5000}
\end{multline}
where $\Sv^2 = \Sb^2 \va^2 + \Sa^2\vb^2$, and what is of interest here
is that the same factors are found in the argument of the
exponential. To know the global response, we need to bring everything
together. By injecting the expressions of the
arguments~\eqref{eq:4000} and~\eqref{eq:5000}, we obtain
\begin{multline*}
  \frac{1}{2 \pi} S \frac{\sqrt{\pi\tau} \Sigma}{\sqrt{2}}
  \Exp{\frac{n\Te}{\tau}-\frac{1}{2}\frac{b -
      a^2}{\Sigma^2\Sa^2\Sb^2\tau}} \\ \left( 1 + \erf \left( \frac{n\Te
        + a}{\sqrt{2 \tau}\Sigma\Sa\Sb} \right) \right) = \\
  \frac{S}{2 \sqrt{2\pi}\Sv}\Exp{-\frac{\left( \oa + n\Te\va
      \right)^2}{2 \Sa^2} - \frac{\left( \obeta + n\Te\vb\right)^2}{2
      \Sb^2}} \\ \erfcx \left( \frac{\Sa\Sb}{\sqrt{2}\tau \Sv} -
    \frac{\Sb\va(\oa + n\Te \va)}{\sqrt{2}\Sa\Sv} -
    \frac{\Sa\vb(\obeta + n\Te\vb)}{\sqrt{2}\Sb\Sv} \right)
\end{multline*}
with, similarly for $\alpha$ and $\beta$: $\Sab^2 = \sab^2 + \Smc^2$,
which finishes the integration of~\eqref{eq:15} over time.

\section{Direct model computation algorithm}
\label{Ann:CalculModelFacto}

This part gives some more details on the concrete computation of a
model output $\Hb \xb$ of Section~\ref{Sec:ModelFacto}. First of all,
there are four different impulse responses whatever the number of
scans. For scans in the same direction, the response is the
same. Thus we can construct four different convolution matrices
$\Hb_i$ for $i=1,2,3,4$ and apply four different discrete convolutions
to the coefficients $\xb$.

We can also deduce the structure of the transpose of the model $\Hb^t
= \Hb_c^t\Pb^t$. The matrix $\Pb^{t}$ is a data summation / zero
padding matrix (addition of the data that possess the same pointing
while setting the other coefficients to zero), and $\Hb_c^t$
corresponds to a convolution with the space reversal impulse
responses.

The product by $\Pb^{t}$ is very similar to the construction of a
naive map except that the data are added rather than averaged. Also,
the operation is done by velocity and not globally. Finally, the
products by $\Hb_c$ and $\Hb_c^t$ are convolutions computed by FFT.

\bibliographystyle{aa}
\bibliography{biben,revuedef,revueabr,base,biblio}

\begin{thebibliography}{49}
\expandafter\ifx\csname natexlab\endcsname\relax\def\natexlab#1{#1}\fi

\bibitem[{{Abergel} {et~al.}(2010){Abergel}, {Arab}, {Compi{\`e}gne}, {Kirk},
  {Ade}, {Anderson}, {Andr{\'e}}, {Baluteau}, {Bernard}, {Blagrave},
  {Bontemps}, {Boulanger}, {Cohen}, {Cox}, {Dartois}, {Davis}, {Emery},
  {Fulton}, {Gry}, {Habart}, {Huang}, {Joblin}, {Jones}, {Lagache}, {Lim},
  {Madden}, {Makiwa}, {Martin}, {Miville-Desch{\^e}nes}, {Molinari}, {Moseley},
  {Motte}, {Naylor}, {Okumura}, {Pinheiro Gon{\c c}alves}, {Polehampton},
  {Rodon}, {Russeil}, {Saraceno}, {Sauvage}, {Sidher}, {Spencer}, {Swinyard},
  {Ward-Thompson}, {White}, \& {Zavagno}}]{Abergel10}
{Abergel}, A., {Arab}, H., {Compi{\`e}gne}, M., {et~al.} 2010, \aap, 518, L96+

\bibitem[{Andrews \& Hunt(1977)}]{Andrews77}
Andrews, H.~C. \& Hunt, B.~R. 1977, Digital Image Restoration (Englewood
  Cliffs, \sca{nj}: Prentice-Hall)

\bibitem[{{Boulanger} {et~al.}(1996){Boulanger}, {Abergel}, {Bernard},
  {Burton}, {Desert}, {Hartmann}, {Lagache}, \& {Puget}}]{Boulanger1996}
{Boulanger}, F., {Abergel}, A., {Bernard}, J., {et~al.} 1996, \aap, 312, 256

\bibitem[{Cantalupo {et~al.}(2010)Cantalupo, Borrill, Jaffe, Kisner, \&
  Stompor}]{Cantalupo10}
Cantalupo, C.~M., Borrill, J.~D., Jaffe, A.~H., Kisner, T.~S., \& Stompor, R.
  2010, The Astrophysical Journal Supplement, 187, 212

\bibitem[{Champagnat {et~al.}(2009)Champagnat, Le~Besnerais, \&
  Kulcs\'ar}]{champagnat09}
Champagnat, F., Le~Besnerais, G., \& Kulcs\'ar, C. 2009, Journal of the Optical
  Society of America A, 26, 1730

\bibitem[{Charbonnier {et~al.}(1997)Charbonnier, Blanc-F\'eraud, Aubert, \&
  Barlaud}]{Charbonnier97}
Charbonnier, P., Blanc-F\'eraud, L., Aubert, G., \& Barlaud, M. 1997,
  \uppercase{ieee} {T}rans. {I}mage {P}rocessing, 6, 298

\bibitem[{Clements {et~al.}(2006)Clements, Chanial, Bendo, Xu, Schulz, Waskett,
  Sibthorpe, \& Laurent}]{clements06}
Clements, D., Chanial, P., Bendo, G., {et~al.} 2006, \textsc{spire} Mapmaking
  Algorithm Review Report, Tech. rep., Astrophysics group at Imperial College
  London

\bibitem[{{de Graauw} {et~al.}(2010){de Graauw}, {Helmich}, {Phillips},
  {Stutzki}, {Caux}, {Whyborn}, {Dieleman}, {Roelfsema}, {Aarts}, {Assendorp},
  {Bachiller}, {Baechtold}, {Barcia}, {Beintema}, {Belitsky}, {Benz}, {Bieber},
  {Boogert}, {Borys}, {Bumble}, {Ca{\"i}s}, {Caris}, {Cerulli-Irelli},
  {Chattopadhyay}, {Cherednichenko}, {Ciechanowicz}, {Coeur-Joly}, {Comito},
  {Cros}, {de Jonge}, {de Lange}, {Delforges}, {Delorme}, {den Boggende},
  {Desbat}, {Diez-Gonz{\'a}lez}, {di Giorgio}, {Dubbeldam}, {Edwards},
  {Eggens}, {Erickson}, {Evers}, {Fich}, {Finn}, {Franke}, {Gaier}, {Gal},
  {Gao}, {Gallego}, {Gauffre}, {Gill}, {Glenz}, {Golstein}, {Goulooze},
  {Gunsing}, {G{\"u}sten}, {Hartogh}, {Hatch}, {Higgins}, {Honingh}, {Huisman},
  {Jackson}, {Jacobs}, {Jacobs}, {Jarchow}, {Javadi}, {Jellema}, {Justen},
  {Karpov}, {Kasemann}, {Kawamura}, {Keizer}, {Kester}, {Klapwijk}, {Klein},
  {Kollberg}, {Kooi}, {Kooiman}, {Kopf}, {Krause}, {Krieg}, {Kramer},
  {Kruizenga}, {Kuhn}, {Laauwen}, {Lai}, {Larsson}, {Leduc}, {Leinz}, {Lin},
  {Liseau}, {Liu}, {Loose}, {L{\'o}pez-Fernandez}, {Lord}, {Luinge}, {Marston},
  {Mart{\'{\i}}n-Pintado}, {Maestrini}, {Maiwald}, {McCoey}, {Mehdi}, {Megej},
  {Melchior}, {Meinsma}, {Merkel}, {Michalska}, {Monstein}, {Moratschke},
  {Morris}, {Muller}, {Murphy}, {Naber}, {Natale}, {Nowosielski}, {Nuzzolo},
  {Olberg}, {Olbrich}, {Orfei}, {Orleanski}, {Ossenkopf}, {Peacock}, {Pearson},
  {Peron}, {Phillip-May}, {Piazzo}, {Planesas}, {Rataj}, {Ravera}, {Risacher},
  {Salez}, {Samoska}, {Saraceno}, {Schieder}, {Schlecht}, {Schl{\"o}der},
  {Schm{\"u}lling}, {Schultz}, {Schuster}, {Siebertz}, {Smit}, {Szczerba},
  {Shipman}, {Steinmetz}, {Stern}, {Stokroos}, {Teipen}, {Teyssier}, {Tils},
  {Trappe}, {van Baaren}, {van Leeuwen}, {van de Stadt}, {Visser}, {Wildeman},
  {Wafelbakker}, {Ward}, {Wesselius}, {Wild}, {Wulff}, {Wunsch}, {Tielens},
  {Zaal}, {Zirath}, {Zmuidzinas}, \& {Zwart}}]{deGraauw10}
{de Graauw}, T., {Helmich}, F.~P., {Phillips}, T.~G., {et~al.} 2010, \aap, 518,
  L6+

\bibitem[{Demoment(1989)}]{demoment89}
Demoment, G. 1989, \uppercase{ieee} {T}rans. {A}coust. {S}peech, {S}ignal
  {P}rocessing, \sca{assp}-37, 2024

\bibitem[{Elad \& Feuer(1999)}]{Elad99}
Elad, M. \& Feuer, A. 1999, \uppercase{ieee} {T}rans. {I}mage {P}rocessing, 8,
  387

\bibitem[{Farsiu {et~al.}(2004)Farsiu, Robinson, Elad, \& Milanfar}]{farsiu04}
Farsiu, S., Robinson, M., Elad, M., \& Milanfar, P. 2004, Image Processing,
  IEEE Transactions on, 13, 1327

\bibitem[{Giovannelli \& Coulais(2005)}]{Giovannelli05}
Giovannelli, J.-F. \& Coulais, A. 2005, {A}stron. {A}strophys., 439, 401

\bibitem[{{Griffin} {et~al.}(2008){Griffin}, {Swinyard}, {Vigroux}, {Abergel},
  {Ade}, {Andr{\'e}}, {Baluteau}, {Bock}, {Franceschini}, {Gear}, {Glenn},
  {Huang}, {Griffin}, {King}, {Lellouch}, {Naylor}, {Oliver}, {Olofsson},
  {Perez-Fournon}, {Page}, {Rowan-Robinson}, {Saraceno}, {Sawyer}, {Wright},
  {Zavagno}, {Abreu}, {Bendo}, {Dowell}, {Dowell}, {Ferlet}, {Fulton},
  {Hargrave}, {Laurent}, {Leeks}, {Lim}, {Lu}, {Nguyen}, {Pearce},
  {Polehampton}, {Rizzo}, {Schulz}, {Sidher}, {Smith}, {Spencer}, {Valtchanov},
  {Woodcraft}, {Xu}, \& {Zhang}}]{Griffin08}
{Griffin}, M., {Swinyard}, B., {Vigroux}, L., {et~al.} 2008, in Presented at
  the Society of Photo-Optical Instrumentation Engineers (SPIE) Conference,
  Vol. 7010, Society of Photo-Optical Instrumentation Engineers (SPIE)
  Conference Series

\bibitem[{Griffin(2006)}]{griffin06}
Griffin, M.~J. 2006, Revised Photometer sensitivity model, working version
  after sensitivity review meeting

\bibitem[{Griffin(2007)}]{griffin07}
---. 2007, The {SPIRE} Analogue Signal Chain and Photometer Detector Data
  Processing Pipeline, Tech. rep., University of Wales Cardiff

\bibitem[{{Griffin} {et~al.}(2010){Griffin}, {Abergel}, {Abreu}, {Ade},
  {Andr{\'e}}, {Augueres}, {Babbedge}, {Bae}, {Baillie}, {Baluteau}, {Barlow},
  {Bendo}, {Benielli}, {Bock}, {Bonhomme}, {Brisbin}, {Brockley-Blatt},
  {Caldwell}, {Cara}, {Castro-Rodriguez}, {Cerulli}, {Chanial}, {Chen},
  {Clark}, {Clements}, {Clerc}, {Coker}, {Communal}, {Conversi}, {Cox},
  {Crumb}, {Cunningham}, {Daly}, {Davis}, {de Antoni}, {Delderfield}, {Devin},
  {di Giorgio}, {Didschuns}, {Dohlen}, {Donati}, {Dowell}, {Dowell}, {Duband},
  {Dumaye}, {Emery}, {Ferlet}, {Ferrand}, {Fontignie}, {Fox}, {Franceschini},
  {Frerking}, {Fulton}, {Garcia}, {Gastaud}, {Gear}, {Glenn}, {Goizel},
  {Griffin}, {Grundy}, {Guest}, {Guillemet}, {Hargrave}, {Harwit}, {Hastings},
  {Hatziminaoglou}, {Herman}, {Hinde}, {Hristov}, {Huang}, {Imhof}, {Isaak},
  {Israelsson}, {Ivison}, {Jennings}, {Kiernan}, {King}, {Lange}, {Latter},
  {Laurent}, {Laurent}, {Leeks}, {Lellouch}, {Levenson}, {Li}, {Li},
  {Lilienthal}, {Lim}, {Liu}, {Lu}, {Madden}, {Mainetti}, {Marliani}, {McKay},
  {Mercier}, {Molinari}, {Morris}, {Moseley}, {Mulder}, {Mur}, {Naylor},
  {Nguyen}, {O'Halloran}, {Oliver}, {Olofsson}, {Olofsson}, {Orfei}, {Page},
  {Pain}, {Panuzzo}, {Papageorgiou}, {Parks}, {Parr-Burman}, {Pearce},
  {Pearson}, {P{\'e}rez-Fournon}, {Pinsard}, {Pisano}, {Podosek}, {Pohlen},
  {Polehampton}, {Pouliquen}, {Rigopoulou}, {Rizzo}, {Roseboom}, {Roussel},
  {Rowan-Robinson}, {Rownd}, {Saraceno}, {Sauvage}, {Savage}, {Savini},
  {Sawyer}, {Scharmberg}, {Schmitt}, {Schneider}, {Schulz}, {Schwartz},
  {Shafer}, {Shupe}, {Sibthorpe}, {Sidher}, {Smith}, {Smith}, {Smith},
  {Spencer}, {Stobie}, {Sudiwala}, {Sukhatme}, {Surace}, {Stevens}, {Swinyard},
  {Trichas}, {Tourette}, {Triou}, {Tseng}, {Tucker}, {Turner}, {Vaccari},
  {Valtchanov}, {Vigroux}, {Virique}, {Voellmer}, {Walker}, {Ward}, {Waskett},
  {Weilert}, {Wesson}, {White}, {Whitehouse}, {Wilson}, {Winter}, {Woodcraft},
  {Wright}, {Xu}, {Zavagno}, {Zemcov}, {Zhang}, \& {Zonca}}]{Griffin10}
{Griffin}, M.~J., {Abergel}, A., {Abreu}, A., {et~al.} 2010, \aap, 518, L3+

\bibitem[{Griffin {et~al.}(2002)Griffin, Bock, \& Gear}]{griffin02}
Griffin, M.~J., Bock, J.~J., \& Gear, W.~K. 2002, {A}pplied {O}ptics, 41, 6543

\bibitem[{Hardie {et~al.}(1997)Hardie, Barnard, \& Armstrong}]{Hardie97}
Hardie, R.~C., Barnard, K.~J., \& Armstrong, E.~E. 1997, \uppercase{ieee}
  {T}rans. {I}mage {P}rocessing, 6, 1621

\bibitem[{Idier(2008)}]{idier08}
Idier, J., ed. 2008, Bayesian Approach to Inverse Problems (London: ISTE Ltd
  and John Wiley \& Sons Inc.)

\bibitem[{K{\"u}nsch(1994)}]{Kunsch94}
K{\"u}nsch, H.~R. 1994, {A}nn. {I}nst. {S}tat. {M}ath., 46, 1

\bibitem[{Leshem {et~al.}(2008)Leshem, Christou, Jeffs, Kuruoglu, \& van~der
  Veen}]{IEEE-SP-ASTRO08}
Leshem, A., Christou, J., Jeffs, B.~D., Kuruoglu, E., \& van~der Veen, A.~J.
  2008, IEEE Journal of Selected Topics in Signal Processing, 2

\bibitem[{Leshem {et~al.}(2010)Leshem, Kamalabadi, Kuruoglu, \& van~der
  Veen}]{IEEE-SPM-ASTRO10}
Leshem, A., Kamalabadi, F., Kuruoglu, E., \& van~der Veen, A.-J. 2010, Signal
  Processing Magazine, 27

\bibitem[{{Miville-Desch{\^e}nes} {et~al.}(2010){Miville-Desch{\^e}nes},
  {Martin}, {Abergel}, {Bernard}, {Boulanger}, {Lagache}, {Anderson},
  {Andr{\'e}}, {Arab}, {Baluteau}, {Blagrave}, {Bontemps}, {Cohen},
  {Compiegne}, {Cox}, {Dartois}, {Davis}, {Emery}, {Fulton}, {Gry}, {Habart},
  {Huang}, {Joblin}, {Jones}, {Kirk}, {Lim}, {Madden}, {Makiwa}, {Menshchikov},
  {Molinari}, {Moseley}, {Motte}, {Naylor}, {Okumura}, {Pinheiro Gon{\c
  c}alves}, {Polehampton}, {Rod{\'o}n}, {Russeil}, {Saraceno}, {Schneider},
  {Sidher}, {Spencer}, {Swinyard}, {Ward-Thompson}, {White}, \&
  {Zavagno}}]{MAMD10}
{Miville-Desch{\^e}nes}, M., {Martin}, P.~G., {Abergel}, A., {et~al.} 2010,
  \aap, 518, L104+

\bibitem[{Molina \& Ripley(1989)}]{Molina89}
Molina, R. \& Ripley, B.~D. 1989, {J}. {A}ppl. {S}tatistics, 16, 193

\bibitem[{Mugnier {et~al.}(2004)Mugnier, Fusco, \& Conan}]{Mugnier04}
Mugnier, L., Fusco, T., \& Conan, J.-M. 2004, {J}. {O}pt. {S}oc. {A}mer., 21,
  1841

\bibitem[{Nguyen {et~al.}(2001)Nguyen, Milanfar, \& Golub}]{Nguyen01}
Nguyen, N., Milanfar, P., \& Golub, G. 2001, \uppercase{ieee} {T}rans. {I}mage
  {P}rocessing, 10, 573

\bibitem[{Nocedal \& Wright(2000)}]{Nocedal00}
Nocedal, J. \& Wright, S.~J. 2000, Numerical Optimization, Series in Operations
  Research (New York: Springer Verlag)

\bibitem[{Orieux(2009)}]{Orieux09b}
Orieux, F. 2009, PhD thesis, Universit\'e Paris-Sud 11

\bibitem[{Orieux {et~al.}(2010)Orieux, Giovannelli, \& Rodet}]{orieux10a}
Orieux, F., Giovannelli, J.-F., \& Rodet, T. 2010, J. Opt. Soc. Am. A, 27, 1593

\bibitem[{Orieux {et~al.}(2009)Orieux, Rodet, \& Giovannelli}]{orieux09a}
Orieux, F., Rodet, T., \& Giovannelli, J.-F. 2009, in Proc. of IEEE
  International Conference on Image Processing (ICIP 2009), Cairo, Egypt

\bibitem[{{Ott}(2010)}]{2010ASPC..434..139O}
{Ott}, S. 2010, in Astronomical Society of the Pacific Conference Series, Vol.
  434, Astronomical Data Analysis Software and Systems XIX, ed. {Y.~Mizumoto,
  K.-I.~Morita, \& M.~Ohishi}, 139

\bibitem[{Park {et~al.}(2003)Park, Park, \& Kang}]{Park03}
Park, S.~C., Park, M.~K., \& Kang, M.~G. 2003, \uppercase{ieee} {T}rans.
  {S}ignal {P}rocessing {M}ag., 21

\bibitem[{Patanchon {et~al.}(2008)Patanchon, Ade, Bock, Chapin, Devlin, Dicker,
  Griffin, Gundersen, Halpern, Hargrave, Hughes, Klein, Marsden, Martin,
  Mauskopf, Netterfield, Olmi, Pascale, Rex, Scott, Semisch, Truch, Tucker,
  Tucker, Viero, \& Wiebe}]{Patanchon07}
Patanchon, G., Ade, P. A.~R., Bock, J.~J., {et~al.} 2008, The Astrophysical
  Journal, 681, 708

\bibitem[{Patti {et~al.}(1997)Patti, Sezan, \& Tekalp}]{Patti97}
Patti, A.~J., Sezan, M.~I., \& Tekalp, A.~M. 1997, \uppercase{ieee} {T}rans.
  {I}mage {P}rocessing, 6, 1064

\bibitem[{{Pilbratt} {et~al.}(2010){Pilbratt}, {Riedinger}, {Passvogel},
  {Crone}, {Doyle}, {Gageur}, {Heras}, {Jewell}, {Metcalfe}, {Ott}, \&
  {Schmidt}}]{Pilbratt10}
{Pilbratt}, G.~L., {Riedinger}, J.~R., {Passvogel}, T., {et~al.} 2010, \aap,
  518, L1+

\bibitem[{{Poglitsch} {et~al.}(2010){Poglitsch}, {Waelkens}, {Geis},
  {Feuchtgruber}, {Vandenbussche}, {Rodriguez}, {Krause}, {Renotte}, {van
  Hoof}, {Saraceno}, {Cepa}, {Kerschbaum}, {Agn{\`e}se}, {Ali}, {Altieri},
  {Andreani}, {Augueres}, {Balog}, {Barl}, {Bauer}, {Belbachir}, {Benedettini},
  {Billot}, {Boulade}, {Bischof}, {Blommaert}, {Callut}, {Cara}, {Cerulli},
  {Cesarsky}, {Contursi}, {Creten}, {De Meester}, {Doublier}, {Doumayrou},
  {Duband}, {Exter}, {Genzel}, {Gillis}, {Gr{\"o}zinger}, {Henning},
  {Herreros}, {Huygen}, {Inguscio}, {Jakob}, {Jamar}, {Jean}, {de Jong},
  {Katterloher}, {Kiss}, {Klaas}, {Lemke}, {Lutz}, {Madden}, {Marquet},
  {Martignac}, {Mazy}, {Merken}, {Montfort}, {Morbidelli}, {M{\"u}ller},
  {Nielbock}, {Okumura}, {Orfei}, {Ottensamer}, {Pezzuto}, {Popesso},
  {Putzeys}, {Regibo}, {Reveret}, {Royer}, {Sauvage}, {Schreiber}, {Stegmaier},
  {Schmitt}, {Schubert}, {Sturm}, {Thiel}, {Tofani}, {Vavrek}, {Wetzstein},
  {Wieprecht}, \& {Wiezorrek}}]{Poglitsch10}
{Poglitsch}, A., {Waelkens}, C., {Geis}, N., {et~al.} 2010, \aap, 518, L2+

\bibitem[{Robert \& Casella(2000)}]{Robert04}
Robert, C.~P. \& Casella, G. 2000, {M}onte-{C}arlo Statistical Methods,
  Springer Texts in Statistics (New York, \sca{ny}: Springer)

\bibitem[{Rochefort {et~al.}(2006)Rochefort, Champagnat, Le~Besnerais, \&
  Giovannelli}]{rochefort06}
Rochefort, G., Champagnat, F., Le~Besnerais, G., \& Giovannelli, J.-F. 2006,
  \uppercase{ieee} {T}rans. {I}mage {P}rocessing, 15, 3325

\bibitem[{Rodet {et~al.}(2008)Rodet, Orieux, Giovannelli, \& Abergel}]{rodet08}
Rodet, T., Orieux, F., Giovannelli, J.-F., \& Abergel, A. 2008,
  \uppercase{ieee} {J}. of {S}elec. {T}opics in {S}ignal {P}roc., 2, 802

\bibitem[{Shewchuk(1994)}]{Shewchuck94}
Shewchuk, J.~R. 1994, An Introduction to the Conjugate Gradient Method Without
  the Agonizing Pain, Tech. rep., Carnegie Mellon University

\bibitem[{Sibthorpe {et~al.}(2009)Sibthorpe, Chanial, \& Griffin}]{sibthorpe09}
Sibthorpe, B., Chanial, P., \& Griffin, M.~J. 2009, ArXiv e-prints

\bibitem[{Sibthorpe \& Griffin(2006)}]{sibthorpe06}
Sibthorpe, B. \& Griffin, M.~J. 2006, Spire Photometer Simulator, Tech. rep.,
  University of Wales Cardiff

\bibitem[{Sudiwala {et~al.}(2002)Sudiwala, Griffin, \& Woodcraft}]{sudiwala02}
Sudiwala, R.~V., Griffin, M.~J., \& Woodcraft, A.~L. 2002, International
  Journal of Infrared and Millimeter Waves, 23, 545

\bibitem[{Thi\'ebaut(2008)}]{Thiebaut08}
Thi\'ebaut, E. 2008, in proc. {SPIE}: Astronomical Telescopes and
  Instrumentation, Vol. 7013, 70131--I

\bibitem[{Tikhonov \& Arsenin(1977)}]{Tikhonov77}
Tikhonov, A. \& Arsenin, V. 1977, Solutions of Ill-Posed Problems (Washington,
  \sca{dc}: Winston)

\bibitem[{Twomey(1962)}]{Twomey62}
Twomey, S. 1962, J. Assoc. Comp. Mach., 10, 97

\bibitem[{Vandewalle {et~al.}(2007)Vandewalle, Sbaiz, Vandewalle, \&
  Vetterli}]{Vandewalle07}
Vandewalle, P., Sbaiz, L., Vandewalle, J., \& Vetterli, M. 2007,
  \uppercase{ieee} {T}rans. {S}ignal {P}rocessing, 55, 3687

\bibitem[{{Wieprecht} {et~al.}(2009){Wieprecht}, {Schreiber}, {de Jong},
  {Jacobson}, {Liu}, {Morien}, {Wetzstein}, {Ali}, {Frayer}, {Lutz}, {Okumura},
  {Popesso}, \& {Sauvage}}]{Wieprecht09}
{Wieprecht}, E., {Schreiber}, J., {de Jong}, J., {et~al.} 2009, in Astronomical
  Society of the Pacific Conference Series, Vol. 411, Astronomical Society of
  the Pacific Conference Series, ed. {D.~A.~Bohlender, D.~Durand, \&
  P.~Dowler}, 531--+

\bibitem[{Woods {et~al.}(2006)Woods, Galatsanos, \& Katsaggelos}]{Woods06}
Woods, N.~A., Galatsanos, N.~P., \& Katsaggelos, A.~K. 2006, \uppercase{ieee}
  {T}rans. {I}mage {P}rocessing, 15, 201

\end{thebibliography}

\end{document}
